\documentclass[preprint,12pt,authoryear]{elsarticle}

\usepackage{amssymb}
\usepackage{amsthm}
\usepackage{amsmath}
\RequirePackage[colorlinks,citecolor=blue,urlcolor=blue]{hyperref}

\usepackage[english]{babel}
\usepackage{stackrel}
\usepackage{amsfonts}
\usepackage{amssymb}
\usepackage{epstopdf}
\usepackage{caption}
\usepackage{subfigure}   
\usepackage{stmaryrd}
\usepackage{mathtools}
\usepackage{bbold}

\numberwithin{equation}{section}
\theoremstyle{plain}
\newtheorem{proposition}{Proposition}[section]
\begin{document}

\begin{frontmatter}



\title{A new Universal Resample Stable Bootstrap-based Stopping Criterion in PLS Components Construction}
%\runtitle{A new Bootstrap-based Stopping Criterion}


\author[adr1,adr2]{J\'er\'emy Magnanensi\corref{cor1}}\ead{magnanensi@math.unistra.fr}
\address[adr1]{Institut de Recherche Math\'ematique Avanc\'ee, UMR 7501, LabEx IRMIA\\
Universit\'e de Strasbourg et CNRS, France}
\address[adr2]{Laboratoire de Biostatistique et Informatique M\'edicale, Facult\'e de M\'edecine, EA3430\\
 Universit\'e de Strasbourg et CNRS, France}
\cortext[cor1]{Corresponding author, 7, Rue Ren\'e Descartes 67084 Strasbourg Cedex, Universit\'e de Strasbourg et CNRS, France} 
\author[adr1]{Fr\'ed\'eric Bertrand}\ead{fbertrand@math.unistra.fr}
%\address{Institut de Recherche Math\'ematique Avanc\'ee, UMR 7501, LabEx IRMIA\\
 %Universit\'e de Strasbourg et CNRS, France\\ \printead{e2}}
\author[adr1]{Myriam Maumy-Bertrand}\ead{mmaumy@math.unistra.fr}
%\address{Institut de Recherche Math\'ematique Avanc\'ee, UMR 7501, LabEx IRMIA\\
 %Universit\'e de Strasbourg et CNRS, France\\ \printead{e3}}
\author[adr2]{Nicolas Meyer}\ead{nmeyer@unistra.fr}
%\address{Laboratoire de Biostatistique et Informatique M\'edicale, Facult\'e de M\'edecine, EA3430\\
% Universit\'e de Strasbourg, France\\ \printead{e4}}

\begin{abstract}
We develop a new robust stopping criterion in Partial Least Squares Regressions (PLSR) components construction characterised by a high level of stability. This new criterion is defined as a universal one since it is suitable both for PLSR and its extension to Generalized Linear Regressions (PLSGLR). This criterion is based on a non-parametric bootstrap process and has to be computed algorithmically. It allows to test each successive components on a preset significant level $\alpha$. In order to assess its performances and robustness with respect to different noise levels, we perform intensive datasets simulations, with a preset and known number of components to extract, both in the case $n>p$ ($n$ being the number of subjects and $p$ the number of original predictors), and for datasets with $n<p$.  We then use \textit{t}-tests to compare the performance of our approach to some others classical criteria. The property of robustness is particularly tested through resampling processes on a real allelotyping dataset. Our conclusion is that our criterion presents also better global predictive performances, both in the PLSR and PLSGLR (Logistic and Poisson) frameworks. 

\end{abstract}


\end{frontmatter}

\section{Introduction}
\label{intr}

Modelling relations by performing usual linear regressions between a univariate response $\mathbf{y}=\left(y_1,\ldots,y_n\right)^T\in\mathbb{R}^{n\times 1}$, $\left(.\right)^T$ representing the transpose, and highly correlated predictors $\mathbf{X}=\left(\mathbf{x}_1,\ldots,\mathbf{x}_p\right)\in\mathbb{R}^{n\times p}$, with $n$ the number of subjects and $p$ the number of predictors, or on datasets including more predictors than subjects,  is not suitable or even possible. However, with the huge technological and computer science advances, providing consistent analysis of such kinds of datasets has become a major challenge, particularly in domains such as medicine, biology or chemistry. To deal with such datasets, statistical methods have been developed, especially the PLS Regression (PLSR) which was first introduced by \citet{wold1983multivariate} and \citet{Wold:1984} and described precisely by \citet{hoskuldsson1988pls} and \citet{wold2001pls}.

PLSR consists in building $K\leqslant \mbox{rk}(\mathbf{X})$ orthogonal ``latent'' variables $\mathbf{T}_K=\left(\mathbf{t}_1,\ldots,\mathbf{t}_K\right)$, also called components, in such a way that $\mathbf{T}_K$ describes optimally the common information space between $\mathbf{X}$ and $\mathbf{y}$. Thus, these components are build up as linear combinations of the original predictors vectors, in order to maximise $\mbox{Cov}\left(\mathbf{y},\mathbf{t}_k\left|\mathbf{T}_{k-1}\right.\right)$ so that:
\begin{equation}
   \mathbf{t}_k=\mathbf{X}\mathbf{w}_k^{*}=\stackrel[j=1]{p}{\sum{}}w^{*}_{jk}\mathbf{x}_j,\:1\leqslant k\leqslant K
\end{equation}
where $\mathbf{w}_k^{*}=\left({w_{1k}}^{*},\ldots,{w_{pk}}^{*}\right)^T$ is the vector of predictors weights, depending on $\mathbf{y}$, in the $k^{th}$ component \citep{wold2001pls}.

Let $K$ be the number of components. The final regression model is thus the following one:
\begin{align}
   \mathbf{y}&=\stackrel[k=1]{K}{\sum{}}c_{k}\mathbf{t}_k+\epsilon \\
	           &=\stackrel[k=1]{K}{\sum{}}c_{k}\left(\stackrel[j=1]{p}{\sum{}}w^{*}_{jk}\mathbf{x}_j\right)+\epsilon\\
						 &=\stackrel[j=1]{p}{\sum{}}\beta_j^{PLS}\mathbf{x}_j+\epsilon,
				\label{eq3}
\end{align}
with $\mathbf{\epsilon}=\left(\epsilon_1,\ldots,\epsilon_n\right)^{T}$ the $n$ by 1 error vector $\mathbf{\epsilon}_i$, verifying $\mathbb{E}\left(\epsilon\left|\mathbf{T}_K\right.\right)=0_n$, $\mathbb{V}\mbox{ar}\left(\epsilon\left|\mathbf{T}_K\right.\right)=\sigma_\epsilon^2\times Id_n$ and $\left(c_1,\ldots,c_K\right)$ the coefficients of regression of $\mathbf{y}$ on $\mathbf{T}_K$.

This model is not a linear one since the hat matrix $\mathbf{H}_K=\mathbf{T}_K\left(\mathbf{T}_K^T\mathbf{T}_K\right)^{-1}\mathbf{T}_K^{T}$ does not only depend on $\mathbf{X}$ but also on $\mathbf{y}$. Thus, in almost all cases:
\begin{equation}
\mathbb{E}\left(\mathbf{y}\left|\mathbf{X}\right.\right)\neq\stackrel[k=1]{K}{\sum{}}c_{k}\left(\stackrel[j=1]{p}{\sum{}}w^{*}_{jk}\mathbf{x}_j\right).
\end{equation}

An extension to Generalized Linear Regression models, noted PLSGLR, has been developed by \citet{bastien2005pls}, with the aim of taking into account the specific distribution of $\mathbf{y}$. In this context, the regression model is the following one:\\
\begin{equation}
   g(\theta)=\stackrel[k=1]{K}{\sum{}}c_k\left(\stackrel[j=1]{p}{\sum{}}w^{*}_{jk}\mathbf{x}_j\right),
\end{equation}
with $\theta$ the conditional expected value of $\mathbf{y}$ for a continuous distribution or the probability vector of a discrete law with a finite support. The link function $g$ depends on the distribution of $\mathbf{y}$.

Since $k\leqslant \mbox{rk}\left(\mathbf{X}\right)$ and $\mbox{Corr}\left(\mathbf{t}_{k_1},\mathbf{t}_{k_2}\right)=0,\forall \left(k_1,k_2\right)\in[\![1,\ldots,K]\!]^2$, issues concerning the ill-conditioned matrix $\mathbf{X}$ are solved, so that usual linear regressions can be used to estimate the $c_{k}$ parameters. However, the determination of the optimal number of components $K$, which is equal to the exact dimension of the link between $\mathbf{X}$ and $\mathbf{y}$, is crucial to obtain correct estimations of the original predictors coefficients. Indeed, concluding $K_1<K$ leads to a loss of information so that links between some predictors and $\mathbf{y}$ will not be correctly modelled. Concluding $K_2>K$ involves that useless information in $\mathbf{X}$ will be used to model knowledge in $\mathbf{y}$, which leads to overfitting.

In this article, we will first remind the reader about the most used existing criteria and introduce our adaptation of the so called bootstrapping pairs to the PLSR and PLSGLR as a new stopping criterion in part \ref{2}. We will also describe the global algorithm we developed to compare these different criteria through intensive datasets simulations. The aim is to analyse results we obtain with these different criteria depending on some increasing noise levels we insert both in $\mathbf{X}$, through an additional component, and in $\mathbf{y}$ by adding a supplementary centred variable. Indeed, it is a common situation in domains such as chemistry or medicine, especially in allelotyping datasets, leading to important issues concerning the extraction of a reliable number of components. In part \ref{3}, we will analyse the results we obtain in the PLSR framework before analysing PLSGLR results for the Logistic Regressions (PLS-LR) and the Poisson Regression (PLS-PR) in part \ref{4}. Results are treated for both $n>p$ and $n<p$ cases. In part \ref{5}, we will treat a real allelotyping dataset by comparing the number of components extracted by the different applicable criteria. We will also compare the robustness of our new bootstrap-based criterion through resampling processes, by approximating the distribution of these extracted number of components. Finally, in part \ref{6}, we will discuss the observed advantages and defaults of each criterion.

\section{Criteria compared through simulations}
\label{2}
\subsection{Existing criteria used for comparison} 
\label{21}

\begin{itemize}
\item In PLSR:
\begin{enumerate}
   \item $\mathbf{Q^{2}.}$ This criterion is obtained by Cross-Validation (CV). We decide to check results either by a leave-one-out CV, $i.e.$ $q=n$ with $q$ the number of parts the dataset is divided into, and with $q$ chosen equal to five ($5$-CV) according to results obtained by \citet{trevor2001elements} and \citet{kohavi1995study}.
	For each new component $\mathbf{t}_k$, the following value is calculated:
	\begin{center}
	  $Q^{2}_{k}=1-\frac{PRESS_{k}}{RSS_{k-1}},$
	\end{center}
with $RSS_{k-1}$ the Residual Sum of Squares of the $k-1$ components model and $PRESS_{k}$ the PREdictive residual Sum of Squares, calculated by CV, of the $k$ components model. \citet{Tenenhaus} considers that a new component $\mathbf{t}_k$ improve significantly the prediction of $\mathbf{y}$ if:
 $$\sqrt{PRESS_k}\leqslant0.95\sqrt{RSS_{k-1}} \:\Longleftrightarrow\: Q^{2}_{k}\geqslant0.0975.$$
Results linked to these two criteria will be graphically and respectively reported as \textbf{Q2lv1o} and \textbf{Q2K5}.
	\item $\mathbf{BICdof.}$ As mentionned in the introduction, the hat matrix $\mathbf{H}_K$ does not only depend on $\mathbf{X}$, but also on $\mathbf{y}$, so that corrected degrees of freedom (\textit{dof}) have to be used in the expression of the BIC criterion \citep{schwarzBIC}. \citet{kramer2011degrees} define this \textit{dof} correction in the PLSR framework (without missing data) and apply it to the BIC criterion. We used the \textit{R} package \textit{plsdof}, based on \citet{kramer2011degrees} work, where this criterion is implemented as follow:
	  $$\mbox{BICdof}=RSS/n+log(n)(\gamma/n)\widehat{\sigma_\epsilon}^{2}.$$
		where $\gamma$ is the degree of freedom (\textit{dof}) of the model \eqref{eq3} and $\widehat{\sigma_\epsilon}^{2}$ is defined by \citet{kramer2011degrees}.\\
		The selected model is the one which realize the first local minimum of this BICdof criterion. We also extract results by retaining models realizing the global minimum and will refer to them as \textbf{BICglob}. 
	\end{enumerate}
	
\item In PLSGLR:
\begin{enumerate}
 \item $\mathbf{AIC.}$ The AIC criterion \citep{akaikeAIC} can be computed, whatever is the distribution involved. However, no corrected \textit{dof} have been determined in this PLSGLR framework.
 \item $\mathbf{BIC.}$ As for the AIC, the BIC is calculable without corrected \textit{dof}.
 \item $\mathbf{CV-MClassed.}$ This criterion could only be used for PLS-LR. Through a 5-CV, it determines for each model the number of predicted miss-classed values. The selected model is the one linked to the minimal value of this criterion. 
 \item $\mathbf{p\_val.}$ \citet{bastien2005pls} define a new component $\mathbf{t}_k$ as non-significant if there is not any significant predictors within it. An asymptotic Wald test is used to conclude to the significance of the different predictors.  
\end{enumerate}
\end{itemize}

\subsection{Bootstrap based criterion}
\label{22}
\subsubsection{Motivations, assumption and definitions}
\label{221}

All the criteria described just above have major flaws including arbitrary bounds dependency, results based on asymptotic laws or derived from $q$-CV which naturally depends on the value of $q$ and on the way the group will be randomly composed.\\
For this purpose, we adapted a non-parametric bootstrap technique in order to test directly, with some confidence level $\left(1-\alpha\right)$, the significance of the different coefficients $c_k$ by extracting confidence intervals (CI) for each of them. 

By maximizing $\mbox{Cov}\left(\mathbf{y},\mathbf{t}_k\left|\mathbf{T}_{k-1}\right.\right)$, each new component $\mathbf{t}_k$ represents the best choice of a new dimensional space direction, extracted from the remaining information, explaining at best $\mathbf{y}-\mathbb{E}\left(\mathbf{y}\left|\mathbf{T}_{k-1}\right.\right)$. This process implies the following result.
\begin{proposition}
Let $\mathbf{y}_{(k-1)}$ and $\mathbf{X}_{(k-1)}$ respectively be the deflated response vector and predictors matrix on $k-1$ components. Suppose that,\\
 $\forall k\in [\![1,K]\!], \exists i\in [\![1,p]\!], \mathbf{x}^T_{i,(k-1)}\mathbf{y}_{(k-1)} \ne 0$, \\
then the PLS components building process implies that:

$\forall k\in [\![1,K]\!],\;c_k>0$ and, conditionally to $\mathbf{X}$, $c_k$ is linked to a positive distribution.   
\label{prop1}
\end{proposition}

Bootstrapping pairs $\left(\mathbf{y},\mathbf{X}\right)$ in order to test the significance of these parameters is not suitable since it will approach the conditional distribution given $\mathbf{X}$. Thus, to test each new component, we approach the conditional distribution of these coefficients given $\mathbf{T}_{k}$ by bootstrapping pairs $\left(\mathbf{y},\mathbf{T}_k\right)$.

We define the significance of a new component as resulting from its significance for both $\mathbf{y}$ and $\mathbf{X}$, so that the extracted number of component $k$ is defined as the last one which is significant for both of them.   

\subsubsection{Adapted bootstrapping pairs as a new stopping criterion}
\label{222}

The so-called bootstrapping pairs was introduced by \citet{freedman1981bootstrapping}. This technique only relies on the assumption that the originals pairs $\left(y_i,\mathbf{t}_{i\bullet}\right)$, where $\mathbf{t}_{i\bullet}$ represents the $i^{th}$ row of $\mathbf{T}_k$, are randomly sampled from some unknown $(k+1)$-dimensional distribution. This technique was developed to treat the so-called correlation models, in which predictors are considered as random and $\epsilon$ may be related to them. Thus, it is appropriate to ``estimate the regression plane for a certain population on the basis of a simple random sample'' \citep[p.1219]{freedman1981bootstrapping}.\\ 
Based on our definition of the significance of a new component, we adapted this method by designing a double bootstrapping pairs algorithmic implementation. The first step consists in bootstrapping pairs $\left(\mathbf{X},\mathbf{T}_k\right)$, leading to a maximal number $k_{max}$ of components which can be extracted. The second step consists in bootstrapping pairs $\left(\mathbf{y},\mathbf{T}_k\right)$ to test the significance of each successive component $\mathbf{t}_k$, with $k\leqslant k_{max}$. To avoid confusions between the number of predictors and the coefficients of the regressions of $\mathbf{X}$ on $\mathbf{T}_k$, we set $m$ as the total number of predictors.\\
The algorithm of this double bootstrapping pairs implementation is designed as follow :
\begin{itemize}
\item Bootstrapping $\left(\mathbf{X},\mathbf{T}_k\right)$:
let $k=1$ and $l=1,\ldots,m$.

\begin{enumerate}
 \item Compute the $k$ first components $\left(\mathbf{t}_1,\ldots,\mathbf{t}_k\right)$.
 \item Bootstrap pair $\left(\mathbf{X},\mathbf{T}_k\right)$, returning $R$ bootstrap samples:
\begin{center}
 $\left(\mathbf{X},\mathbf{T}_k\right)^{b_1},\ldots,\left(\mathbf{X},\mathbf{T}_k\right)^{b_R}$.
\end{center}
 \item For each $\left(\mathbf{X},\mathbf{T}_k\right)^{b_r}$, do $m$ Ordinary Least Squares (OLS) regressions:
\begin{center} 
$\mathbf{x}_l^{b_r}=\stackrel[j=1]{k}{\sum{}}\left(\hat{p_{lj}}^{b_r}.\mathbf{t}_j^{b_r}\right)+\hat{\delta}_{lk}^{b_r}$.
\end{center}
	\item $\forall p_{lk}$, construct a $\left(100\times\left(1-\alpha\right)\right)$\% bilateral $BC_a$ CI, noted:
\begin{center}
$\mbox{CI}_l=\left[\mbox{CI}_{l,1}^{k},\mbox{CI}_{l,2}^{k}\right]$.
\end{center}
	\item \textbf{If} $\exists l\in\left\{1,\ldots,m\right\}$, $0\notin \mbox{CI}_l$, \textbf{then} $k=k+1$ and return to step 1. \textbf{Else}, $k_{max}=k-1$.
\end{enumerate}

\item Bootstrapping $\left(\mathbf{y},\mathbf{T}_k\right)$:
let $k=1$. Note that for PLSGLR, a generalized regression is performed at step 3.

\begin{enumerate}
 \item Compute the $k$ first components $\left(\mathbf{t}_1,\ldots,\mathbf{t}_k\right)$.
 \item Bootstrap pair $\left(\mathbf{y},\mathbf{T}_k\right)$, returning $R$ bootstrap samples:
 \begin{center}
 $\left(\mathbf{y},\mathbf{T}_k\right)^{b_1},\ldots,\left(\mathbf{y},\mathbf{T}_k\right)^{b_R}$.
 \end{center}
 \item For each pair $\left(\mathbf{y},\mathbf{T}_k\right)^{b_r}$, do the OLS regression:
 \begin{center}
 $\mathbf{y}^{b_r}=\stackrel[j=1]{k}{\sum{}}\left(\hat{c}_j^{b_r}.\mathbf{t}_j^{b_r}\right)+\hat{\epsilon}_k^{b_r}$.
 \end{center}
	\item Since $c_k>0$, construct a $\left(100\times\left(1-\alpha\right)\right)$\% unilateral $BC_a$ CI:
	\begin{center}
	$\mbox{CI}=\left[\mbox{CI}_1^{k},+\infty\right[$ for $c_k$.
	\end{center}
	\item \textbf{While} $\mbox{CI}_1^{k}>0$ and $k\leqslant k_{max}$, \textbf{do} $k=k+1$, and return to step 1. \textbf{Else}, the final extracted number of components is $K=k-1$.
\end{enumerate}
\end{itemize}

Results linked to this bootstrap-based criterion will be graphically reported as \textbf{BootYT}.

\subsubsection{Illustration of CV issues: first applications on real datasets}
\label{223}

\begin{figure*}[ht]
       \centering
          \subfigure{\includegraphics[trim = 0cm 0cm 0cm 0cm, clip,scale=0.3]{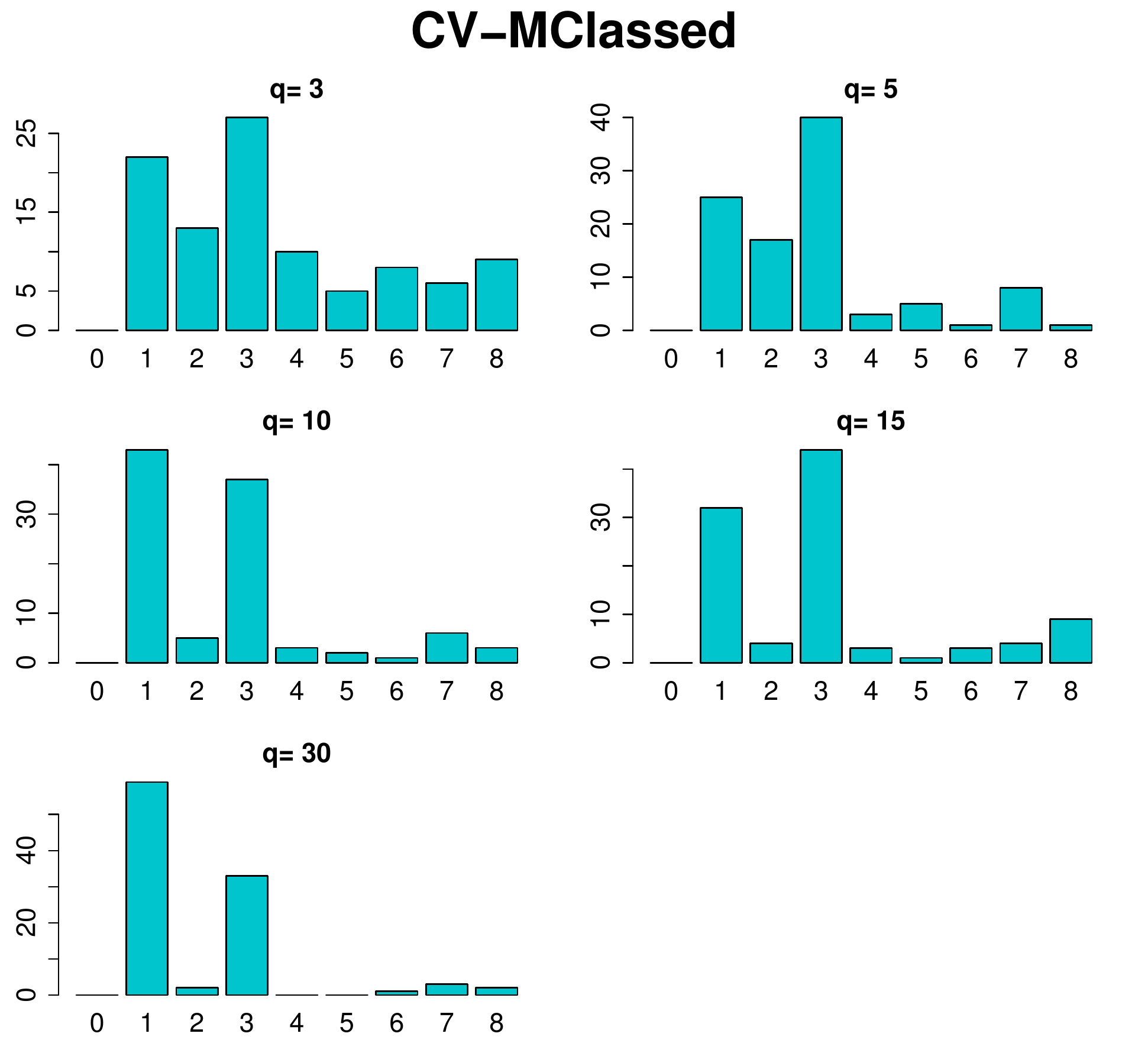}}
          \subfigure{\includegraphics[trim = -2cm 0cm 0cm 0cm, clip,scale=0.3]{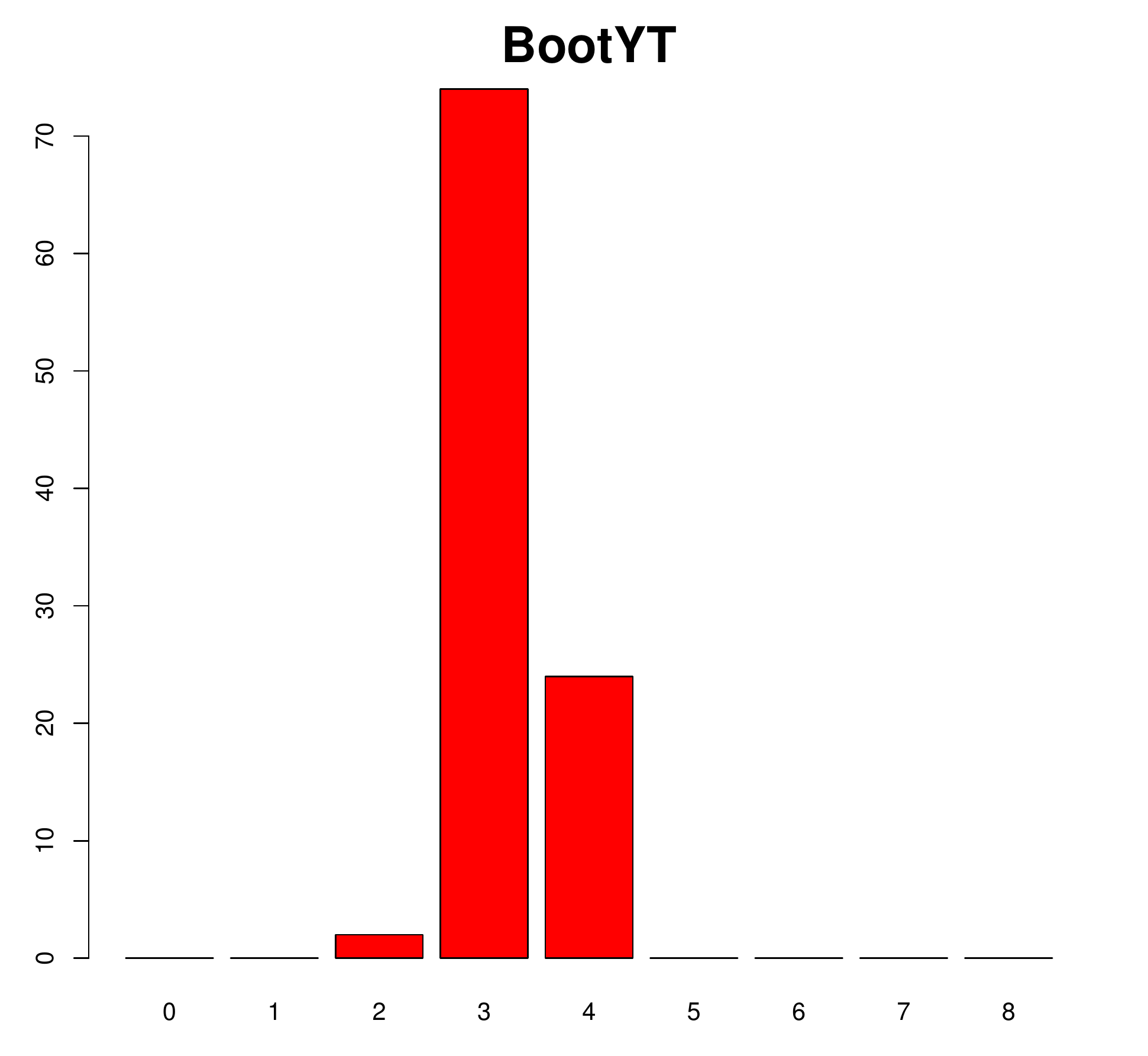}}\\
					\caption{Extracted numbers of component using $q$-folds CV-MClassed (Left) and BootYT (Right) criteria}
   \label{fig:1a}
\end{figure*}

As mentioned by \citet{BoulesteixCV14}, important issues concerning the stability of the $q$-fold CV procedure for the choice of tuning parameters, here the number of components, have been observed. These issues are directly induced by the value of $q$ and by the random character of this resampling-based procedure while splitting the original dataset into two distinct sets, a training one and a test one. To illustrate consequences on the tuning parameter, we treated two real datasets.\\
The first dataset was collected on patients carrying a colon adenocarcinoma. It has 104 observations on 33 binary qualitative explanatory variables and one binary response variable representing the cancer stage according to the Astler-Coller (AB vs CD) classification \citep{astler1954prognostic}. This binary response lead us to perform PLS-Logistic regressions. This dataset, named \textit{aze\_compl}, is available in the $R$ package \textit{plsRglm} \citep{Bertrand2014}.\\
We performed a 100 times the determination of the number of components using the CV-MClassed criterion with $q\in\{3,5,10,15,30\}$. Then, we performed the same process by using our new bootstrap-based criterion. Results are reported in Fig.\ref{fig:1a}. 

Results obtained through $q$-fold CV, with $q\ne n$, and displayed on Fig.\ref{fig:1a} are typical examples of these issues. Depending on the choice of $q$ and on the way the different fold are split, the extracted number of components could be dramatically different. In addition, obtaining a complete distribution of the number of components is mostly impossible, due to the high number of different possibilities for splitting the original datasets into $q$ groups.

\begin{proposition}
Let $n=pq+r,\:0\leqslant r\leqslant q-1$ be the euclidean division of $n$ by $q$. Then, the number of distinct partitions of the original dataset into $r$ $\left(p+1\right)$-element subsets and $\left(q-r\right)$ $p$-element subsets for a CV does not depend on the order of their placement, and is equal to:
\begin{equation}
f(n,q)=\frac{n!}{r!\left(q-r\right)!}\times\left(\frac{1}{\left(p+1\right)!}\right)^r\times\left(\frac{1}{p!}\right)^{q-r}
\end{equation} 
\label{prop2}
\end{proposition}
 
The leave-one-out CV, which is the only complete CV ($f(n,n)=1$, \textit{ie} only one possibility for choosing $n$ folds out of $n$ subjects), concludes to one component to extract. However, it suffers from variance issues concerning the bias-variance tradeoff on the estimation of the prediction error \citep{trevor2001elements} \citep{kohavi1995study}. Our new bootstrap-based criterion is more stable on this dataset and allows the user to choose the number of components, in  this case three, through a more accurate process.

\begin{figure*}[ht]
       \centering
          \subfigure{\includegraphics[trim = 0cm 0cm 0cm 0cm, clip,scale=0.3]{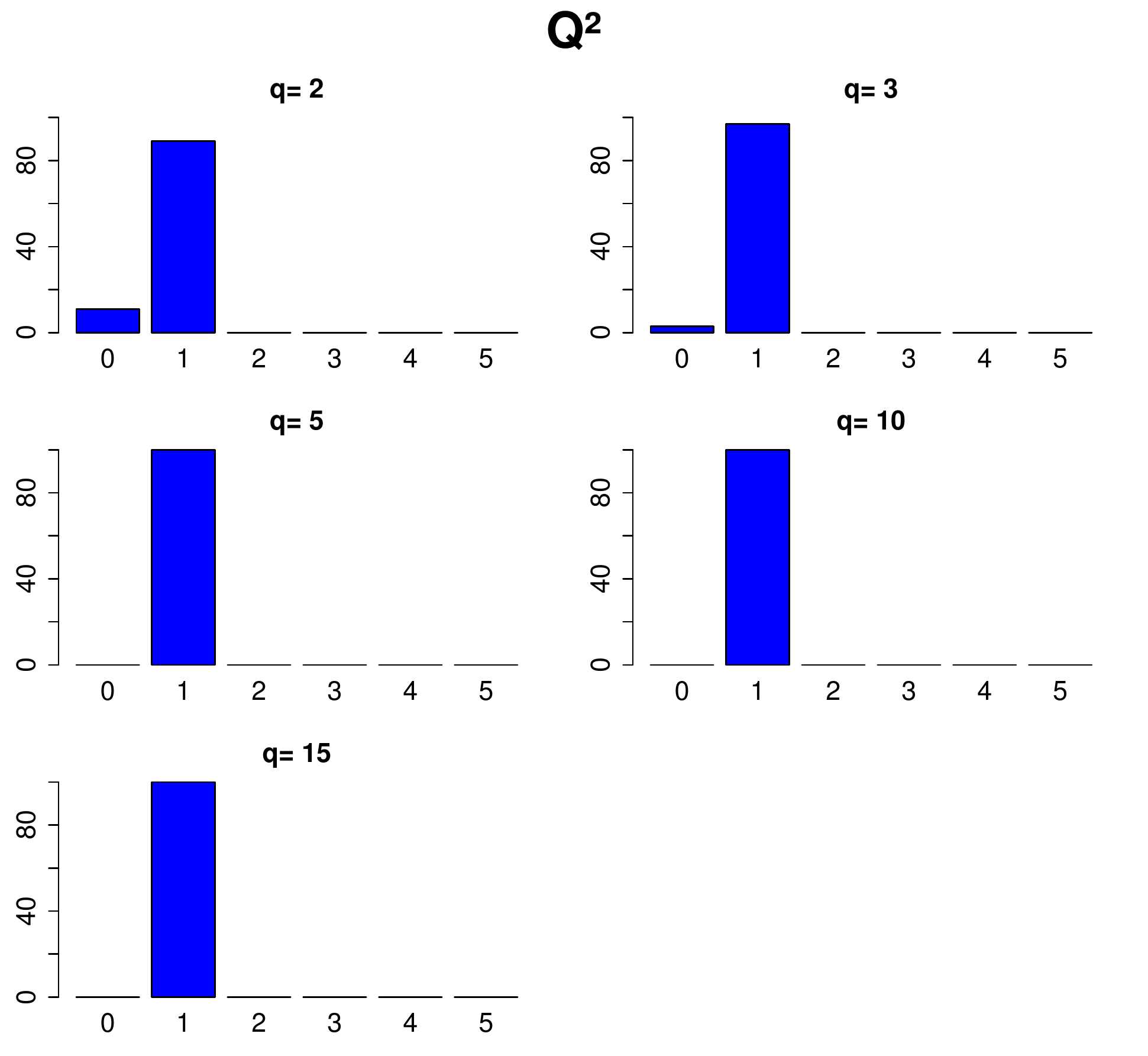}}
          \subfigure{\includegraphics[trim = -2cm 0cm 0cm 0cm, clip,scale=0.3]{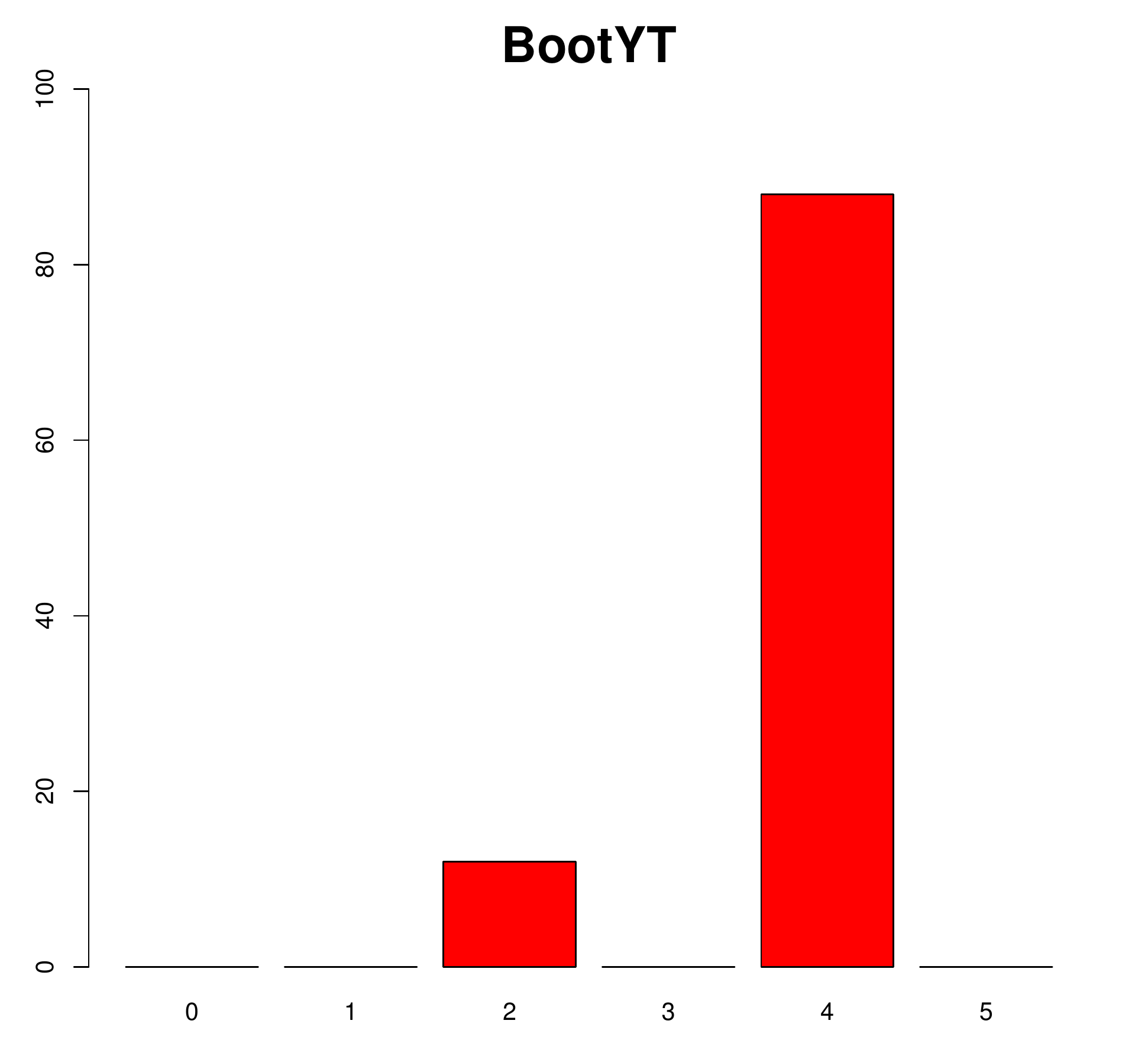}}\\
					\caption{Extracted numbers of component using $q$-folds $Q^2$ (Left) and BootYT (Right) criteria}
   \label{fig:1b}
\end{figure*}

The second example is a benchmark dataset, called ``Processionnaire du Pin'', which is treated in depth by \citet{Tenenhaus}. It has 33 observation on 10 explanatory variables and is also available in the $R$ package \textit{plsRglm} under the name \textit{pine}. More details on this dataset are available in \citep{Tenenhaus}.\\
The same process was applied on this second practical case, with usual PLS regressions. Thus, we compare the $Q^2$ criterion obtained through $q$-fold CV, $q\in\{2,3,5,10,15\}$, and our new bootstrap-based criterion. The $Q^2$ criterion obtained through leave-one-out CV concludes to one significant component. Others results are displayed in Fig.\ref{fig:1b}.

In this case, the $q$-folds CV does not suffer from stability issues, as those seen just before, since the $Q^2$ criterion is much more stable than the minimisation of the number of miss-classed values. However, extracting one component is not recommended. \citet{Tenenhaus}, after a complete analysis of this dataset, proves that four components is the best decision. This under-estimating issue linked to the $Q^2$ criterion will be verified and confirmed in part \ref{31}, by treating results obtained through datasets simulations. Thus, while the $Q^2$ criterion under-estimates this optimal number of components, our new bootstrap-based criterion concludes to four components in more than 80\% of results.  

\subsection{Simulation plan}
\label{23}

To compare these different criteria, datasets simulations have been performed by adapting the \textit{simul\_data\_UniYX} function, available in the $R$ package \textit{plsRglm}.  

Simulations were performed to obtain a three dimensions common space between $\mathbf{X}$ and $\mathbf{y}$, leading to an optimal number of component equal to three. These three components own standard deviations respectively equal to $\sigma_1=10$, $\sigma_2=8$ and $\sigma_3=6$. Simulations were performed under two different cases, both in PLSR and PLSGLR framework. The first one is the $n>p$ situation with $n=200$ and $p\in\Omega_{200}=[\![7,50]\!]$. The second one is the $n<p$ situation where $n=20$ and $p\in\Omega_{20}=[\![25,50]\!]$. For each fixed couple $\left(\sigma_4,\sigma_5\right)$, which respectively represents the standard deviation owned by the useless fourth component present in $\mathbf{X}$ and the random noise standard deviation in $\mathbf{y}$, we simulated 100 datasets with $p_l$ predictors, $l=1,\ldots,100$, obtained by sampling with replacement in $\Omega_n$.The $\mathbf{X}$ matrices are so built around four orthogonal components. Finally, the number of bootstrap samples was fixed to $R=500$. 

\section{PLSR results}
\label{3}
As mentioned in part \ref{23}, the optimal number of components linked to these simulated datasets is equal to three. A four components model will thus be too complex, even if the estimated response will not significantly change since $\mbox{Corr}\left(\mathbf{y},\mathbf{t}_4\right)\simeq0$. Indeed, this fourth component only improves the representation of the original predictors by modelling useless informations in $\mathbf{X}$, but is not helpful for a better estimation of $\mathbf{y}$. All supplementary component is built up with some random noise present in $\mathbf{X}$ in order to explain remaining knowledge in $\mathbf{y}$.

\subsection{PLSR: Case \texorpdfstring{$n>p$}{n>p}}
\label{31}

For this framework, we consider the following values for noises standard deviations:

\begin{center}
$\left(A\right):\left\{
\begin{array}{ll}
\sigma_4\in\left\{0.01,0.21,\ldots,5.81\right\}\cup\left\{6.01,7.01,\ldots,30.01\right\}\\
\sigma_5\in\left\{0.01,0.51,\ldots,20.01\right\}
\end{array}
\right.$
\end{center}

These sets of values lead so to 2255 different couples $\left(\sigma_4,\sigma_5\right)$. Results are so stored in five tables (one per criterion) of dimension $2255\times 100$. Columns correspond to the 100 datasets simulated per couple.

In order to perform a first selection among the existing criteria, we focus on results obtained with:

\begin{center}
$\left(B\right):\left\{
\begin{array}{ll}
\sigma_4\in\left\{0.01,0.21,\ldots,5.81\right\}\\
\sigma_5\in\left\{0.01,0.51,\ldots,20.01\right\}
\end{array}
\right.$
\end{center}

\begin{figure*}[ht]
		\centering
          \subfigure{\includegraphics[trim = 0cm 0cm 0cm 0cm, clip,scale=0.2]{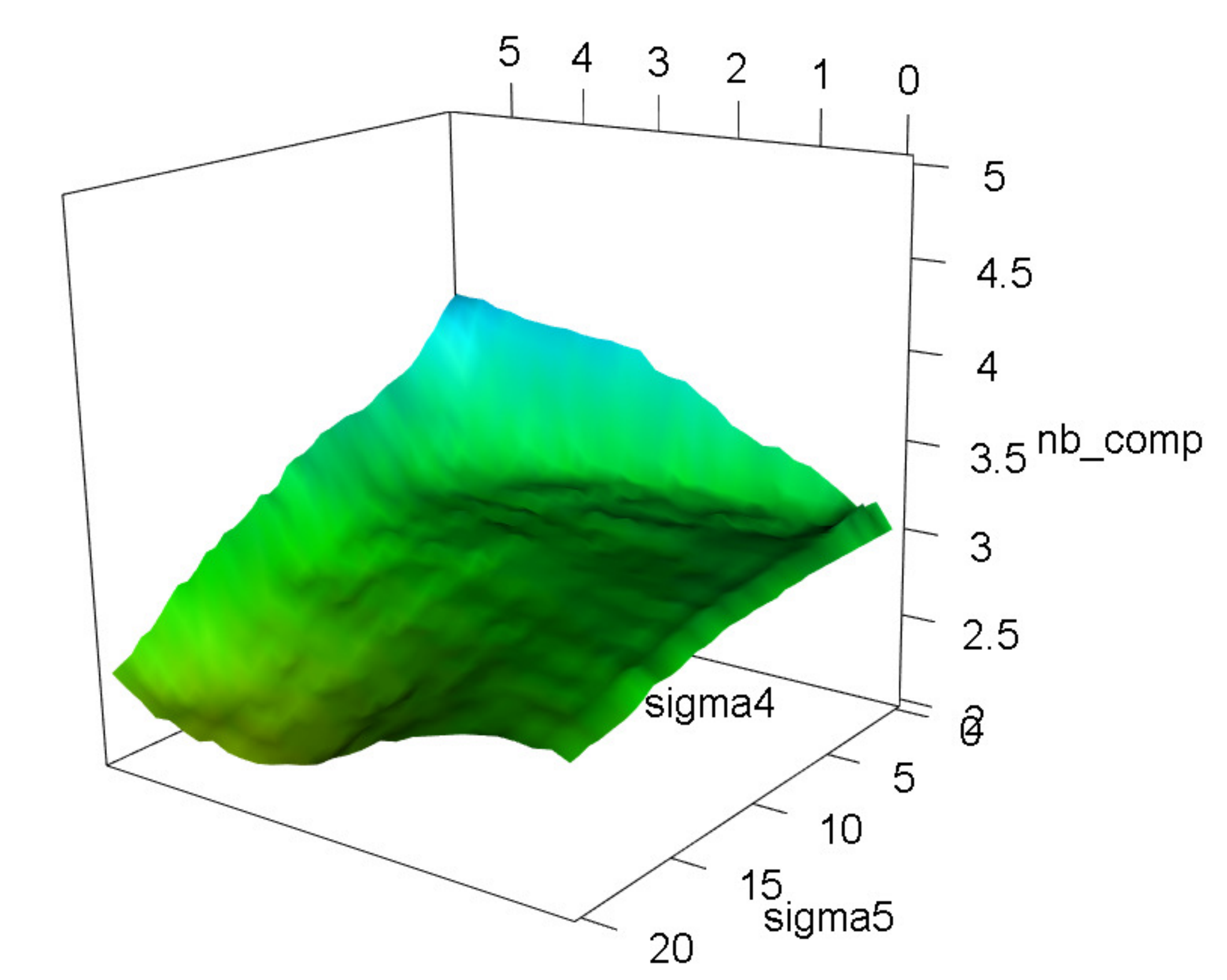}}
          \subfigure{\includegraphics[trim = 0cm 0cm 0cm 0cm, clip,scale=0.2]{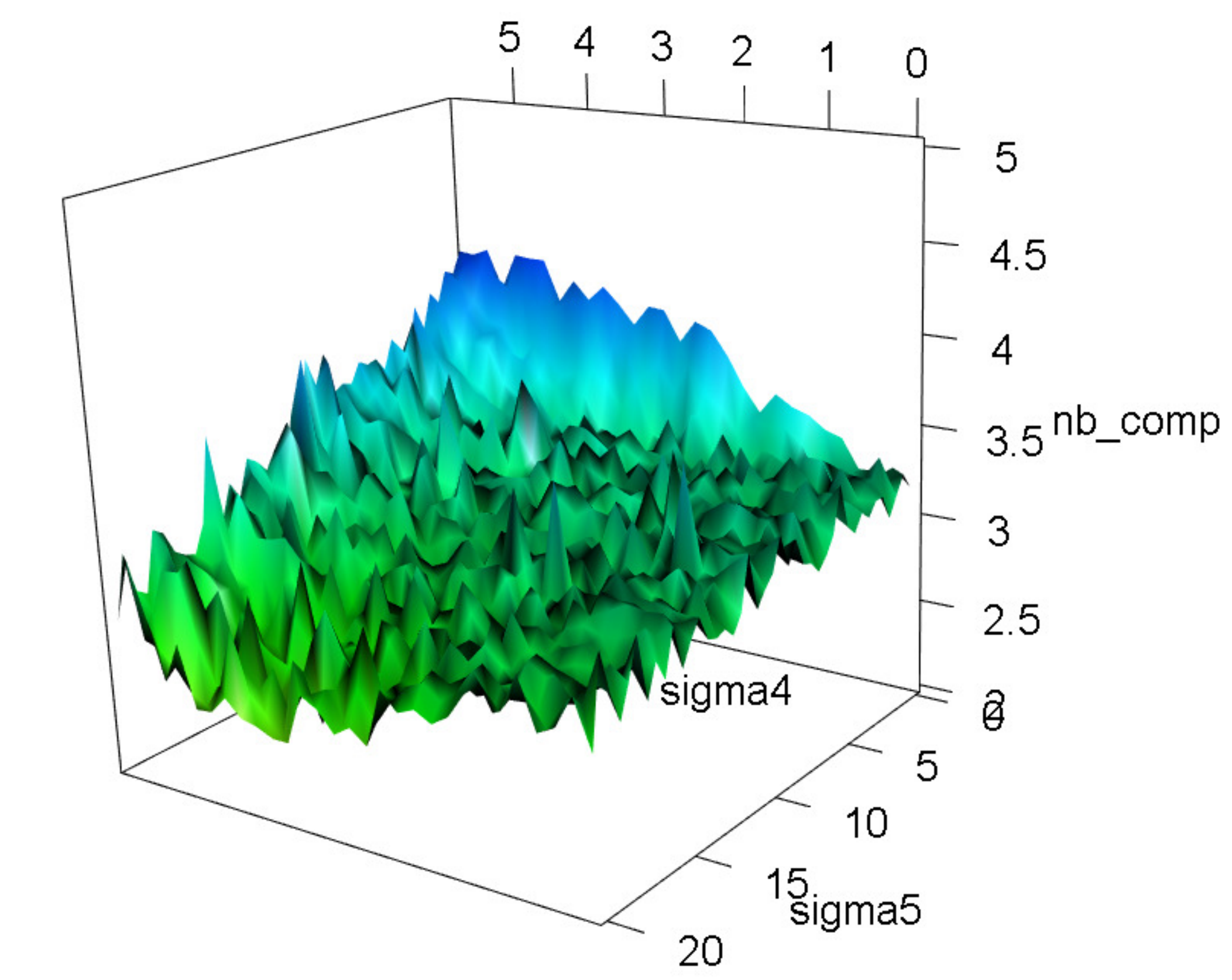}}
    \caption{$n>p$, case $\left(B\right)$; Left: BICdof row means; Right: BICglob row means}		
		\label{fig:2}
\end{figure*}

\begin{figure*}[ht]
		\centering
          \subfigure{\includegraphics[trim = 0cm 0cm 0cm 0cm, clip,scale=0.2]{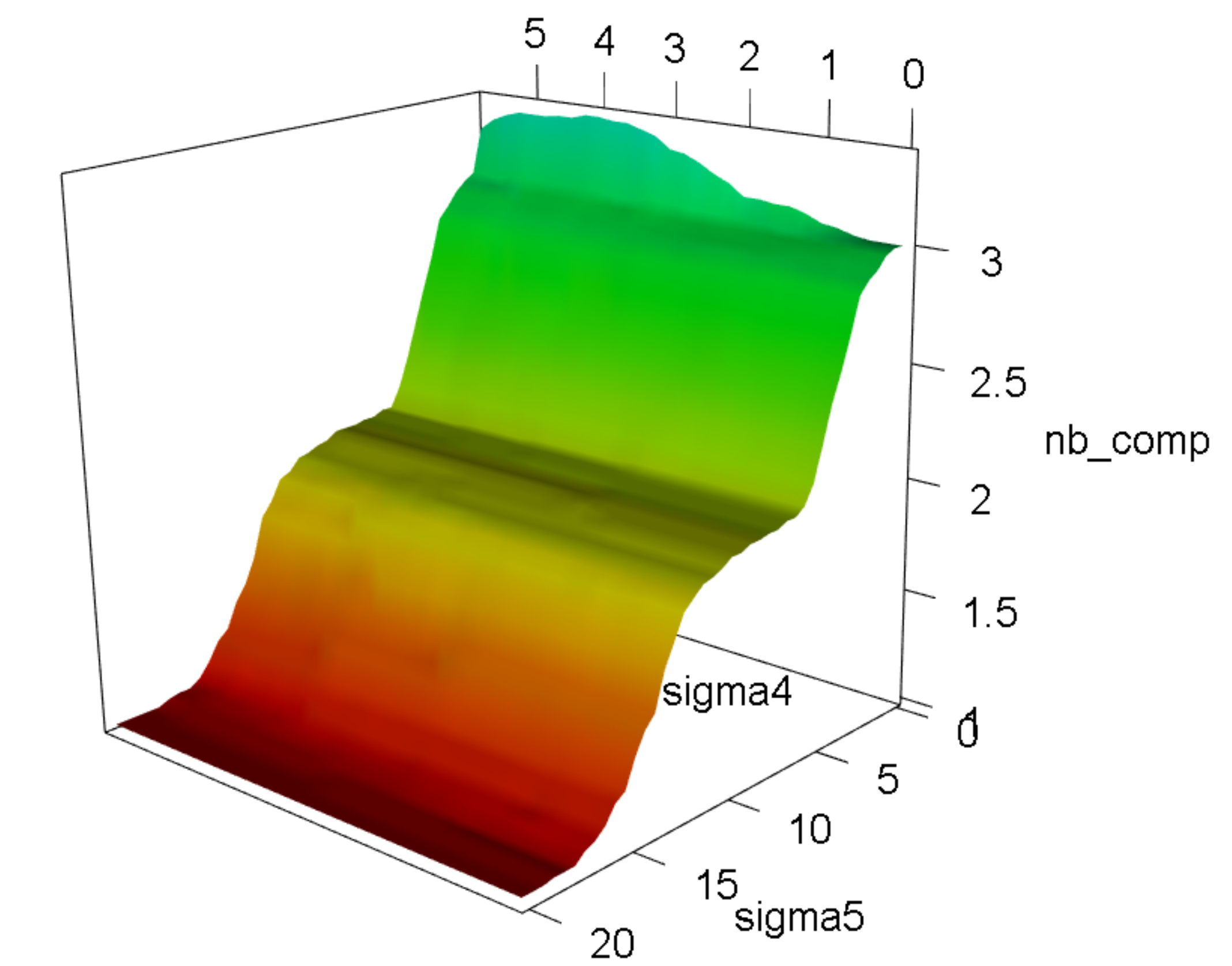}}
          \subfigure{\includegraphics[trim = 0cm 0cm 0cm 0cm, clip,scale=0.2]{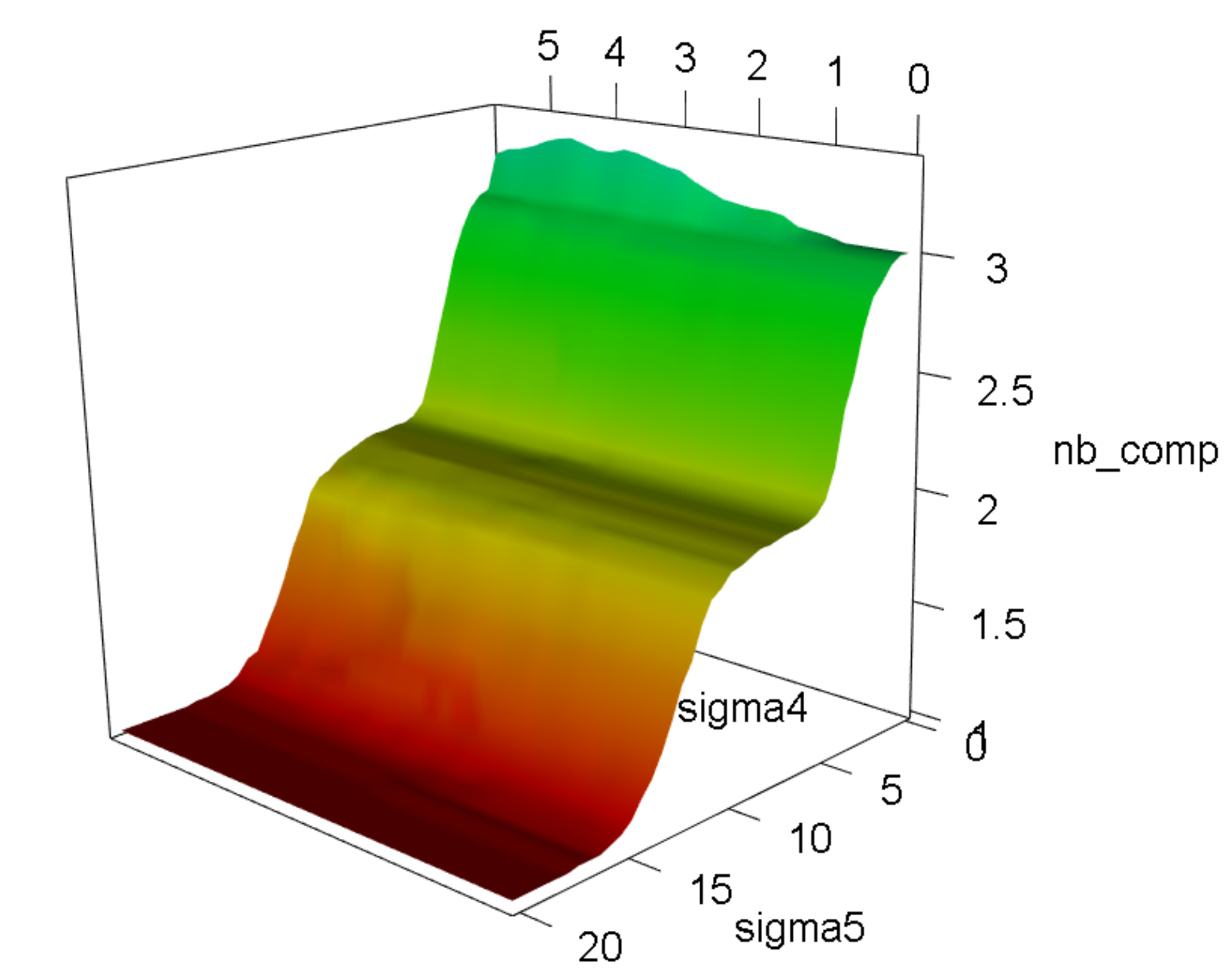}}
    \caption{$n>p$, case $\left(B\right)$; Left: Q2lv1o row means; Right: Q2K5 row means}		
		\label{fig:3}
\end{figure*}

We extract row means for each studied criteria (except for BootYT),  and report them in Fig.\ref{fig:2} and \ref{fig:3} as functions of $\sigma_4$ and $\sigma_5$.  

Based on results mapped out in Fig.\ref{fig:2}, the BICglob criterion has some defaults concerning the stability of its results. This issue is mainly due to the non necessarily increase of the adapted \textit{dof} as a function of the number of components. As a direct consequence, adding a component sometimes lead to smaller \textit{dof} than the previous model ones which leads this criterion to overestimating issues with huge variability levels. Thus, the BICglob criterion has to be avoided. 

Extracting the optimal number of components $K$ as the one which is linked to the model realising the first local minimum of this adapted criterion, permits to focus the comparison on the first models, which are mainly linked to increasing \textit{dof}. Thus, issues linked to the BICglob criterion are solved and results displayed in Fig.\ref{fig:2}, concerning the BICdof criterion, matched to the expected values. The increase of the displayed response surface for the highest values of $\sigma_4$ will be treated while analysing the entire response dataset. 

Based on results displayed in Fig.\ref{fig:3}, the influence of the value of $q$ is negligible on this $Q^2$ criterion so that we will only discuss about results linked to the Q2K5 criterion, as recommended by \citet{trevor2001elements}. This criterion is very sensitive to the increasing noise level in $\mathbf{y}$ so that it globally underestimates the number of components. \citet{Tenenhaus} mentioned this underestimating issue linked to this criterion while analysing the dataset introduced in part \ref{223}. The fact that this criterion performs an expected and optimal estimation of the number of component only for $\sigma_5\simeq0$ is much more harmful for this benchmark criterion. Indeed, real datasets, especially in targets areas as mentioned during the introduction, are never noiseless.\\  

\begin{figure*}[ht]
       \centering
			    \subfigure{\includegraphics[trim = 0cm 0cm 0cm 0cm, clip,scale=0.2]{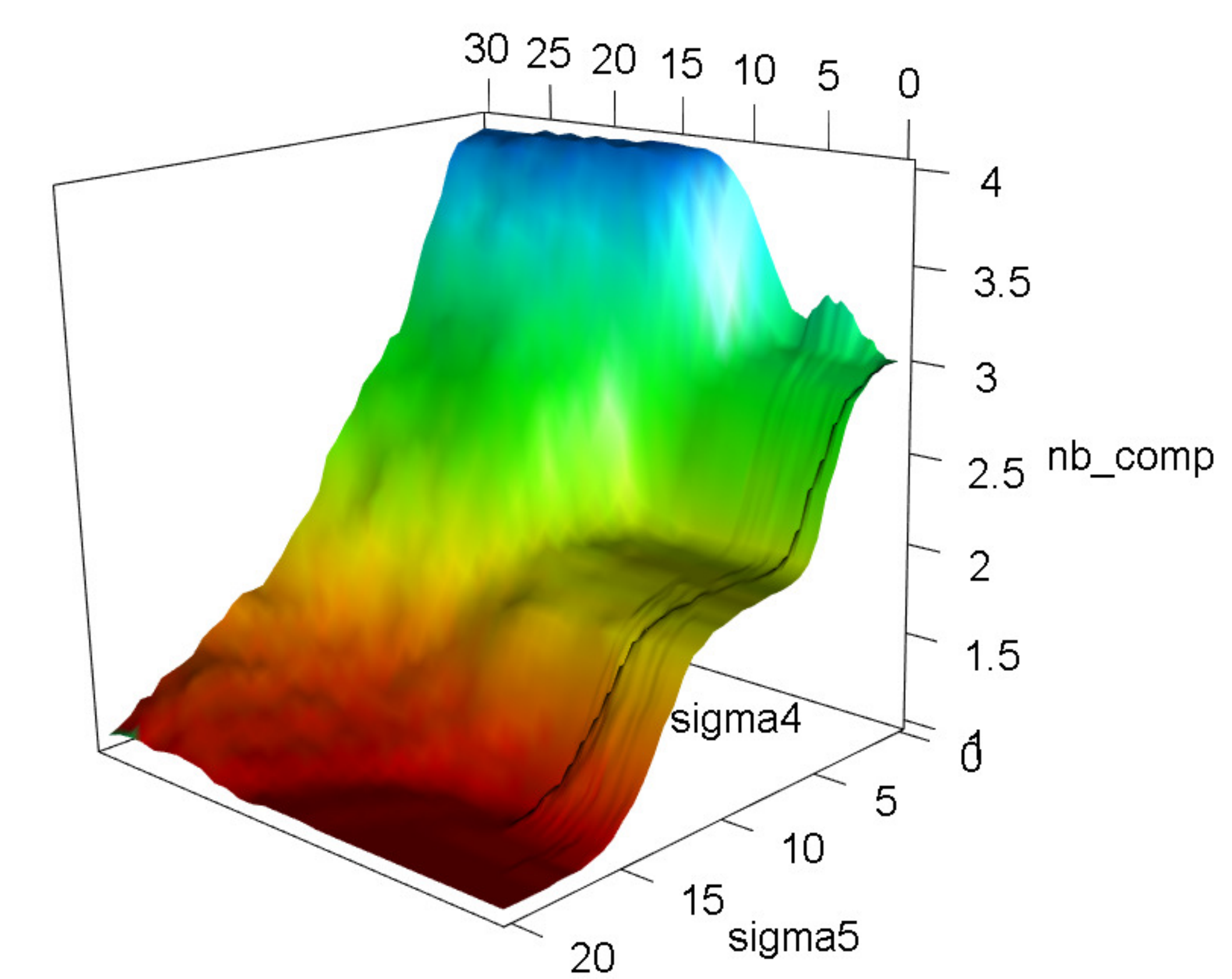}}
          \subfigure{\includegraphics[trim = 0cm 0cm 0cm 0cm, clip,scale=0.2]{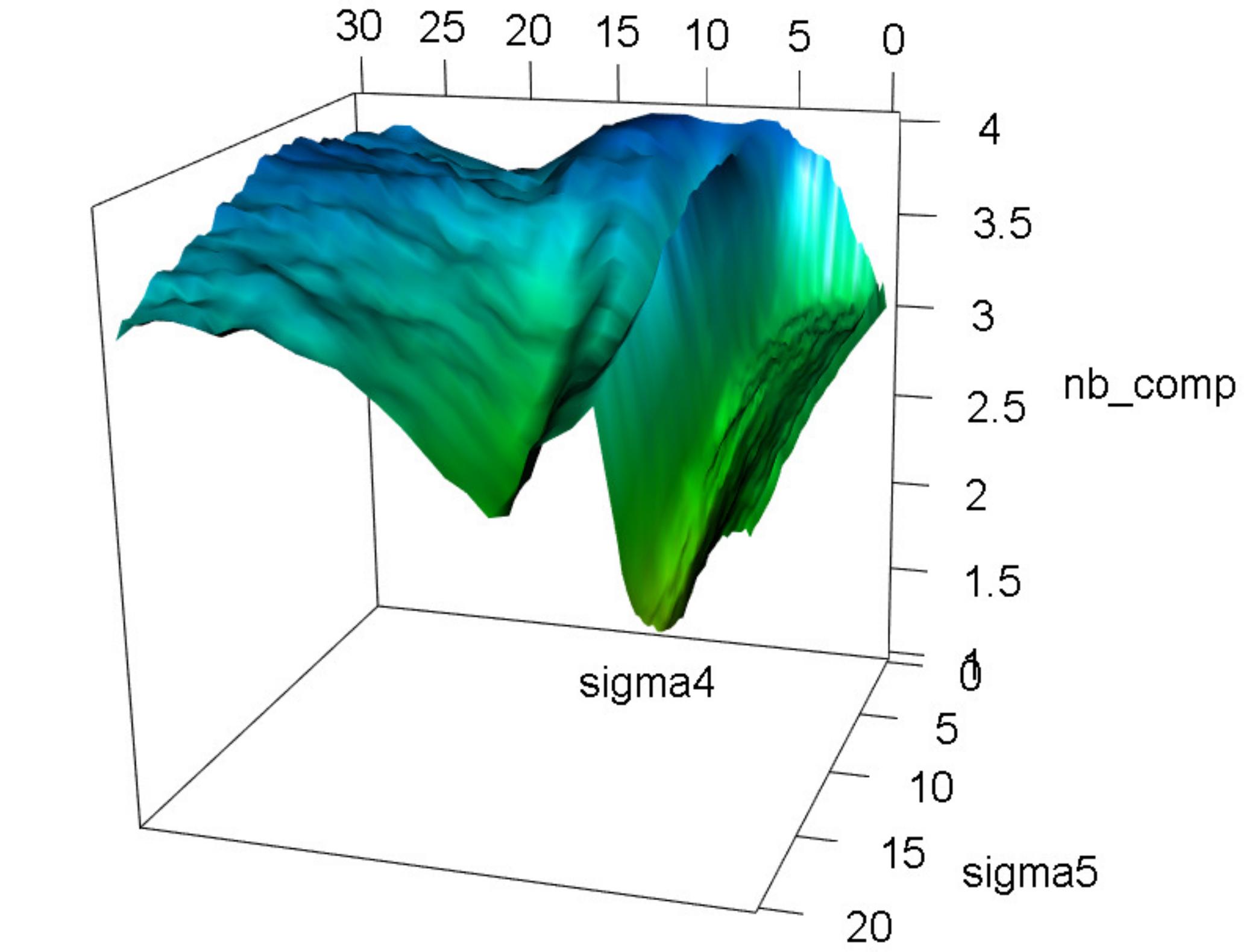}}
          \subfigure{\includegraphics[trim = 0cm 0cm 0cm 0cm, clip,scale=0.2]{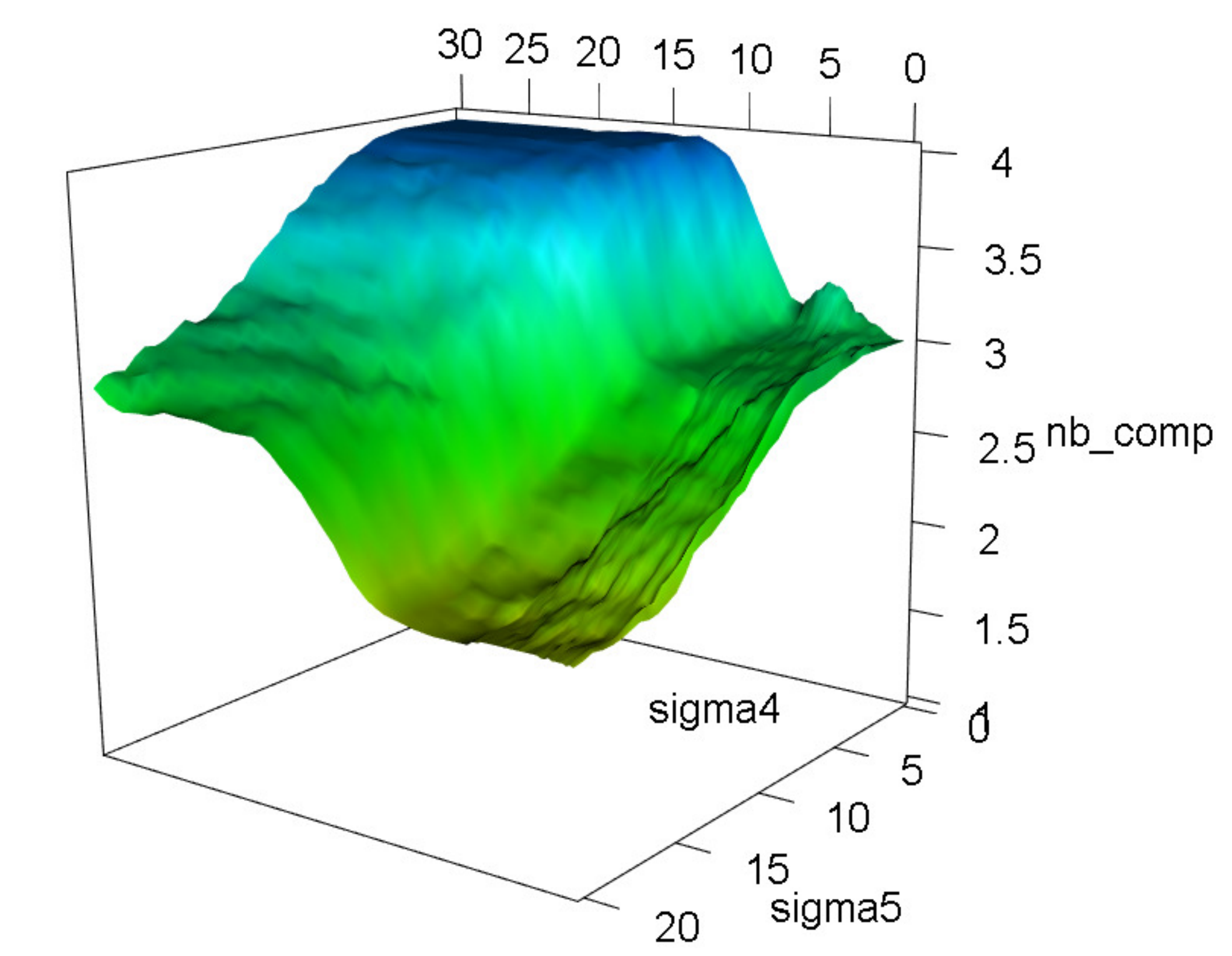}}
    \caption{$n>p$, case $\left(A\right)$; From left to right: Q2K5, BICdof and BootYT row means}
   \label{fig:4}
\end{figure*}

\begin{figure*}[ht]
		\centering
		\subfigure{\includegraphics[trim = 0cm 0.5cm 0cm 0cm, clip, scale=0.22]{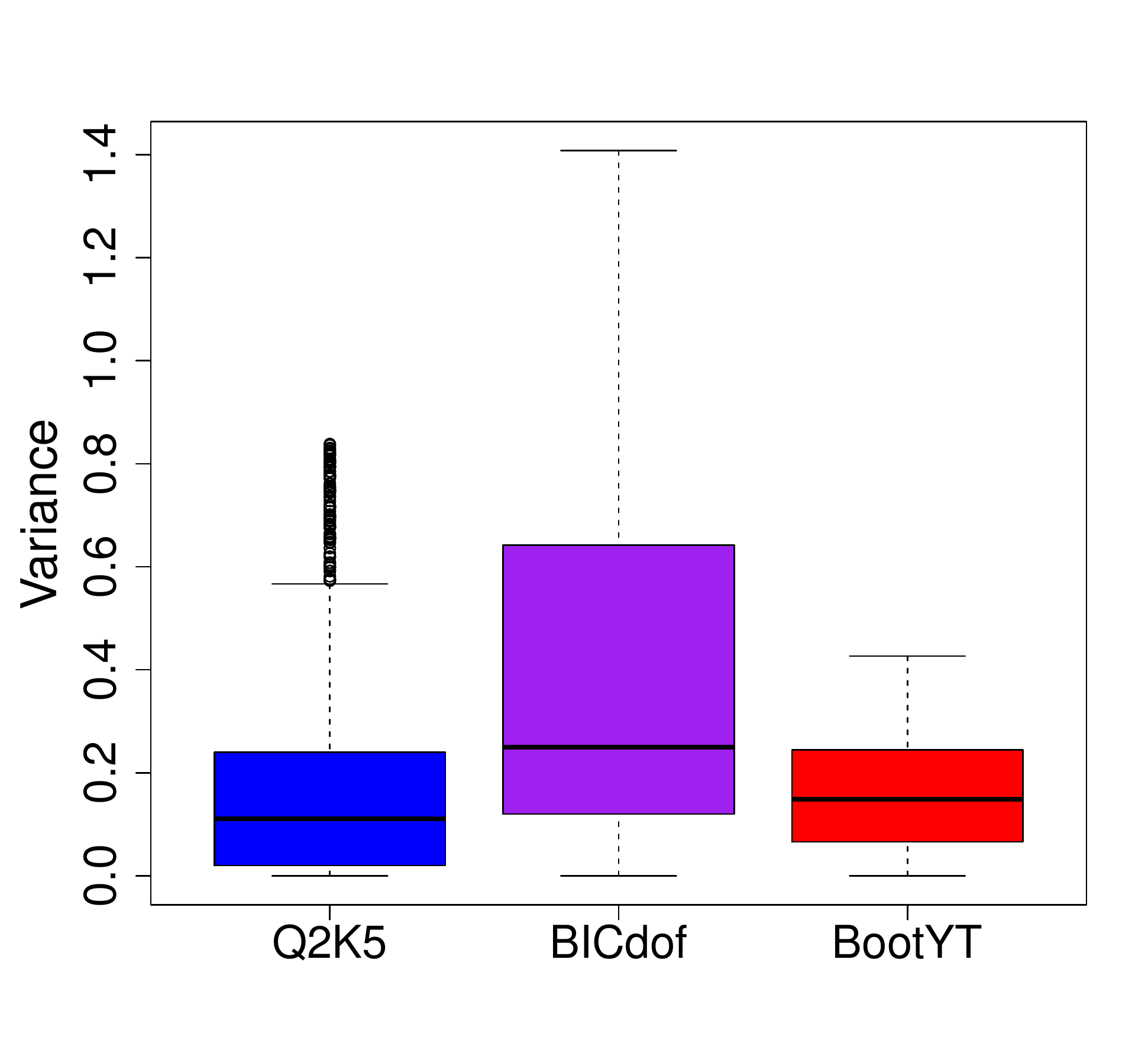}}
		\subfigure{\includegraphics[trim = 0cm 0.5cm 0cm 0cm, clip, scale=0.22]{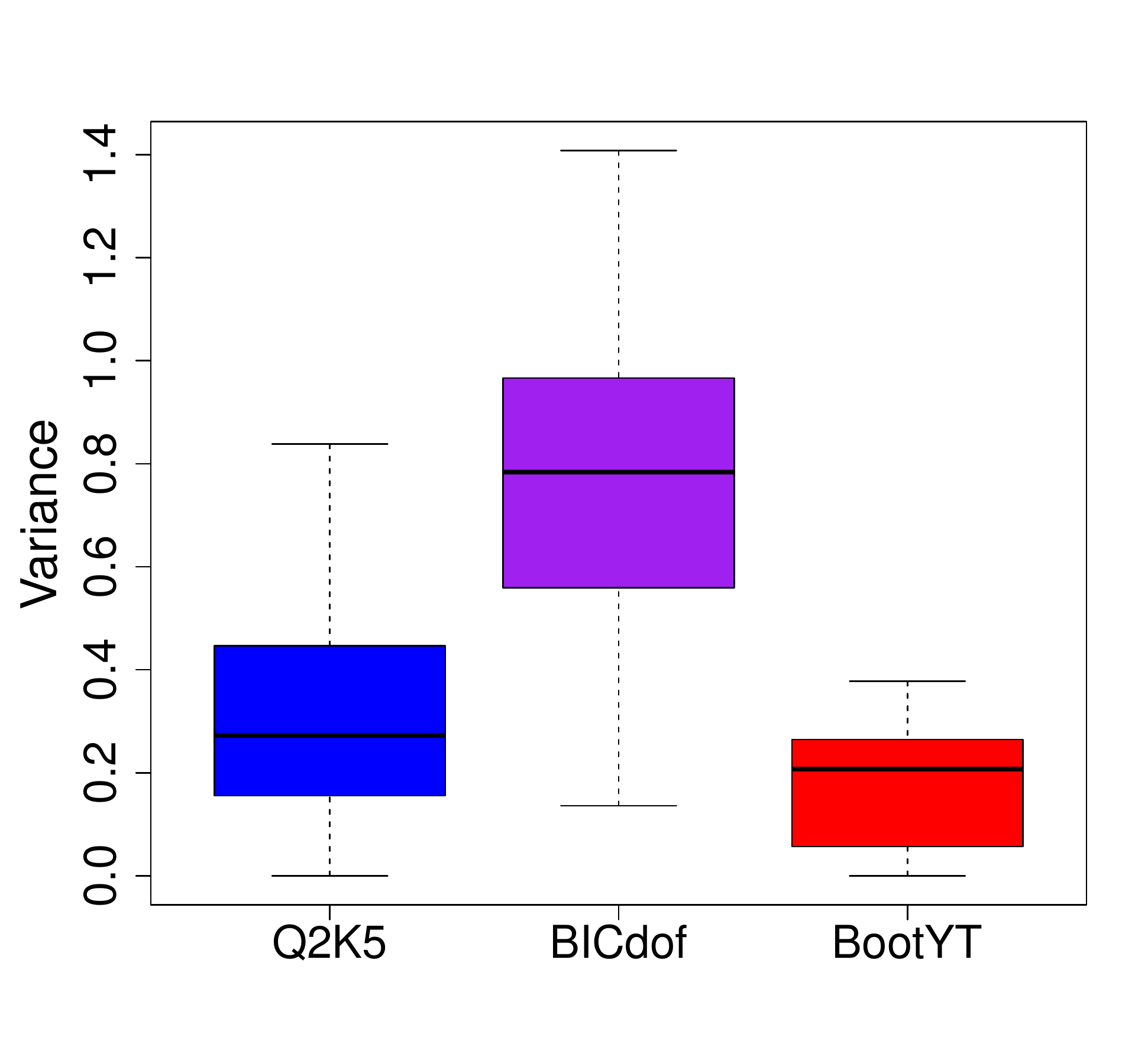}}
		\caption{$n>p$; Left: Boxplots of row variances, case $\left(A\right)$; Right: Boxplots of row variances for $\sigma_4\geqslant15.01$}
   \label{fig:5}
\end{figure*}

We then display three graphical representations of all the 2255 available row means linked to the Q2K5, BICdof and BootYT criteria in Fig.\ref{fig:4}. Each row variances were also extracted and graphically reported in Fig.\ref{fig:5}.  

Considering the extracted number of components as a discriminant factor, we conclude that the $Q^2$ criterion is the less efficient criterion by being the most sensitive one to the increasing value of $\sigma_5$ so that it globally underestimates the number of components. Comparing BICdof and BootYT, or advertising one of them is quite difficult in this large $n$ case. BICdof has a low computational runtime and is the less sensitive one to the increasing value of $\sigma_5$ by retaining 86.37\% of results equal to three or four components. However, referring to Fig.\ref{fig:5}, the variability of results linked to the BICdof is globally higher than the one linked to our new bootstrap based criterion, especially on datasets with large values of $\sigma_4$ $\left(\sigma_4>\sqrt{\stackrel[i=1]{3}{\sum{}}\sigma_i^2}=\sqrt{200}\simeq14.14\right)$. In this particular case, our new bootstrap-based criterion keeps its stability while the median of BICdof results, for instance, is more than tripled (0.25 to 0.79) compared to the one linked to all data. Moreover, the BootYT is more robust than the BICdof to the increasing noise level in $\mathbf{X}$ in terms of extracted number of components. 

\subsection{PLSR: Case \texorpdfstring{$n<p$}{n<p}}
\label{32}
\subsubsection{Row means and variances analysis}
\label{321}

This small training sample size, allows us to consider high-dimensional settings and is interesting since usually least squares regression could not be performed. 

\begin{figure*}[ht]
		\centering
		\subfigure{\includegraphics[trim = 0cm 0cm 0cm 0cm, clip,scale=0.2]{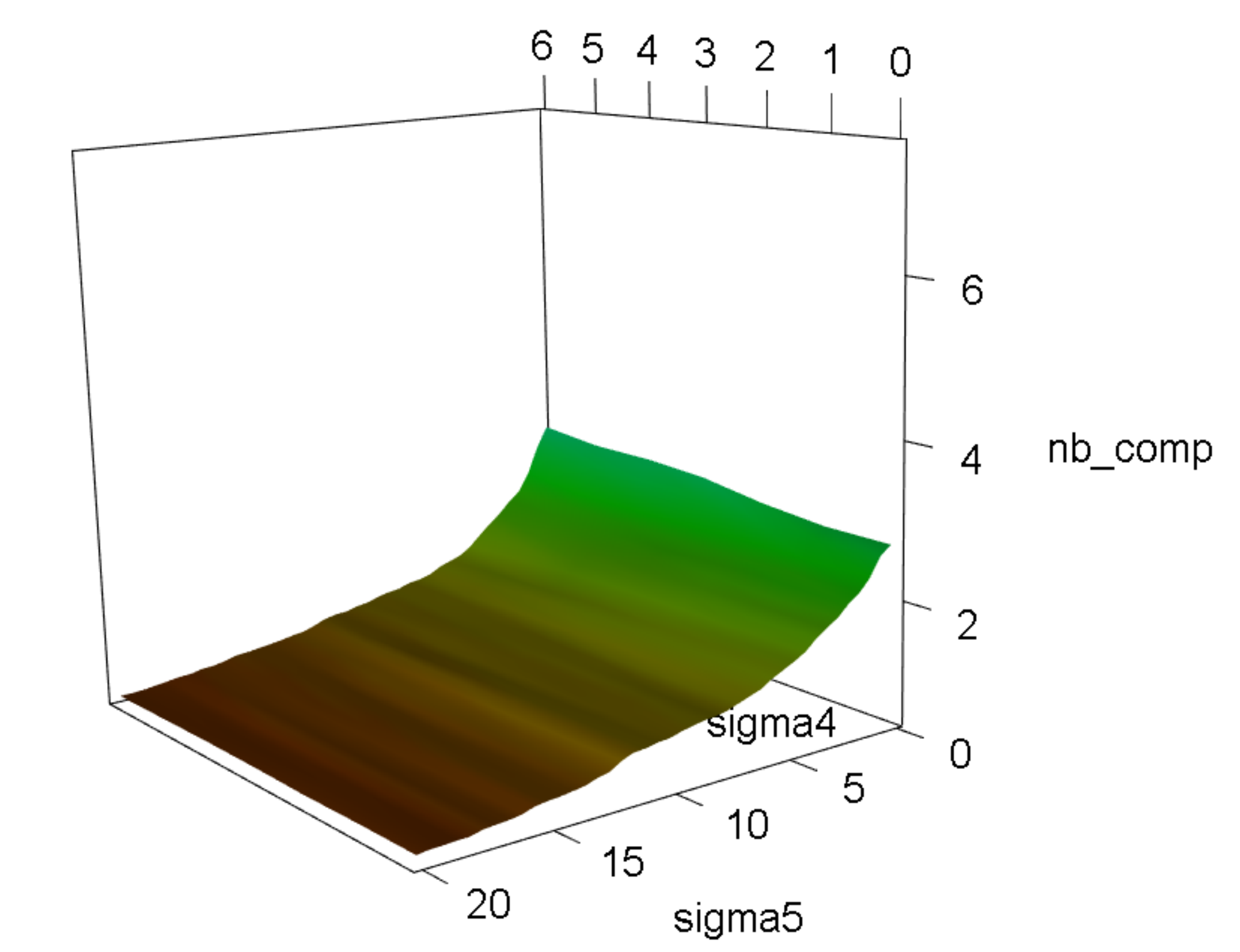}}
		\subfigure{\includegraphics[trim = 0cm 0cm 0cm 0cm, clip, scale=0.2]{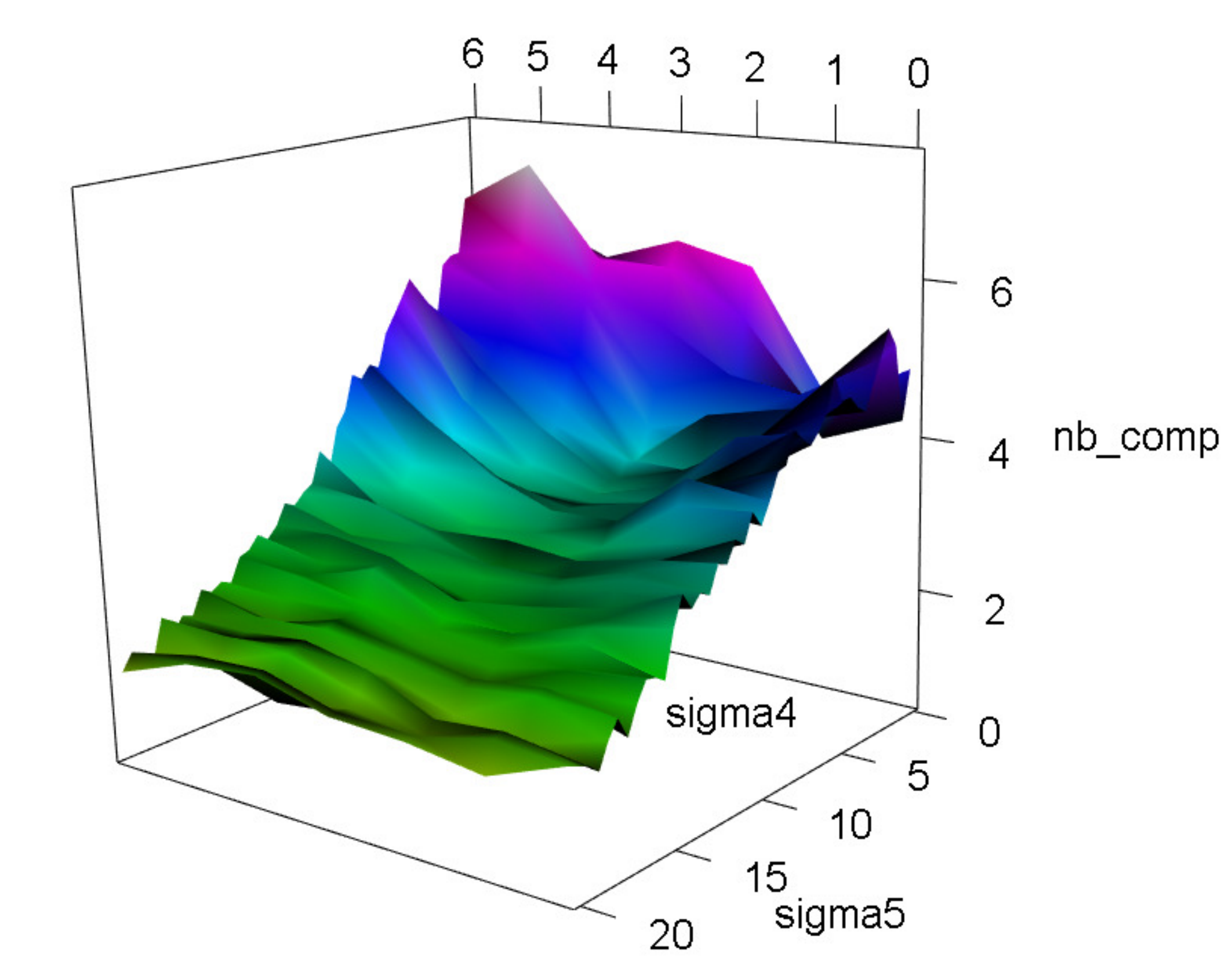}}
		\subfigure{\includegraphics[trim = 0cm 0cm 0cm 0cm, clip,scale=0.2]{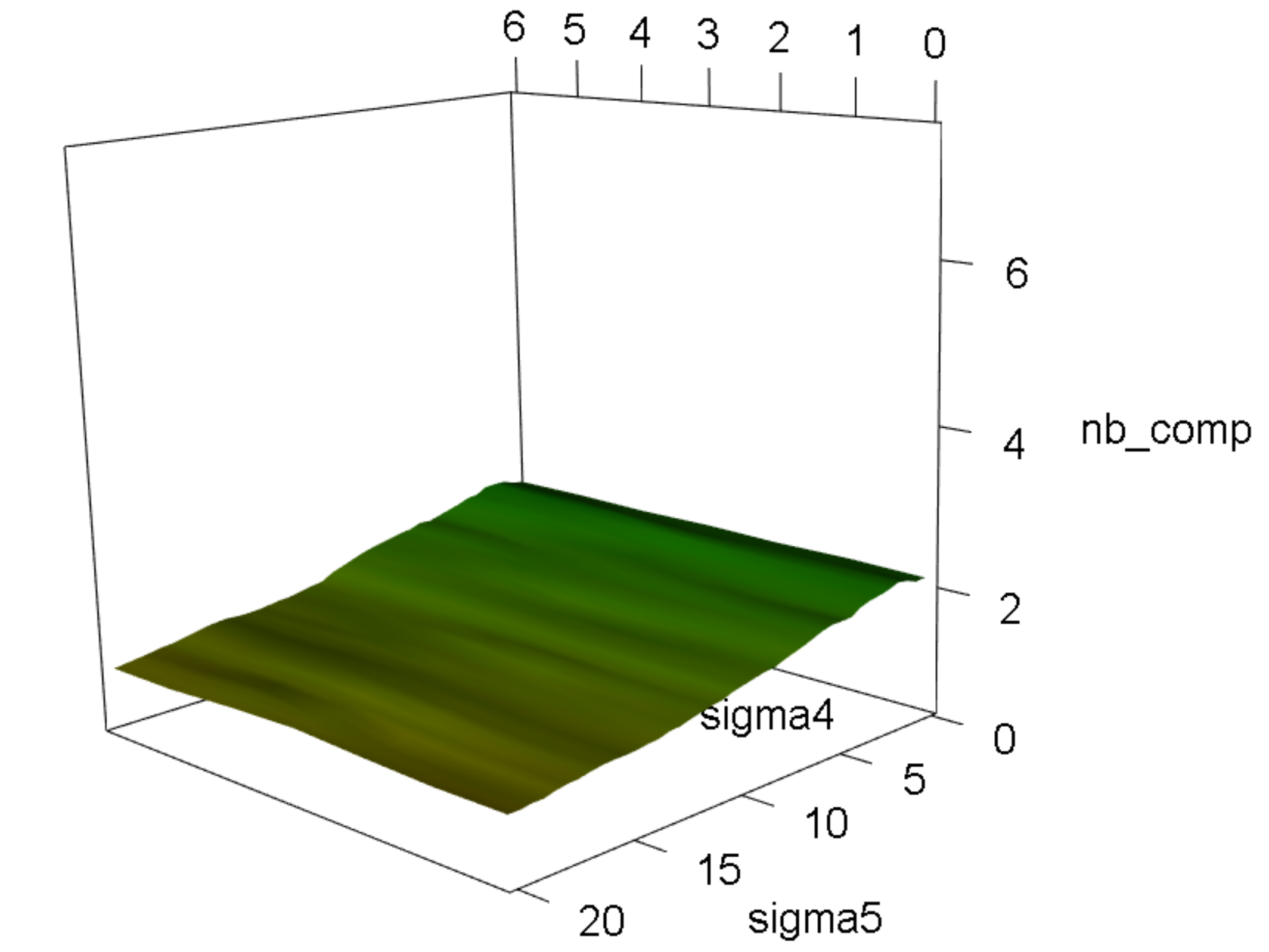}}
		\caption{$n<p$; From left to right: Q2K5, BICdof and BootYT row means}
   \label{fig:6}
\end{figure*}

\begin{figure*}[ht]
		\centering
		\subfigure{\includegraphics[trim = 0cm 0.6cm 0cm 1.6cm, clip, scale=0.215]{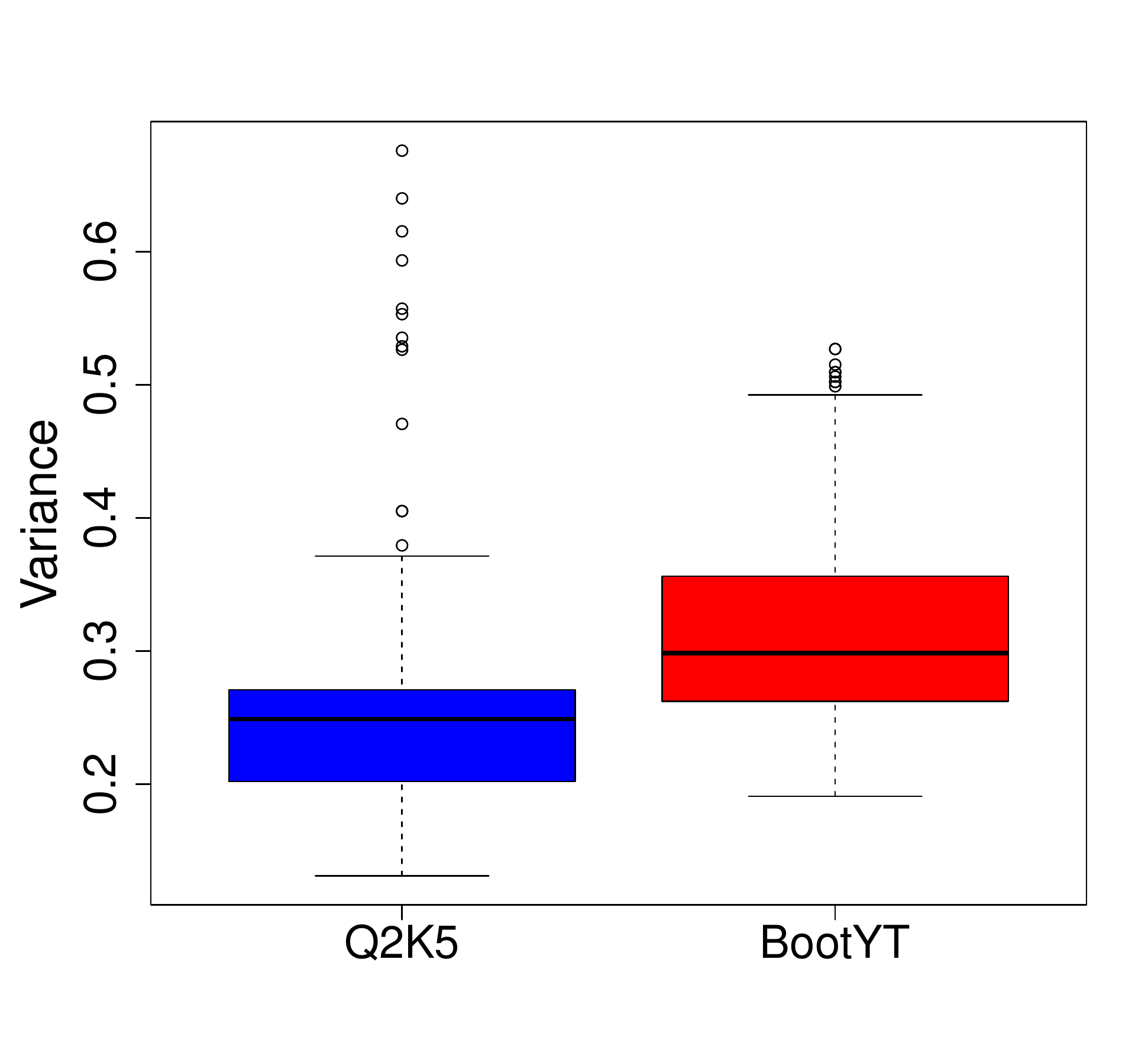}}
		\subfigure{\includegraphics[trim = 0cm 0.6cm 0cm 1.6cm, clip, scale=0.215]{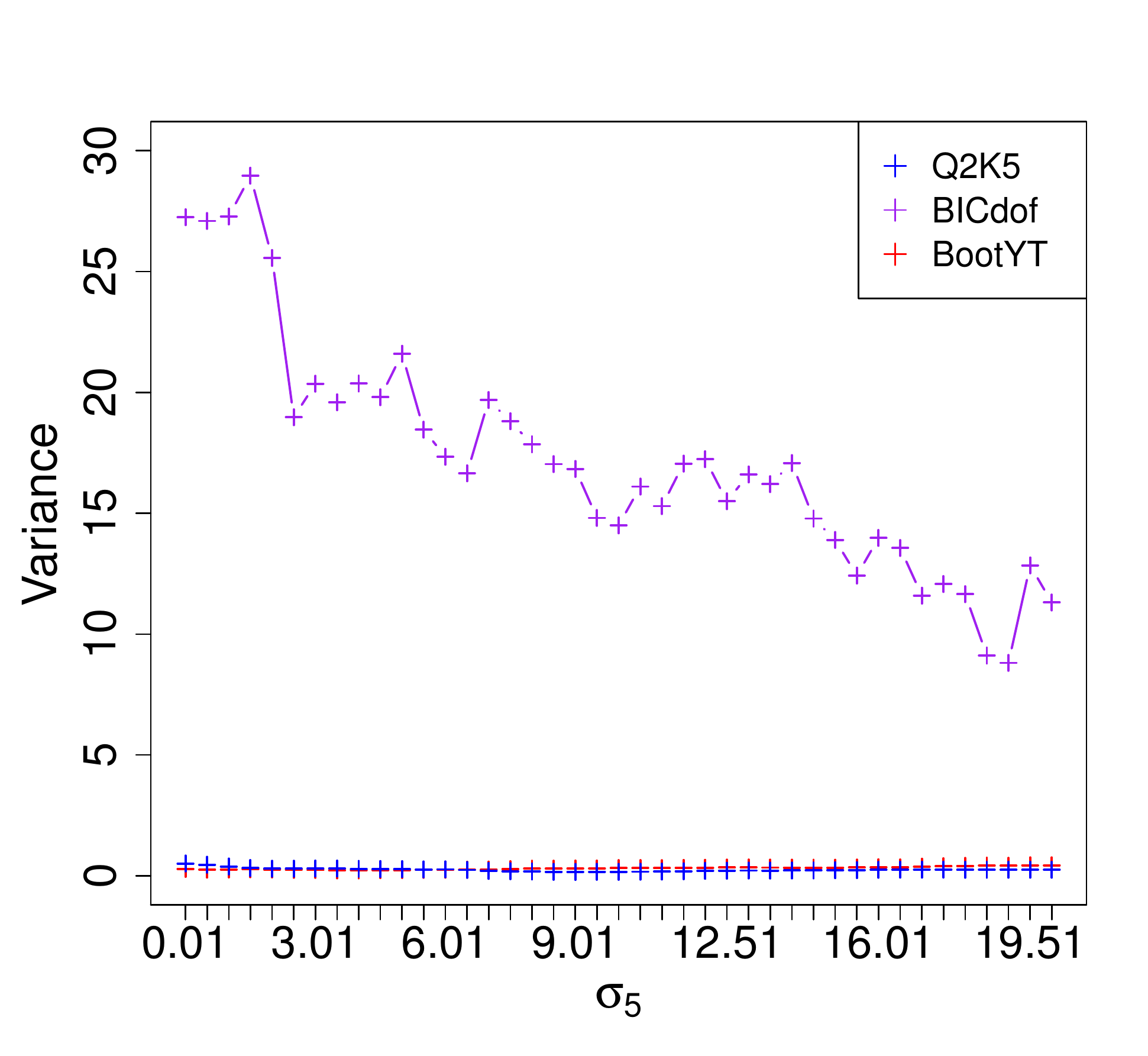}}
		\caption{$n<p$; Left: Boxplots of row variances; Right: Evolution of row variances}
   \label{fig:7}
\end{figure*}

We consider $\sigma_4\in\left\{0.01,1.01,\ldots,6.01\right\}$ and $\sigma_5\in\left\{0.01,0.51,\ldots,20.01\right\}$ leading to 287 different couples $\left(\sigma_4,\sigma_5\right)$. Results are so stored in three tables of dimension $287\times 100$. Row means are represented as a function of $\sigma_4$ and $\sigma_5$ in Fig.\ref{fig:6}. Graphical representations of row variances were also performed in Fig.\ref{fig:7}.

Referring to Fig.\ref{fig:6}, the BICdof suffers from overestimating issues. Moreover, based on results display on Fig.\ref{fig:7}, it returns results linked to out of range values of variance compared to the two others criteria. These two issues, which are mainly due to the extraction of 1678 (5.847\%) results equal to 19 components, added to its its lack of robustness to the increasing noise level $\sigma_5$, lead us to conclude that this criterion should be avoid in this $n<p$ framework.  

Our new boostrap-based criterion underestimates the number of components but own an attractive robustness to the increasing noise level in $\mathbf{y}$ so that it returns results between 1.2 and 2.2 in average. Moreover, it returns results with low variability for fixed couple $\left(\sigma_4,\sigma_5\right)$, contrary to the BICdof criterion. The Q2K5 criterion has a comparable attractive feature of stability but is less robust to noise level in $\mathbf{y}$ than our new bootstrap based criterion, linking the Q2K5 to globally more important underestimating issues.

So, by considering the number of extracted components as a discriminant factor, we conclude that the BootYT criterion is the best one to deal with these $n<p$ datasets.\\

\subsubsection{MSE analysis}
\label{322}
	
	We then assess the predictive performances of each of these three criteria (BootYT, BICdof and Q2K5). Thus, for each of the 28 700 simulated datasets in this case, we simulated 80 more observations as test points. We computed both training and testing Normalized Mean Square Error (NMSE). The normalisation was done by dividing the training or testing MSE of the obtained model with the corresponding MSE linked to the trivial one (constant model equal to the mean of the training data). Furthermore, as mentioned by \citet[p.702]{kramer2011degrees}, ``the large test sample size ensures a reliable estimation of the test error.'' 
	
\begin{figure*}[ht]
       \centering
          \subfigure{\includegraphics[trim = 0cm 0cm 0cm 0cm, clip,scale=0.195]{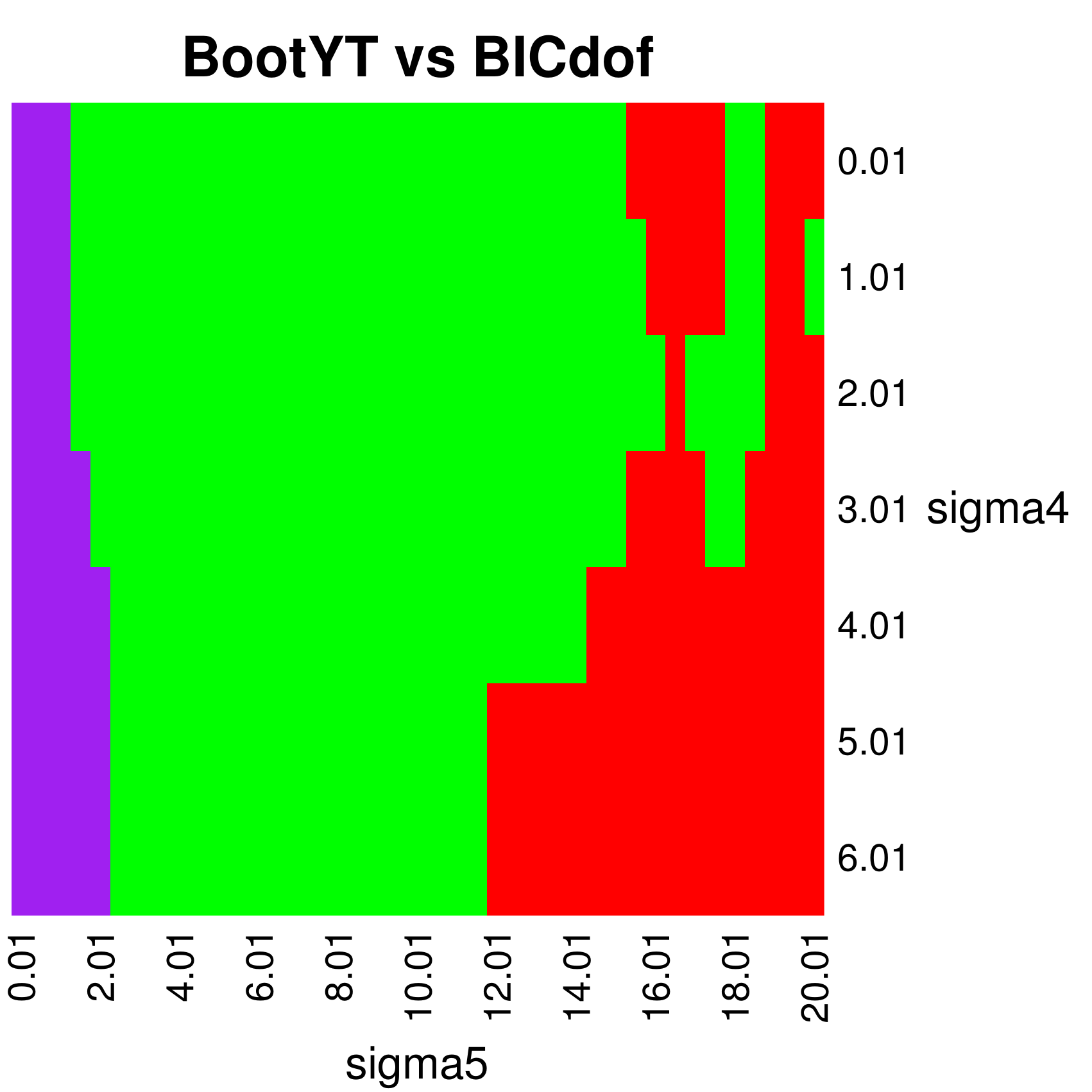}}
          \subfigure{\includegraphics[trim = 0cm 0cm 0cm 0cm, clip,scale=0.195]{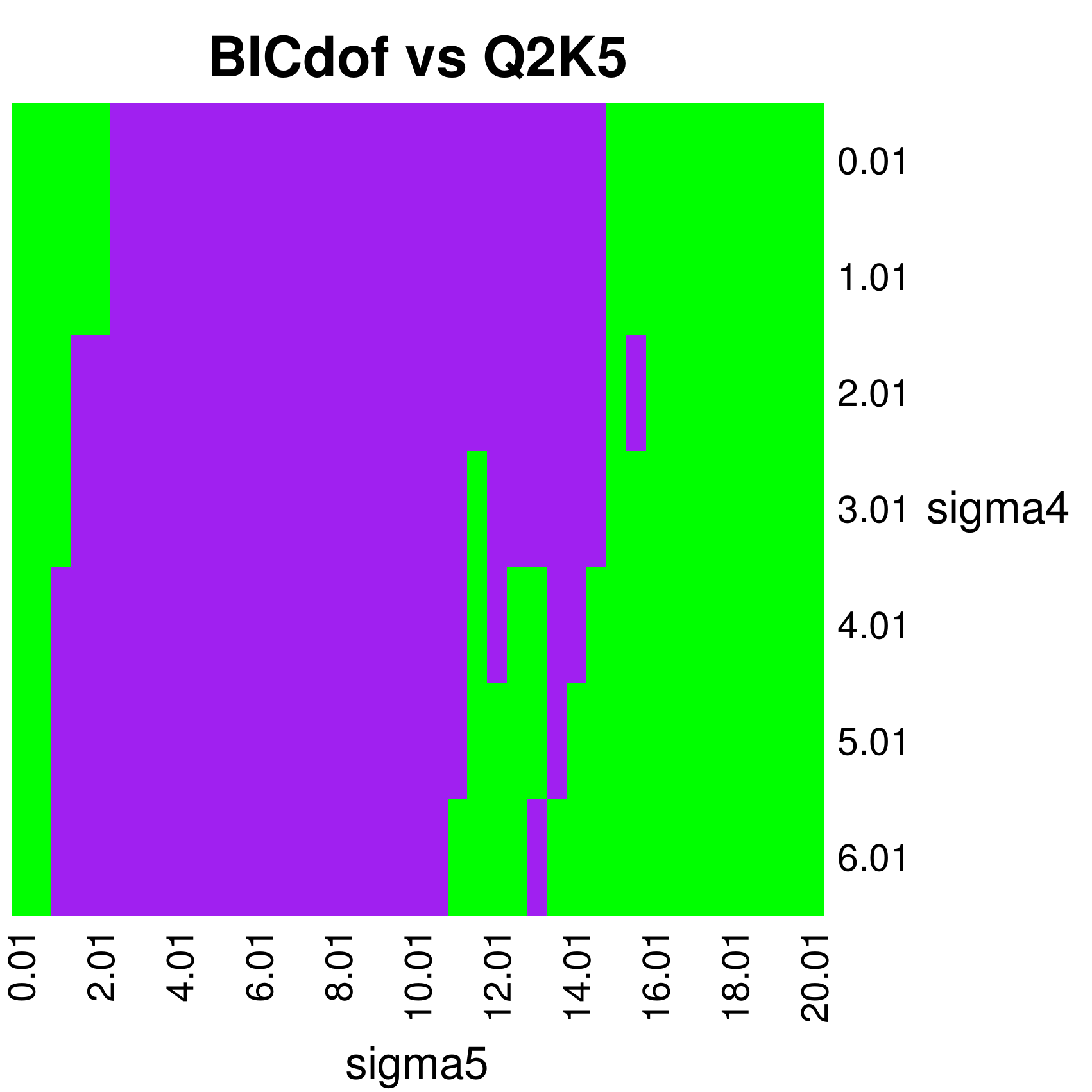}}
					\subfigure{\includegraphics[trim = 0cm 0cm 0cm 0cm, clip,scale=0.195]{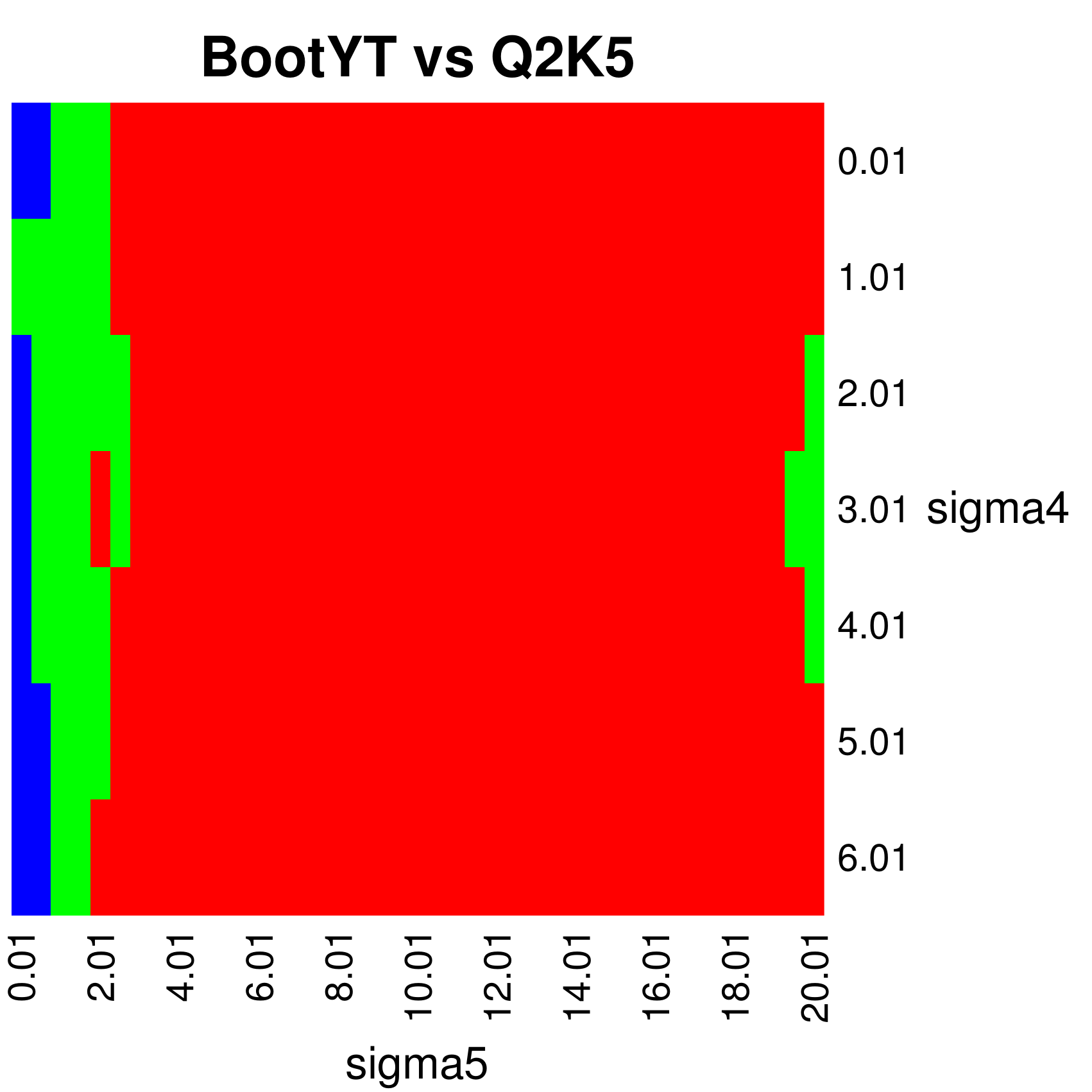}}\\
    \caption{\textit{t}-tests results: BootYT better (red), BICdof better (purple), Q2K5 better (blue), no significant difference (green)}
   \label{fig:9}
\end{figure*}

	We treat these predictive results for each couple of values $\left(\sigma_4,\sigma_5\right)$ by testing the equalities of NMSE means through asymptotic \textit{t}-tests with Welch-Satterthwaite \textit{dof} approximation \citep{welch1947generalization}. All these tests were performed on level $\alpha=0.05$. Results of these \textit{t}-tests are graphically reported in Fig.\ref{fig:9}.
	
Concerning BootYT vs Q2K5, the Q2K5 has a better predictive ability for some very low values of $\sigma_5$. This result is not surprising since, in this case, the Q2K5 criterion returns numbers of components closer to three than BootYT does (Fig.\ref{fig:6}). However, tests results between the BICdof and the Q2K5 criterion are not concluding to a significant better predictive performance of the Q2K5 criterion for small values of $\sigma_5$ despite the BICdof globally overestimates the number of components in this case (Fig.\ref{fig:6}). In fact, due to the small values of $\sigma_5$ and despite the bad estimations of the number of components returned by the BICdof criterion by extracting more than three or four components, testing NMSE react in the same way than the training ones $i.e.$ the higher the extracted number of components is, the lower the predictive NMSE are. This fact lead us to only focus on the extracted number of components when $\sigma_5\simeq0$, leading the Q2K5 criterion to be the best one. 
	
	Finally, in all others cases, the BootYT criterion returns models with, at least, comparable or better predictive performances than the two others. 
	
 	\subsection{PLSR: Conclusion}
	\label{33}
	In the $n>p$ case, the BootYT criterion offers a better robustness to noise in $\mathbf{y}$ than the Q2K5. It is also more robust to the increasing noise level in $\mathbf{X}$ than the BICdof, which moreover has some variance issues for high values of $\sigma_4$. We also conclude the BootYT criterion as a good compromise between the two others criteria, owning their advantages without their drawbacks. Concerning the $n<p$ case, our bootstrap-based criterion is less sensitive than the others to the increasing noise level in $\mathbf{y}$ and is linked to low variance results, leading it to have better predictive performances on datasets with a non-negligible noise level in $\mathbf{y}$.
	
\section{PLSGLR results}	
\label{4}
In this section, we present results on the comparison of our bootstrap-based criterion with 4 other criteria (AIC, BIC, CV-MClassed and p\_val, see part \ref{21}). Note that, in this framework, due to the specific distribution of $\mathbf{y}$ and link-function $g$ we chose, the increase of $\sigma_5$ does not lead to a linear increase of noise level in $\mathbf{y}$, as it did in datasets simulations for PLSR analysis.

\subsection{PLS-LR results}
\label{41}
In this case, the increasing value of $\sigma_5$ does not influence the extracted number of components as much as it did during the PLS analysis. Indeed, due to the used link function: 
\begin{align*}
inv.logit : \mathbb{R}&\longrightarrow\left]0,1\right[\\
x&\longmapsto\frac{1}{1+\exp\left(-x\right)}
\end{align*}
the increasing value of $\sigma_5$ does not imply an increase of the response vector variance, which belong to $\left[0;0.25\right]$, so that the decrease of significant components is not implied. 

Finally, the bijectivity of the $inv.logit$ function insures the presence of three common components between $\mathbf{X}$ and $\mathbf{y}$.

\subsubsection{PLS-LR: Case \texorpdfstring{$n>p$}{n>p}}
\label{411}

\begin{figure*}[ht]
		\centering
          \subfigure{\includegraphics[trim = 0cm 0cm 0cm 0cm, clip,scale=0.22]{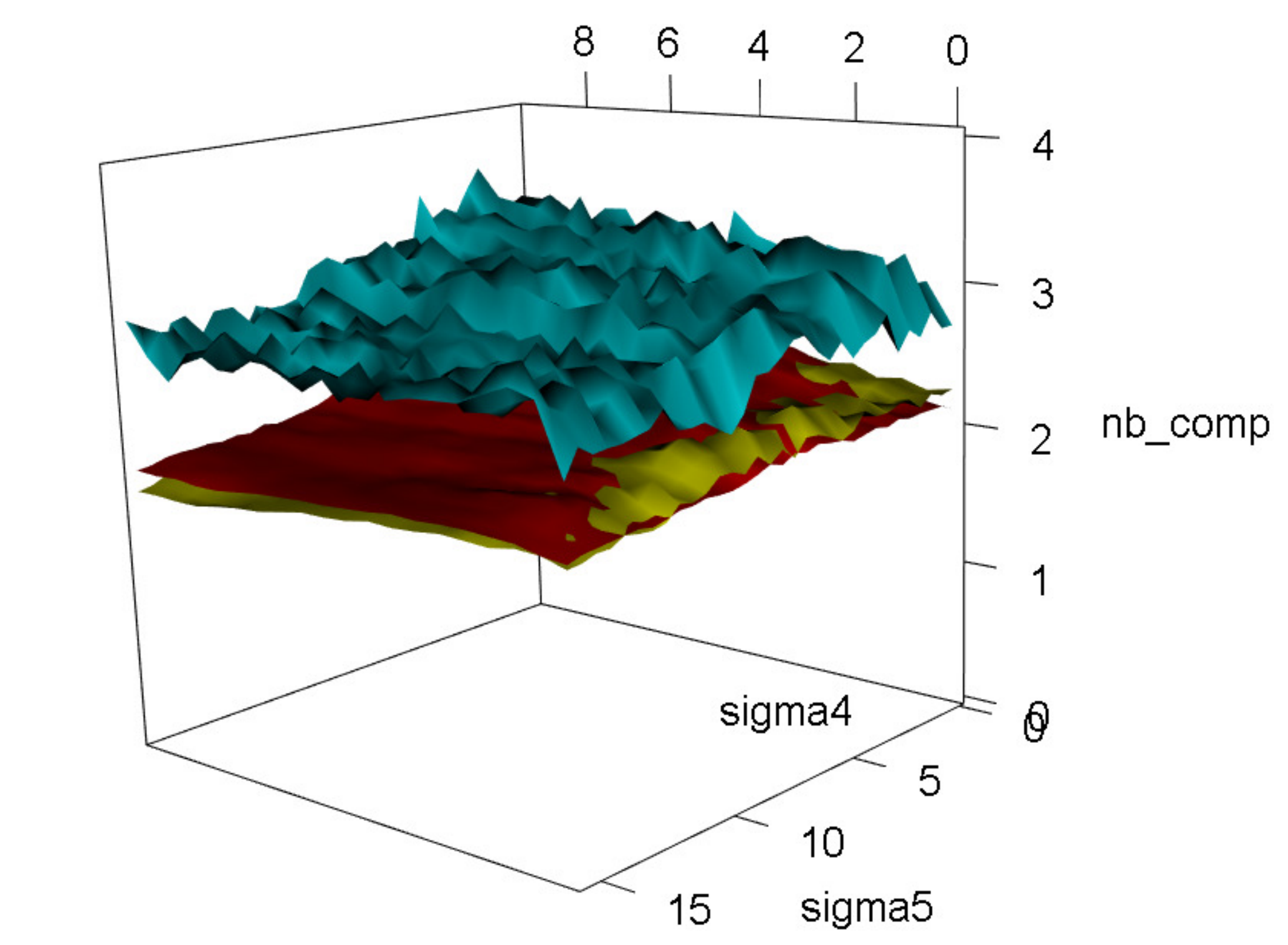}}
          \subfigure{\includegraphics[trim = -2cm 0cm 0cm 0cm, clip,scale=0.22]{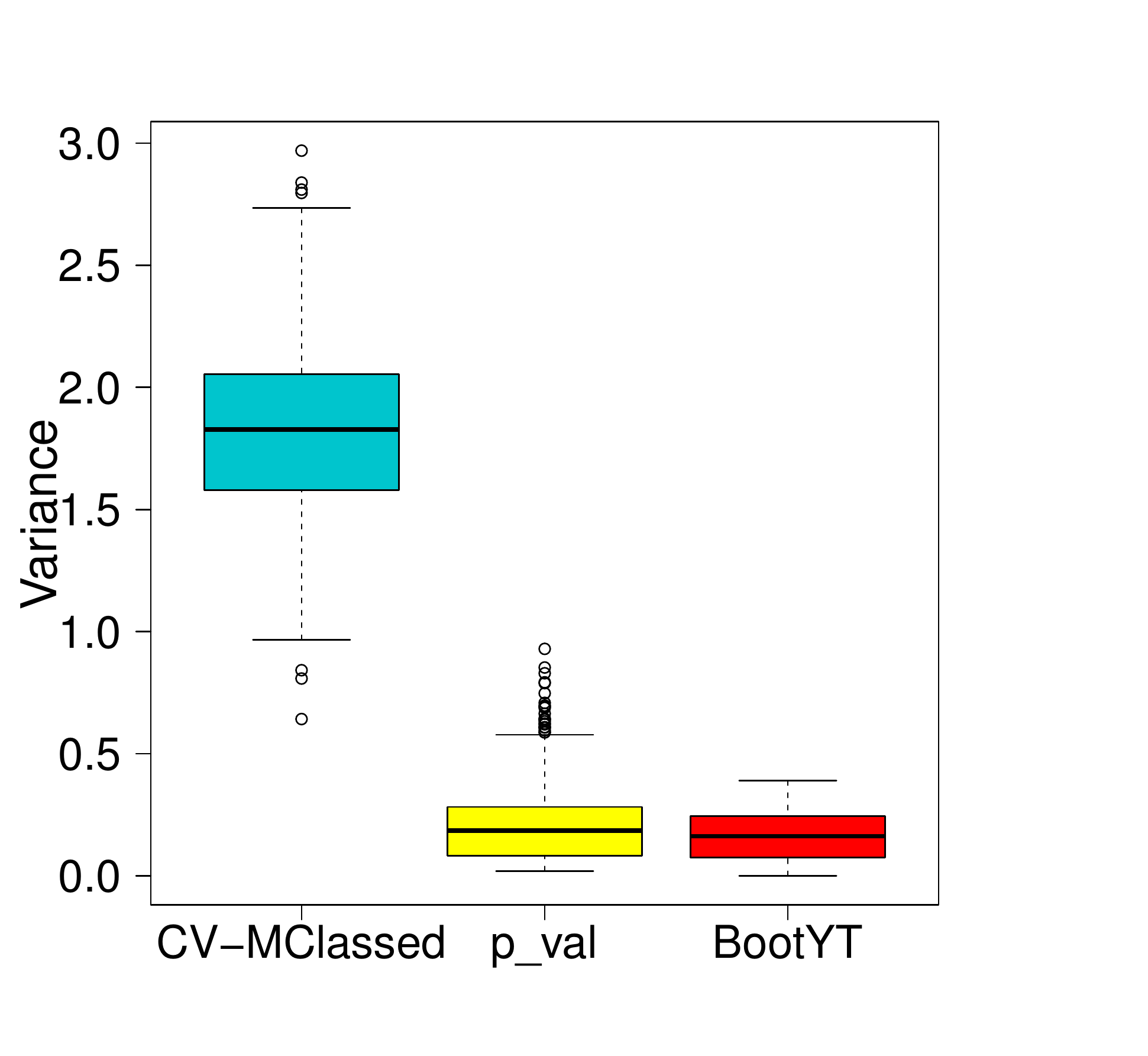}}
    \caption{$n>p$; Left: Row means surfaces; Right: Boxplots of row variances}
   \label{fig:10}
\end{figure*}

In this case, we consider $\sigma_4\in\left\{0.01,0.51,\ldots,9.51\right\}$ and $\sigma_5\in\left\{0.01,0.51,\ldots,15.51\right\}$ leading to 640 different couples $\left(\sigma_4,\sigma_5\right)$. Results are so stored in five tables of dimension $640\times 100$. We graphically report CV-MClassed, p\_val and BootYT row means as functions of $\sigma_4$ and $\sigma_5$ as well as boxplots of their row variances in Fig.\ref{fig:10}.

Based on these graphics, the CV-MClassed performs well in estimating the optimal number of components in average. However, this good property has to be nuanced by the high variances linked to its results and which lead this criterion to be used with caution. The BootYT and p\_val criteria return similar results in this $n>p$ case. Both of them slightly underestimate the optimal number of components but with the advantage of low variances of their results.  

\begin{figure*}[ht]
		\centering
          \includegraphics[trim = 0cm 0cm 0cm 0cm, clip,scale=0.2]{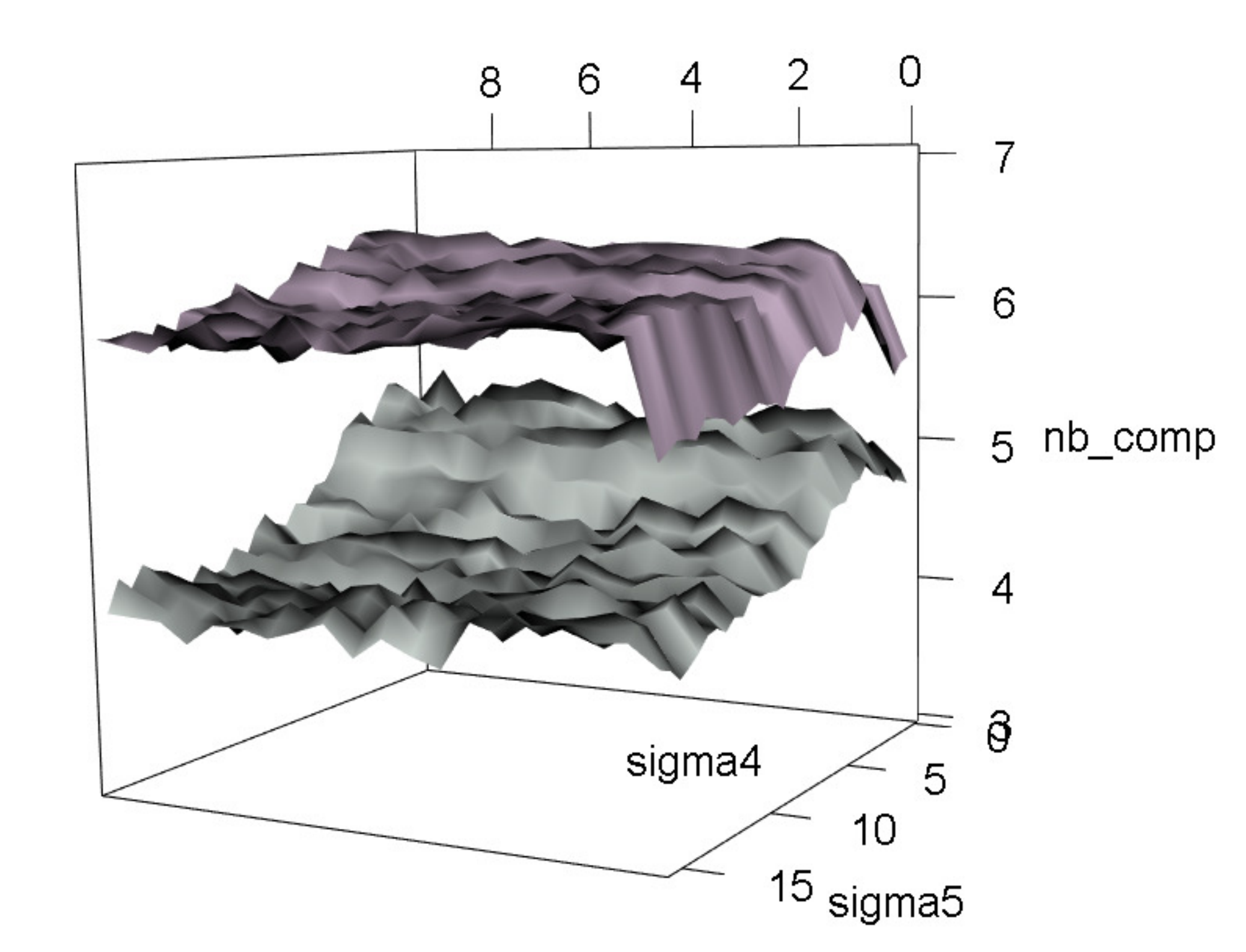}
    \caption{$n>p$; From top to bottom: AIC row means surface. BIC row means surface.}
   \label{fig:11}
\end{figure*}

We also observed results obtained with the non-corrected AIC and BIC criteria and display these results in Fig.\ref{fig:11}. 

The non-corrected \textit{dof} lead the AIC and BIC criteria to globally overestimate the number of components. Thus, these criteria have to be avoided until the development of a \textit{dof} correction in this PLSGLR framework and will not be considered in the $n<p$ case. Note that we only compared AIC and BIC values linked to the $k$-components models with $k\leqslant7$ and retained the one which realize the minimum of the studied criterion. \\

\subsubsection{PLS-LR: Case \texorpdfstring{$n<p$}{n<p}}
\label{412}

In this case, we consider $\sigma_4\in\left\{0.01,0.51,\ldots,9.51\right\}$ and $\sigma_5\in\left\{0.01,0.51,\ldots,9.51\right\}$ leading to 400 different couples $\left(\sigma_4,\sigma_5\right)$. Results are so stored in three tables of dimension $400\times 100$. We set the maximal value of $\sigma_5$ to 9.51, and not to 15.51 as for the $n>p$ case, in order to save computational runtime since an increasing value of $\sigma_5$ does not affect the choice of the number of extracted components.

\begin{figure*}[ht]
		\centering
          \subfigure{\includegraphics[trim = 0cm 0cm 0cm 0cm, clip,scale=0.22]{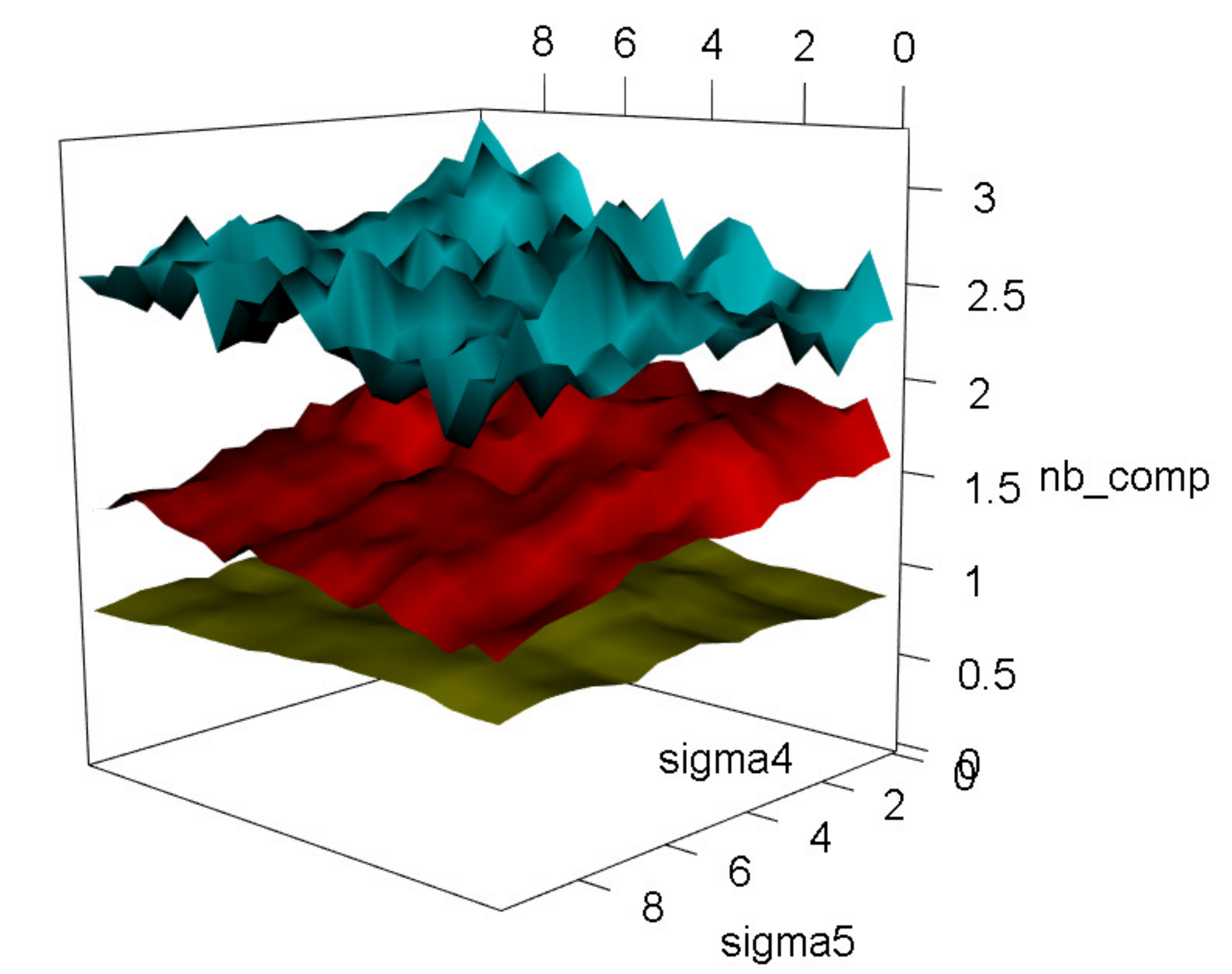}}
          \subfigure{\includegraphics[trim = 0cm 0cm 0cm 0cm, clip,scale=0.22]{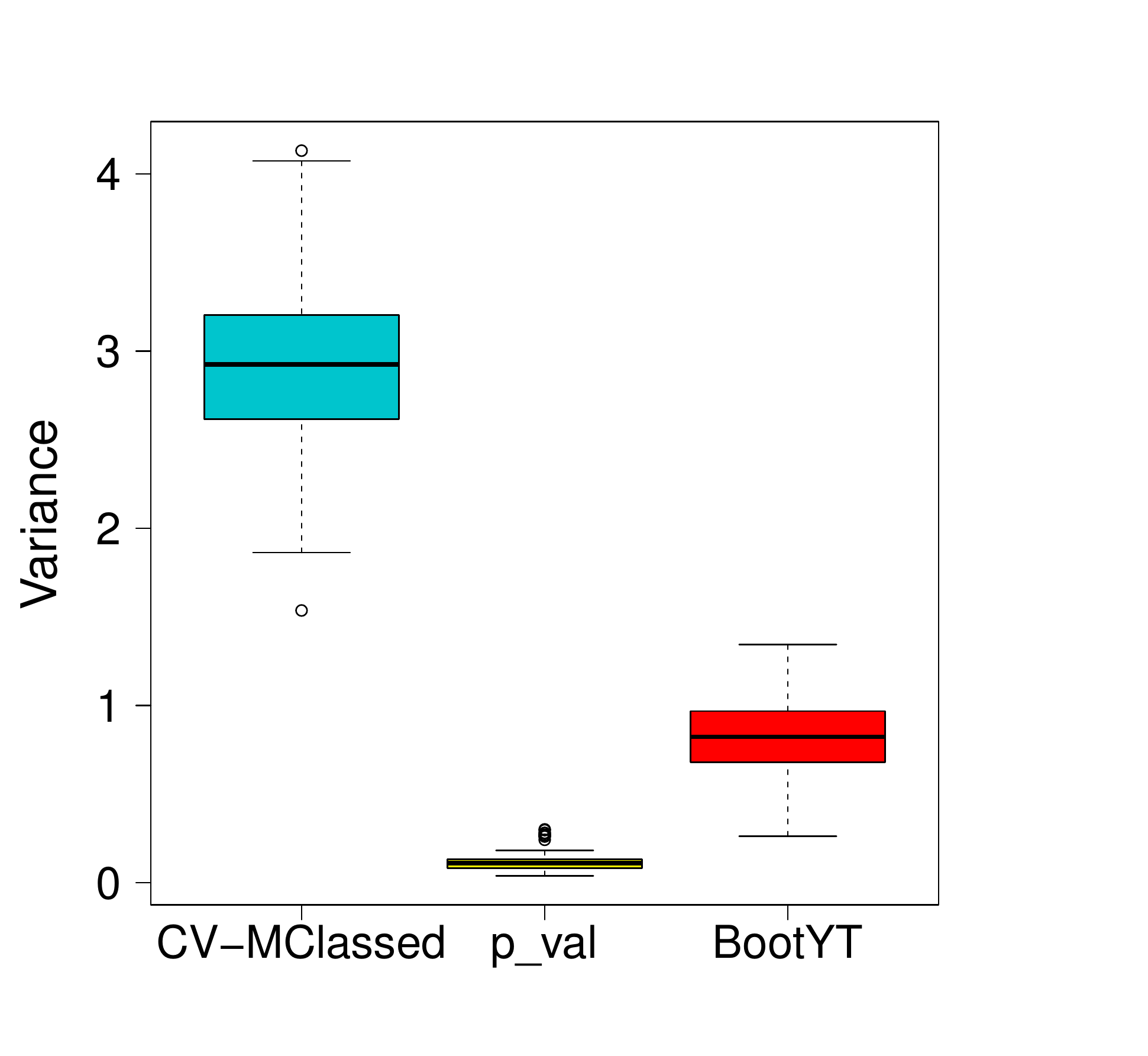}}
    \caption{$n<p$; Left: Row means surfaces; Right: Boxplots of row variances}
   \label{fig:12}
\end{figure*}

We graphically report row means as a function of $\sigma_4$ and $\sigma_5$ as well as boxplots of row variances in Fig.\ref{fig:12}. 

The CV-MClassed criterion maintains the same property of well estimating in average and issue of variability as in the $n>p$ framework. Concerning the two others criteria, we observed a higher underestimating issue linked to the p\_val criterion than for the BootYT one. Furthermore, they both had low variability in results they return.

\subsubsection{PLS-LR: MSE and miss-classed values analysis}
\label{413}

In order to test their predictive performances, we simulated 80 more observations for each simulated datasets (40 000), and computed the testing NMSE linked to each models established by the three criteria. Furthermore, since the binary response obtained by the model is equal to 1 if the estimated response is over 0.5, 0 if not, returning higher NMSE does not necessarily lead to higher number of miss-classed values. Thus, we also computed the number of predictive miss-classed values (\textit{M\_classed}) for each of these three criteria. 

\begin{figure*}[ht]
       \centering
          \subfigure{\includegraphics[trim = 0cm 0cm 0cm 0cm, clip,scale=0.23]{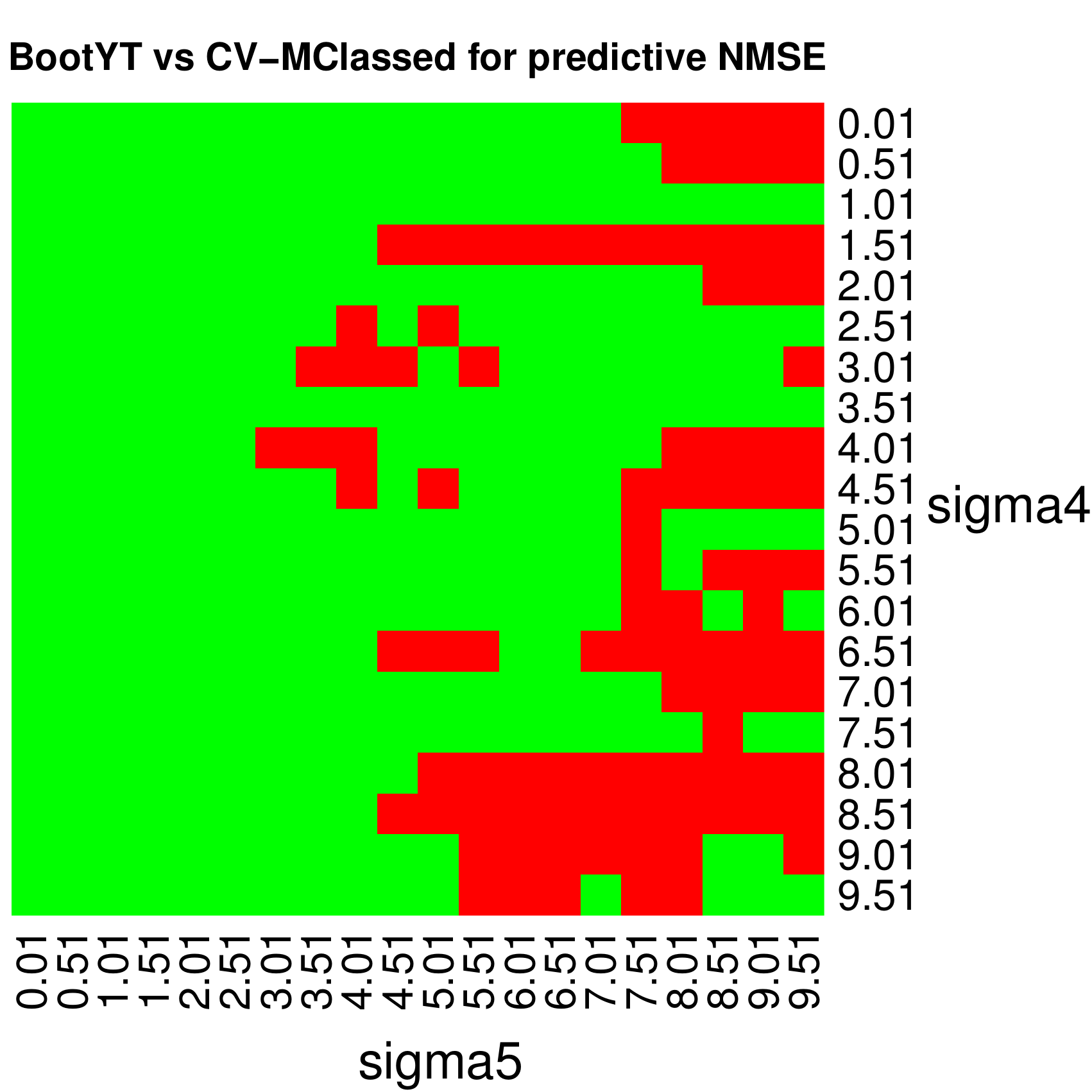}}
          \subfigure{\includegraphics[trim = 0cm 0cm 0cm 0cm, clip,scale=0.23]{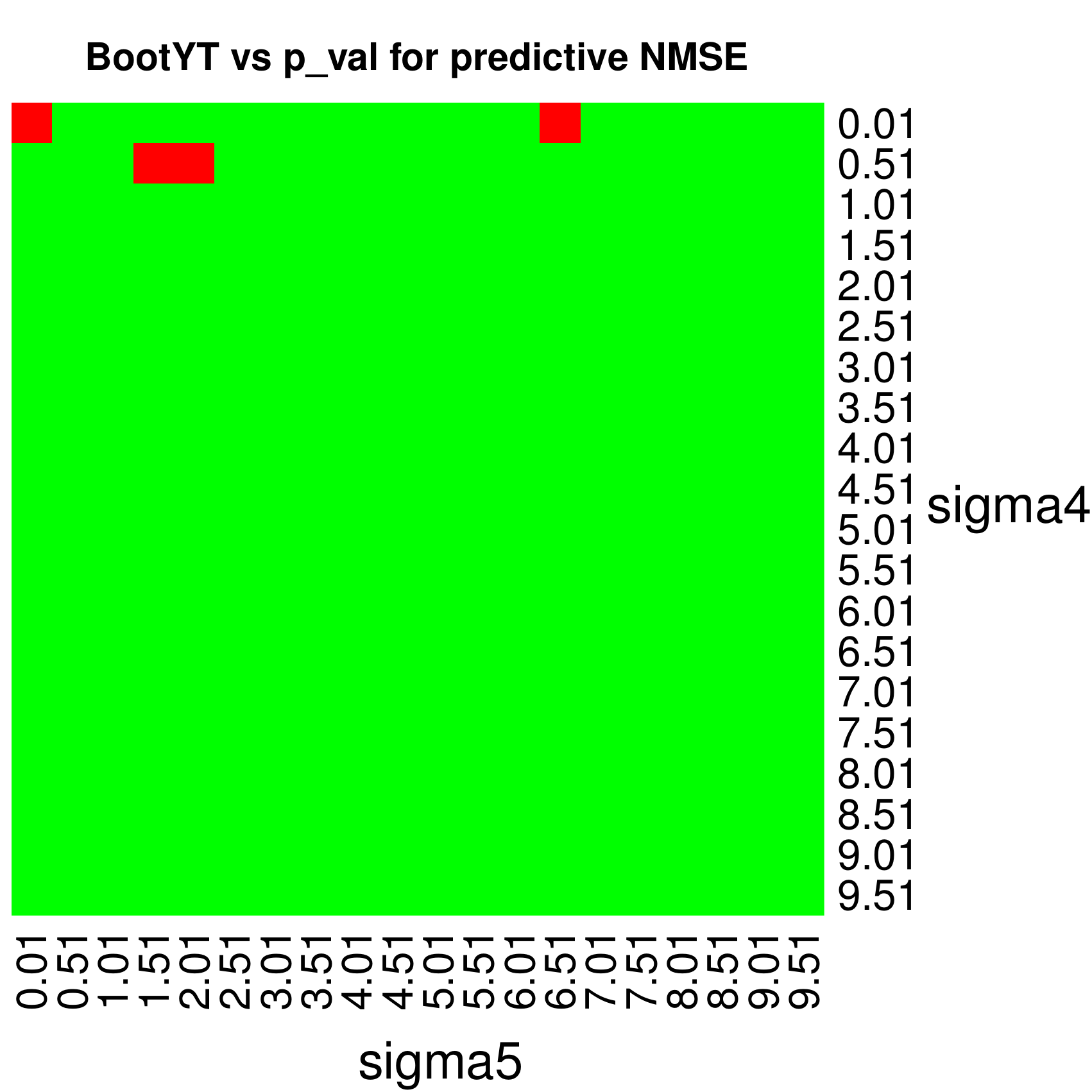}}
					\subfigure{\includegraphics[trim = 0cm 0cm 0cm 0cm, clip,scale=0.23]{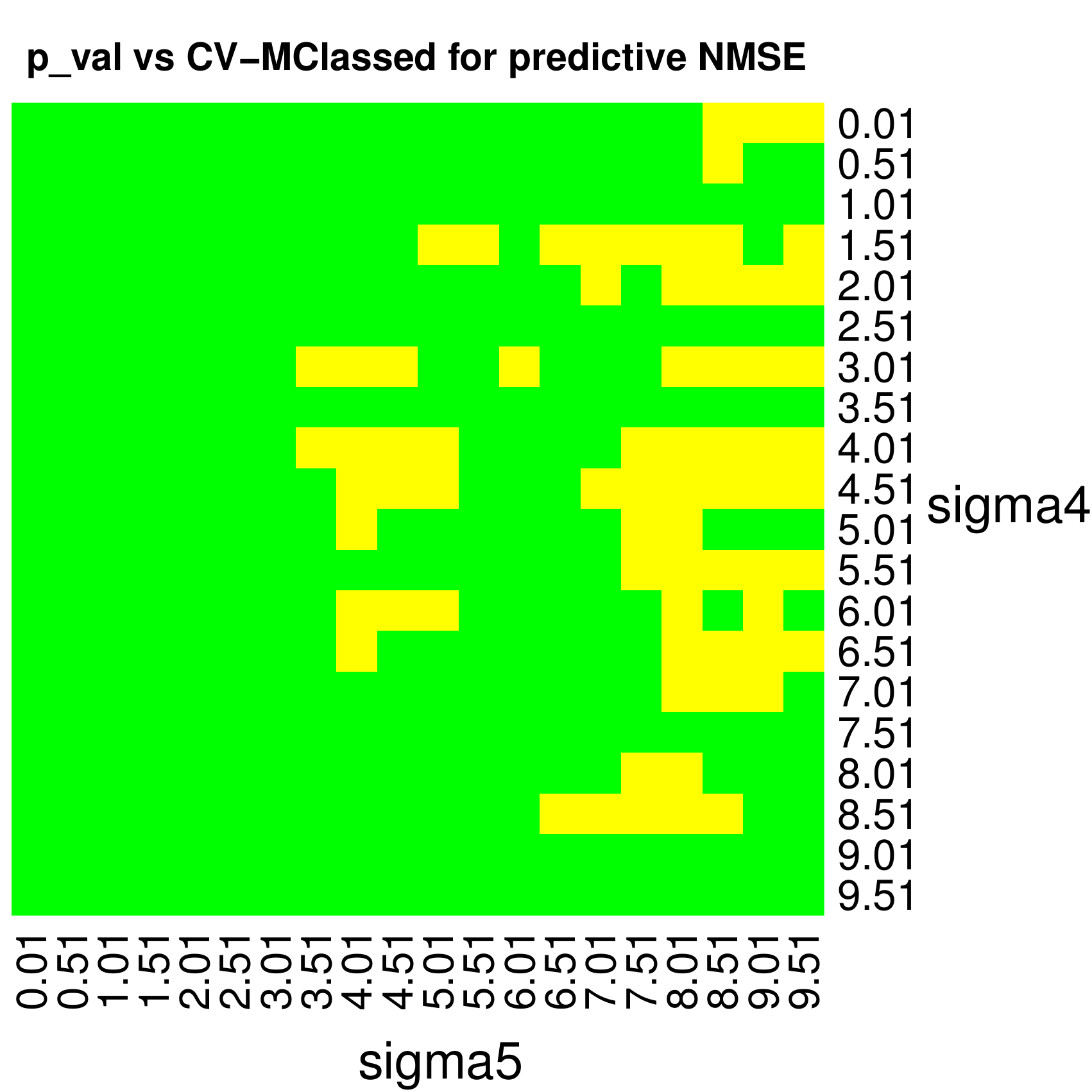}}\\
					\subfigure{\includegraphics[trim = 0cm 0cm 0cm 0cm, clip,scale=0.23]{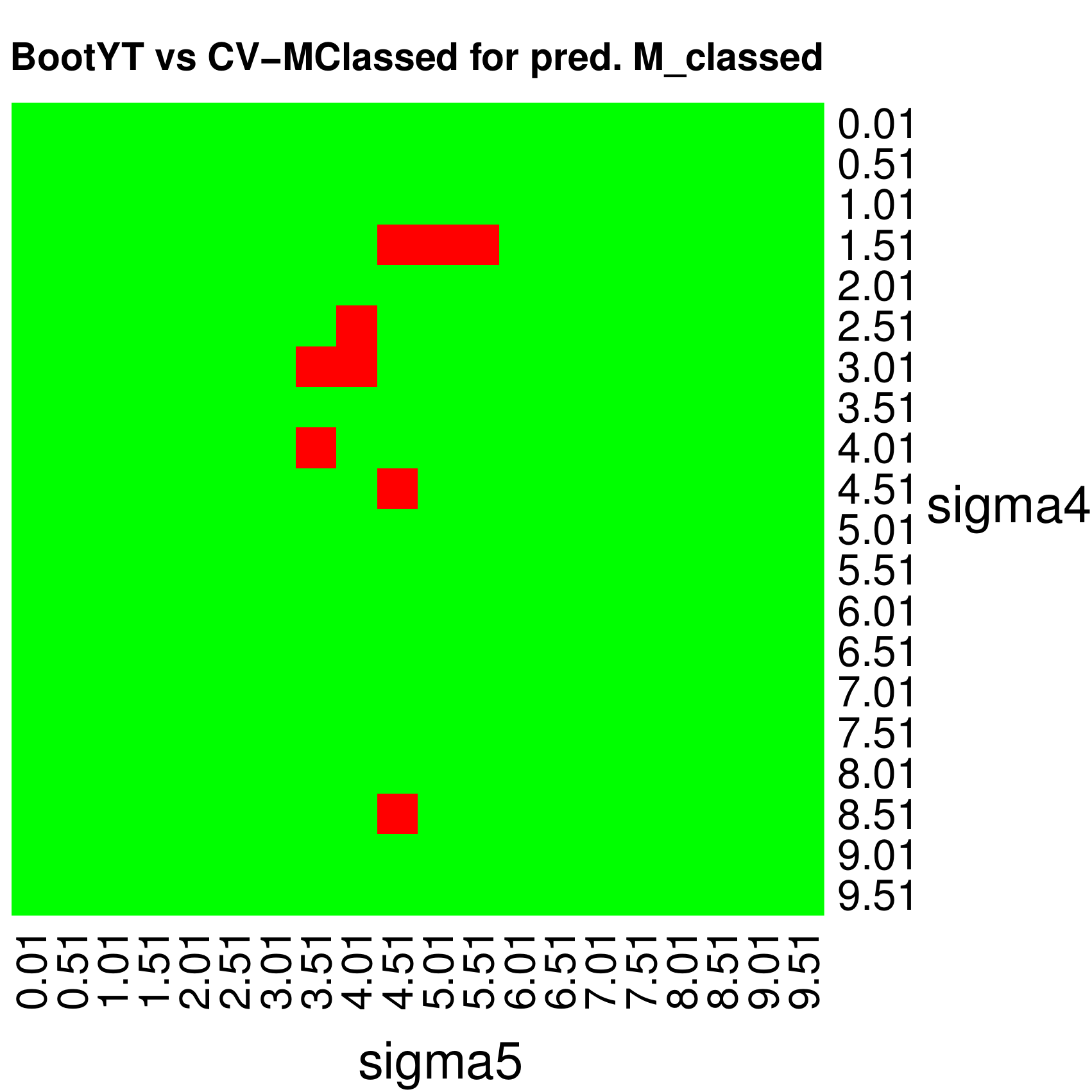}}
          \subfigure{\includegraphics[trim = 0cm 0cm 0cm 0cm, clip,scale=0.23]{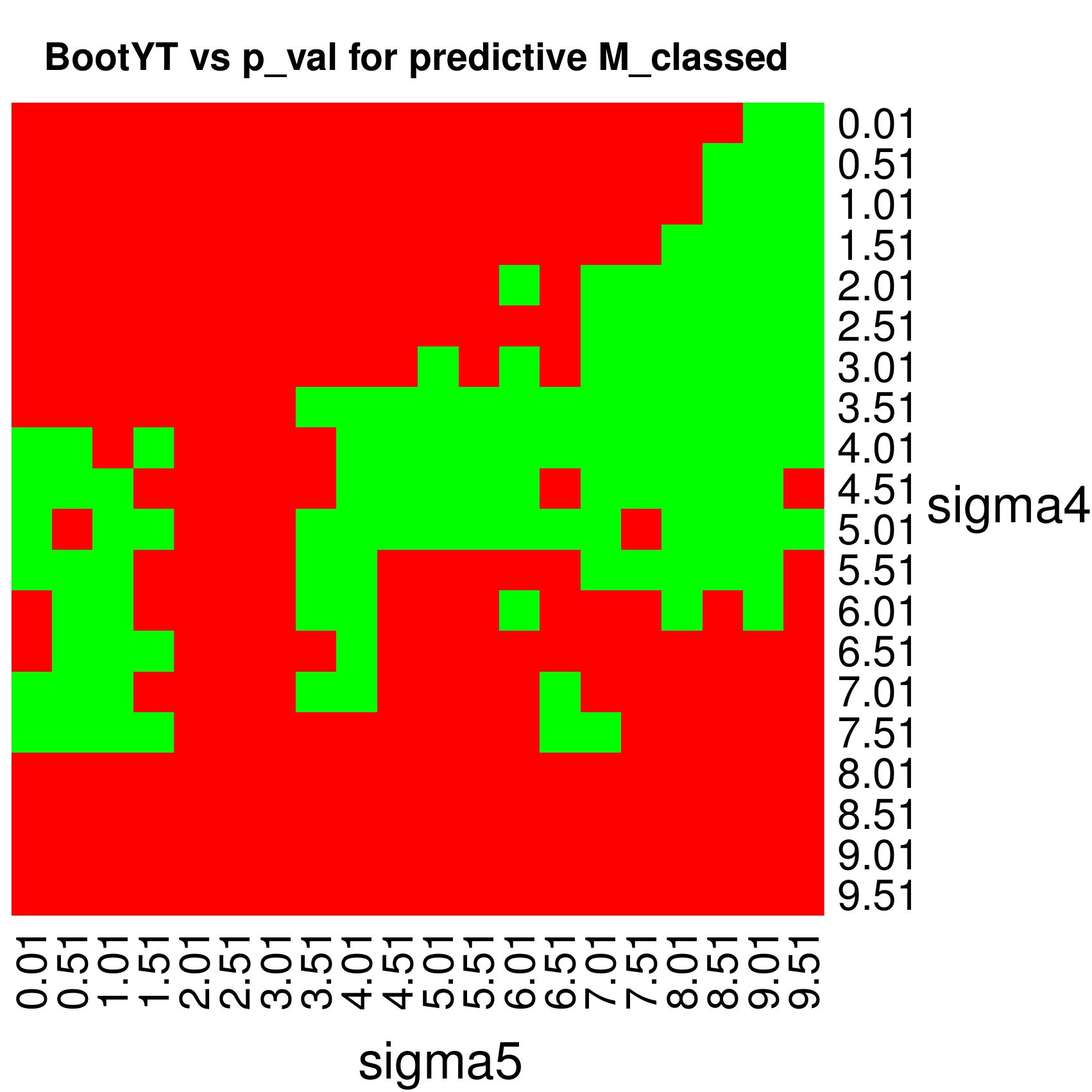}}
					\subfigure{\includegraphics[trim = 0cm 0cm 0cm 0cm, clip,scale=0.23]{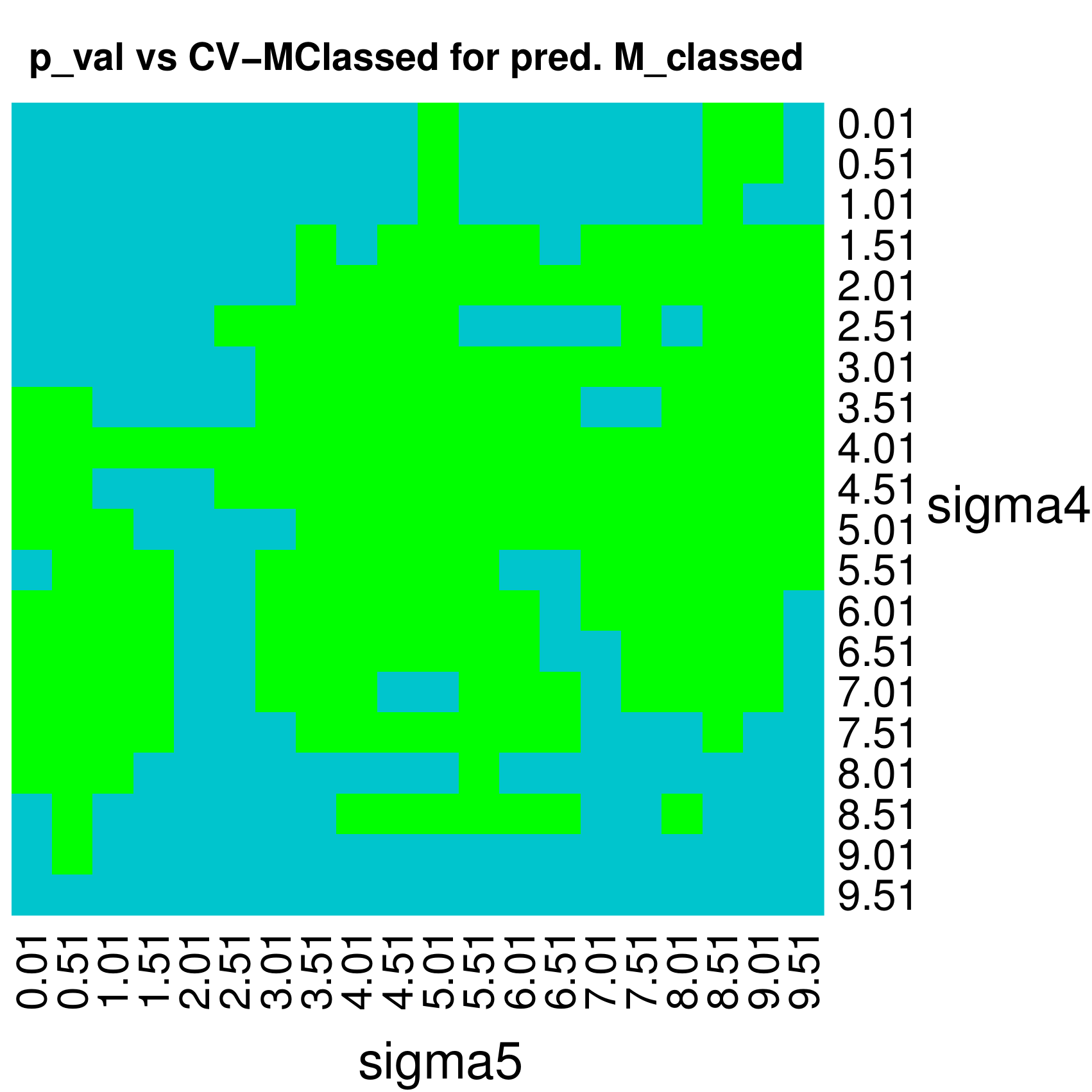}}\\
    \caption{\textit{t}-tests results: BootYT better (red), CV-MClassed better (turquoise), p\_val better (yellow), no significant difference (green)}
   \label{fig:14}
\end{figure*}

In order to compare their predictive performances, \textit{t}-tests were computed for each fixed values of $\left(\sigma_4,\sigma_5\right)$ so that we can observe precisely which criterion has better performances depending on the noise level in $\mathbf{X}$ and $\mathbf{y}$. Results of these tests are graphically reported in Fig.\ref{fig:14}.

The bootstrap-based criterion is never less efficient than the others criteria. If there is globally no significant differences between bootstrapping pairs and the p\_val criterion concerning the predictive NMSE, BootYT is better than this criterion concerning the predictive number of miss-classed values. Then, there is only few cases where bootstrapping pairs is significantly better than the CV-MClassed criterion concerning the predictive number of miss-classed values. But, concerning the predictive NMSE, the BootYT criterion is better than this last one by returning significant smallest NMSE values, especially for high values of $\sigma_5$.

The boostrap-based criterion is also the best one by having, at least, similar predictive performances compared to the two others.

\subsubsection{PLS-LR: Conclusion}
\label{414}

Through these simulations, we can reasonably assume that the bootstrap-based criterion is globally more efficient than the others ones. In the $n>p$ case, it offers a similar stability compared to the p\_val criterion. However, it globally underestimates the optimal number of components when the CV-MClassed criterion retains it on average but with high variabilities. Concerning the $n<p$ case, the BootYT criterion has better predictive performances than the two others studied criteria in terms of predictive NMSE and predictive miss-classed values. It also keeps a quite low variability, which is really important for a future routine implementation. Finally, the AIC and BIC are clearly not adapted since the corrected \textit{dof} are not established yet. 

\subsection{PLS-PR results}
\label{42}
\subsubsection{PLS-PR: Row means analysis}
\label{422}

Results were stored in four tables of dimension $440\times 100$ in the $n>p$ case and $360\times 100$ in the $n<p$ framework. Each of the 240 first rows correspond, in both cases, to results for a fixed couples of values $\left(\sigma_4,\sigma_5\right)$, with $\sigma_4\in\left\{0.01,0.51,\ldots,9.51\right\}$ and $\sigma_5\in\left\{0.01,0.21,\ldots,2.21\right\}$. They so correspond to results linked to simulations with a noise parameter $\sigma_5$ globally lower than the available information standard deviation we approached and which is so approximatively equal to 1.727. 

In the $n>p$ case, the 200 last rows correspond to results for fixed couples of values $\left(\sigma_4,\sigma_5\right)$, with $\sigma_4\in\left\{0.01,0.51,\ldots,9.51\right\}$ and  $\sigma_5\in\left\{2.51,3.01,\ldots,7.01\right\}$. Concerning the $n<p$ case, the last 120 rows correspond to results for fixed couples $\left(\sigma_4,\sigma_5\right)$, with  $\sigma_4\in\left\{0.01,0.51,\ldots,9.51\right\}$ and $\sigma_5\in\left\{2.51,3.01,\ldots,5.01\right\}$.

\begin{figure*}[ht]
       \centering
          \subfigure{\includegraphics[trim = 0cm 0cm 0cm 0cm, clip,scale=0.2]{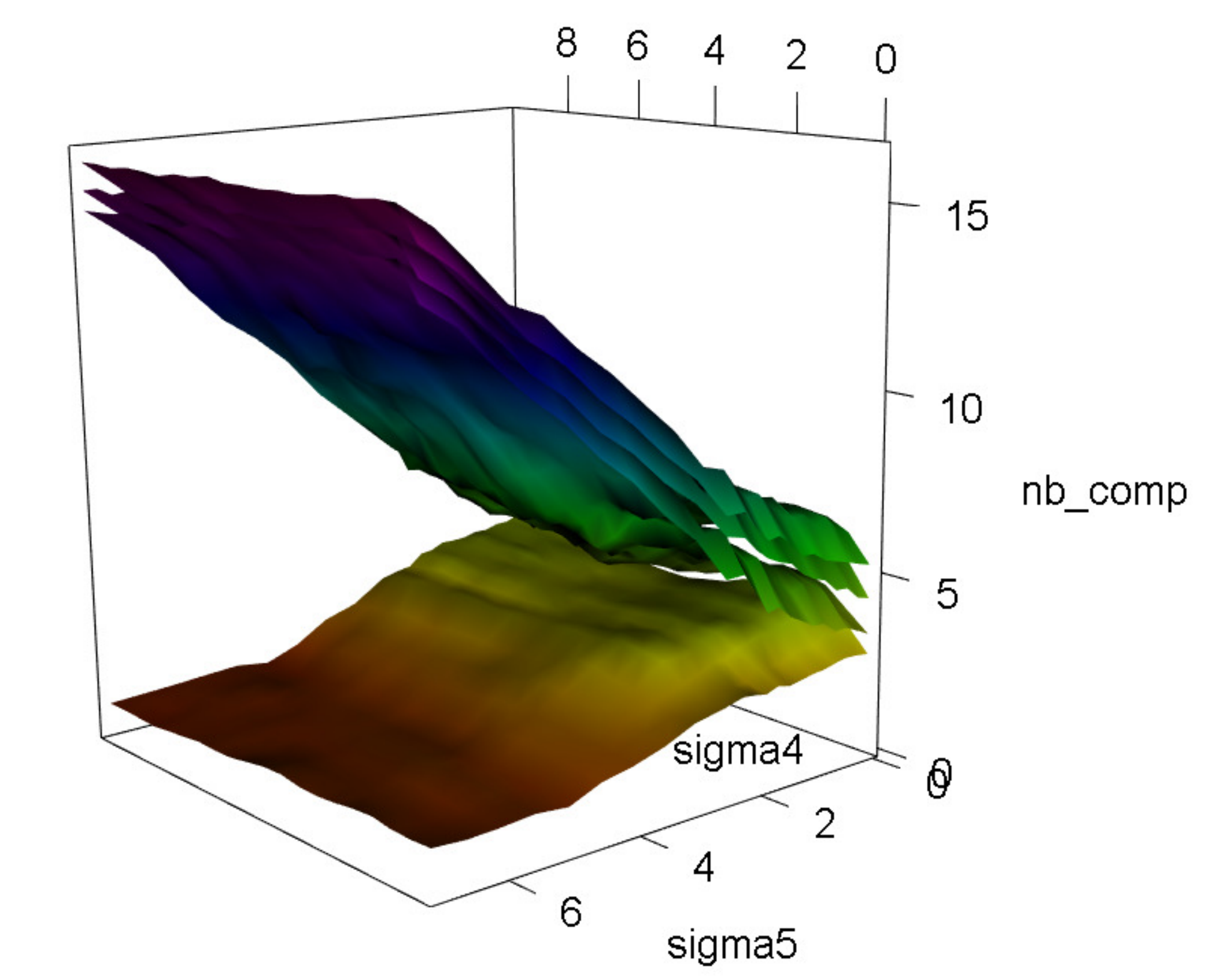}}
          \subfigure{\includegraphics[trim = 0cm 0cm 0cm 0cm, clip,scale=0.2]{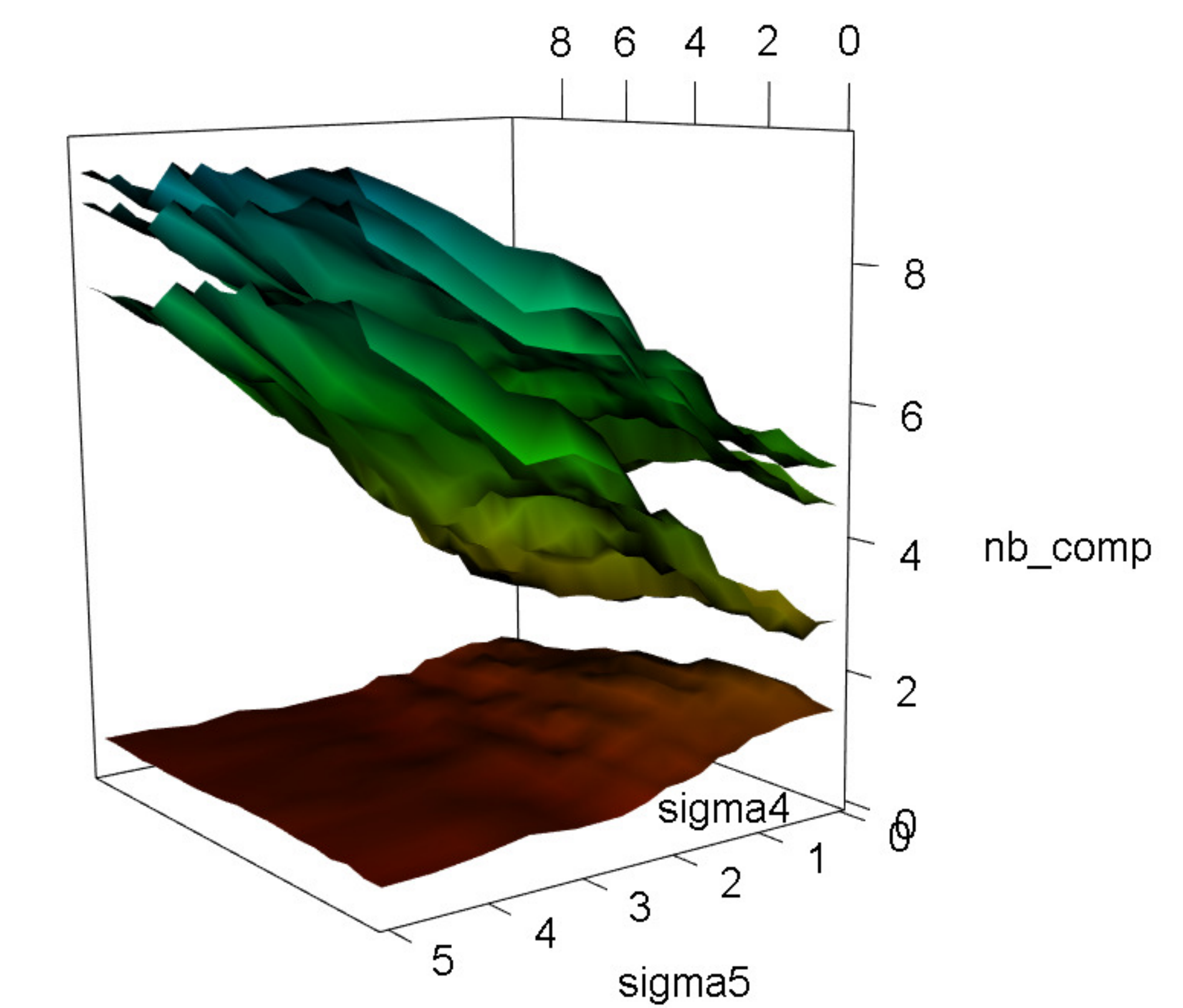}}
				\caption{From top to bottom: AIC, BIC, p\_val and BootYT row means surfaces; Left: $n>p$ case; Right: $n<p$ case}
   \label{fig:16}
\end{figure*}

AIC, BIC, p\_val and BootYT row means are displayed as functions of $\sigma_4$ and $\sigma_5$ in Fig.\ref{fig:16}.

Except for the bootstrap-based criterion, all the criteria return an increasing number of components as $\sigma_5$ increases. These results lead us to conclude that our new bootstrap-based stopping criterion is the only one which is adapted to a Poisson distribution. Based on these graphical representations, no additional analysis of these numbers of components was done. Moreover, we decided to only compare predictive performances of both p\_val and BootYT criteria.

\subsubsection{PLS-PR: MSE analysis}
\label{423}

\begin{figure*}[ht]
		\centering
          \subfigure{\includegraphics[trim = 0cm 0.5cm 0cm 2cm, clip,scale=0.23]{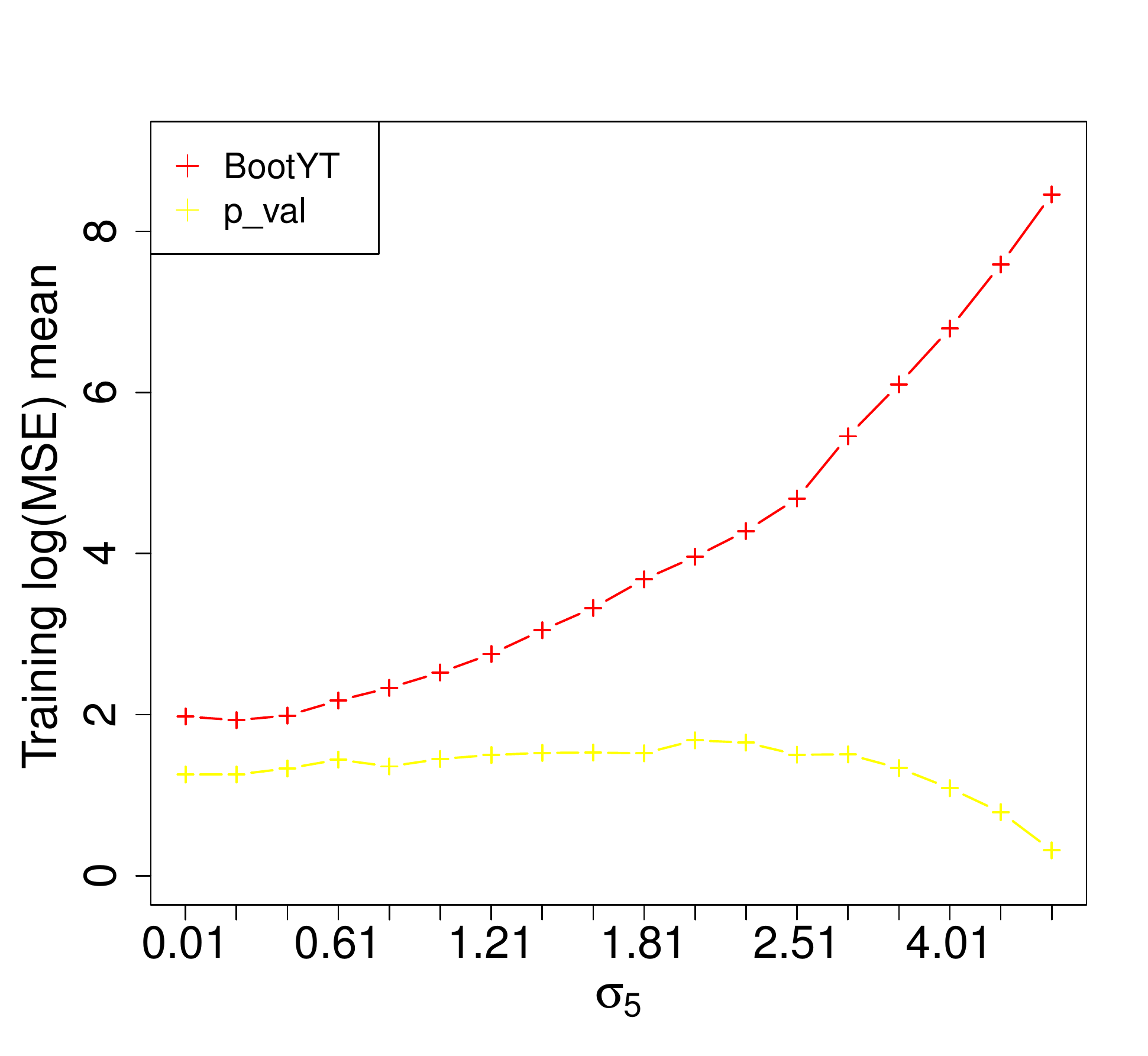}}
					\subfigure{\includegraphics[trim = -6.5cm 0.5cm 0cm 2cm, clip, scale=0.23]{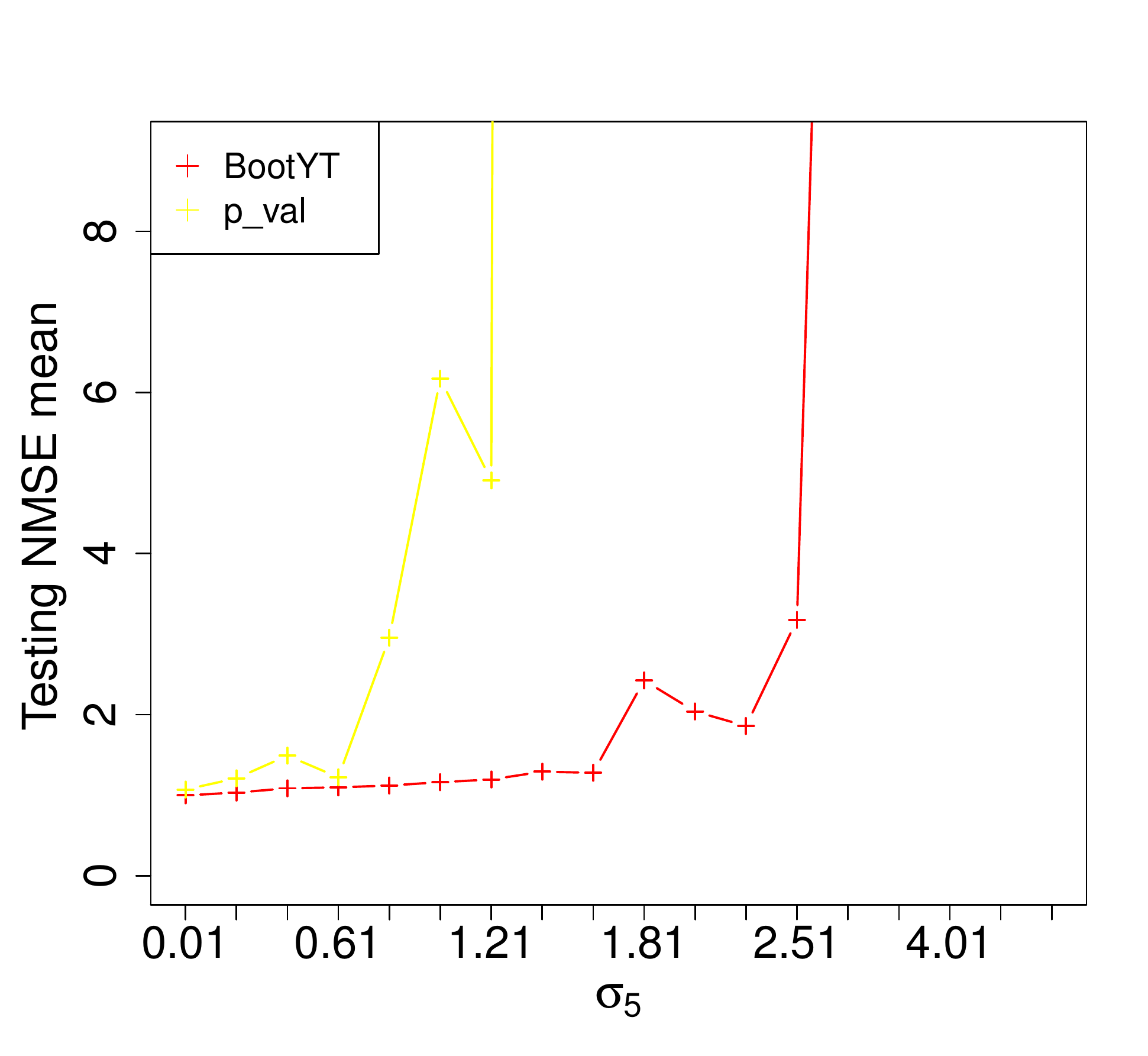}}
    \caption{$n<p$; Left: Evolution of training log(MSE) means; Right: Evolution of testing NMSE means}
   \label{fig:17}
\end{figure*}

We extracted the training log(MSE) as well as the testing NMSE, based on 80 supplementary simulated data, in the $n<p$ case and displayed their means for each value of $\sigma_5$ in Fig.\ref{fig:17}.

The global decrease of the log(MSE) linked to the p\_val models confirms that this criterion will model the random noise in $\mathbf{y}$. On the contrary, the bootstrap based criterion returns a regular increase of the log(MSE), which empirically proves that it better succeeds in separating the real common information from the random noise. The overfitting issue linked to the p\_val criterion has some major consequences on its predictive ability and lead so to higher NMSE than the BootYT ones. Furthermore, the BootYT criterion is more robust than the p\_val one to the increasing noise level in $\mathbf{y}$. It returns consistent testing NMSE for datasets linked to value of $\sigma_5$ lower than the common information standard deviation, \textit{i.e} $\sigma_5<1.727$, and even returns quite reasonable testing NMSE means up to  $\sigma_5=2.51$, while the p\_val criterion returns out of range testing NMSE for $\sigma_5\geqslant 1.01$.
	
\begin{figure}[ht]
       \centering
          \includegraphics[trim = 0cm 0.5cm 0cm 2cm, clip,scale=0.23]{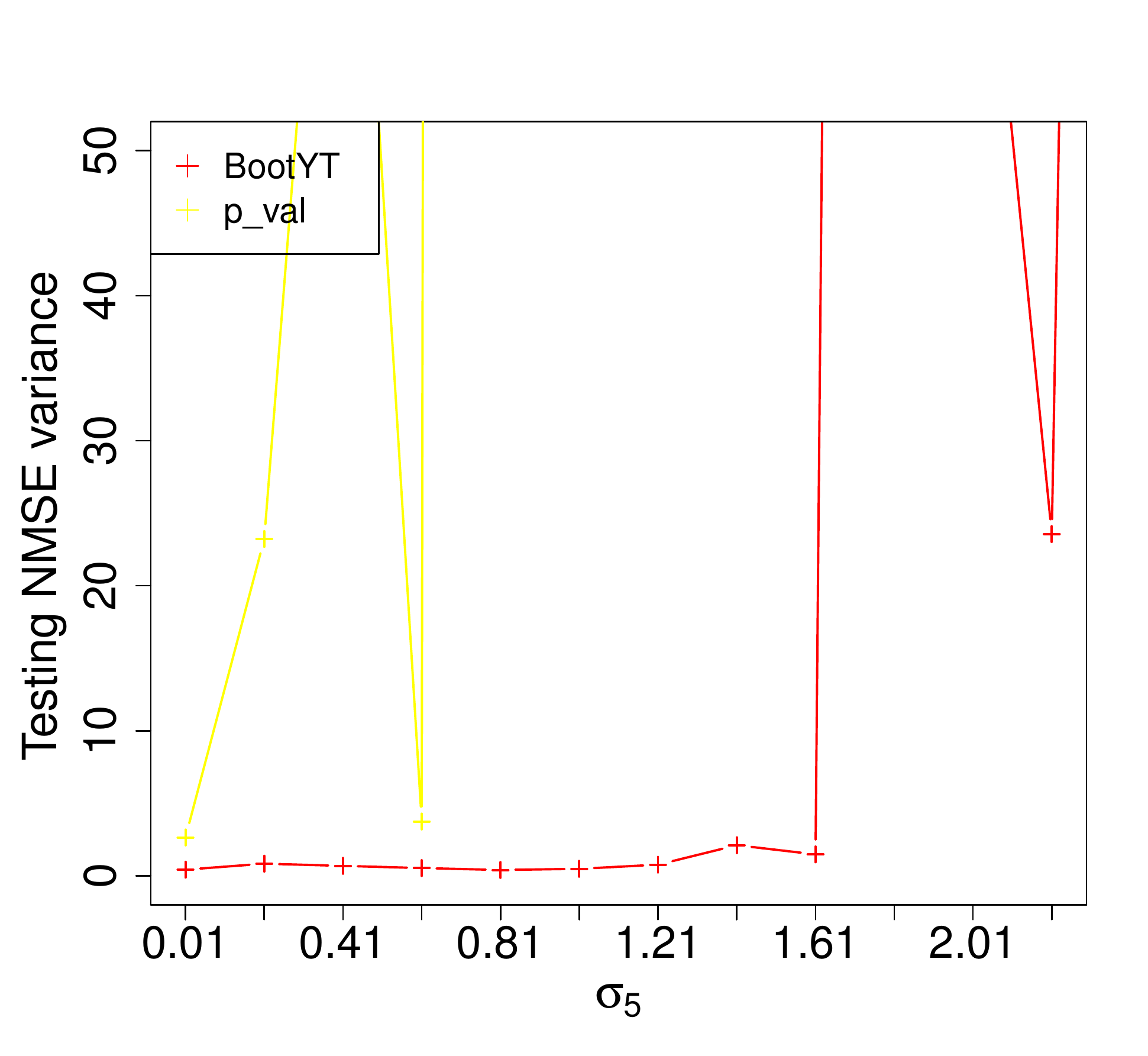}
    \caption{$n<p$; Evolution of testing NMSE variances}
   \label{fig:19}
\end{figure}

We then displayed the evolution of testing NMSE variances in Fig.\ref{fig:19}.
		 
Results obtained by the bootstrap based criterion are linked to acceptable variances while $\sigma_5<1.727$, so that we can postulate that while the random noise standard deviation is lower than the real common information one, this new criterion returns stable results. To the contrary, the p\_val models are linked to high variability of their testing NMSE. This fact is due to their over-complexities we described in part \ref{422}.

However, these out-of-range variances lead also to non-significant means differences while using \textit{t}-tests on these datasets although means differences are notable, based on results displayed in Fig.\ref{fig:17}.\\ 

\begin{figure*}[ht]
       \centering
          \subfigure{\includegraphics[trim = 0cm 0.5cm 0cm 2cm, clip,scale=0.22]{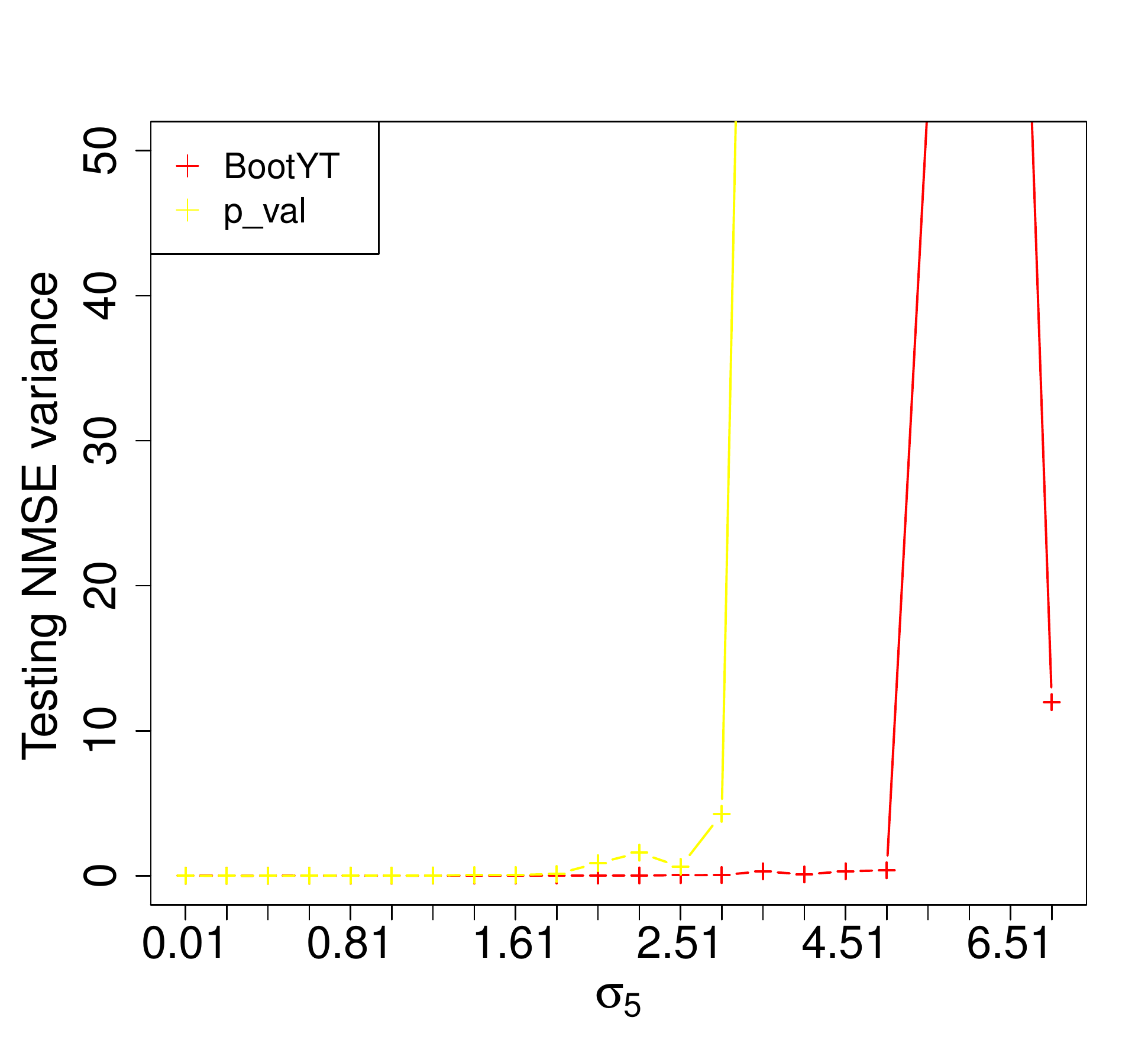}}
          \subfigure{\includegraphics[trim = 0cm 0.5cm 0cm 2cm, clip,scale=0.22]{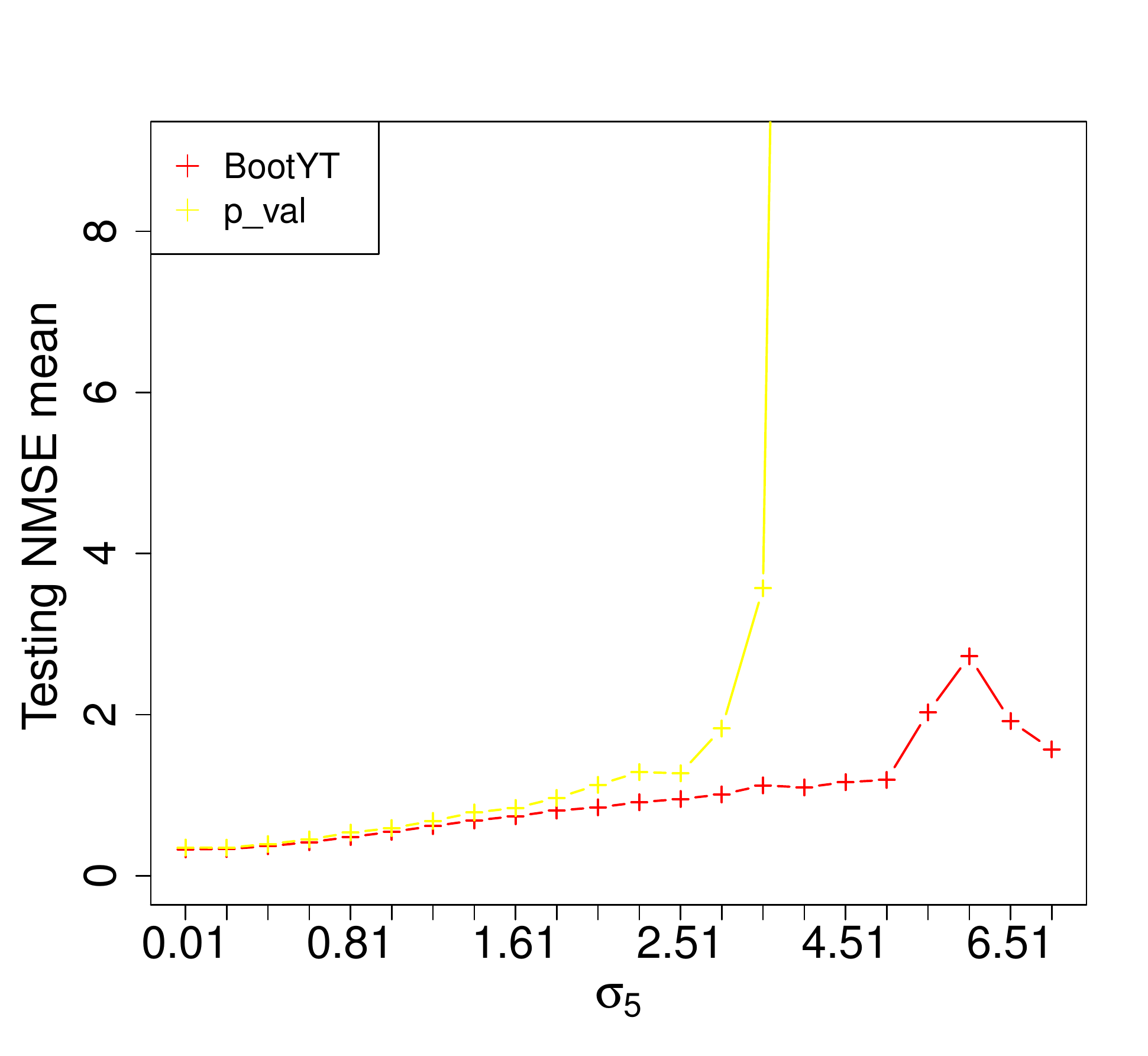}}
					\subfigure{\includegraphics[trim = 0cm 0.5cm 0cm 2cm, clip, scale=0.22]{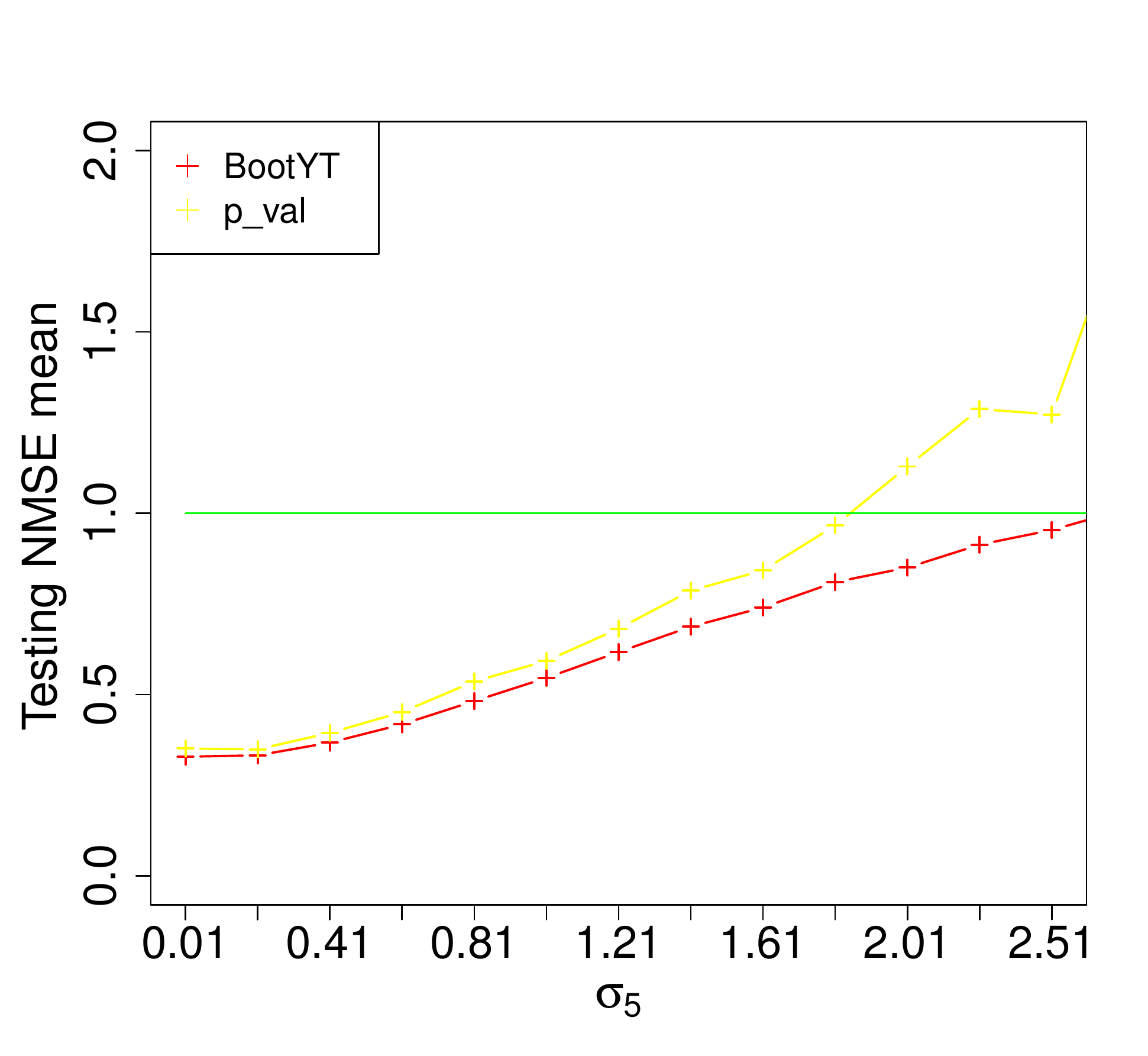}}
    \caption{$n>p$; Left: Evolution of testing NMSE variances; Center: Evolution of testing NMSE means on all data; Right: Evolution of testing NMSE means for $\sigma_5\leqslant 2.51$}
   \label{fig:20}
\end{figure*}

To obtain consistent \textit{t}-tests outcomes to compare the predictive ability of these criteria, we had to use results we obtained in the $n>p$ case. We so simulated 100 supplementaries subjects as testing sets. The evolution of means and variances of testing NMSE for both bootstrap based criterion and p\_val criterion are reported in Fig.\ref{fig:20}.

\begin{figure}[h]
		\centering
    \includegraphics[scale=0.23]{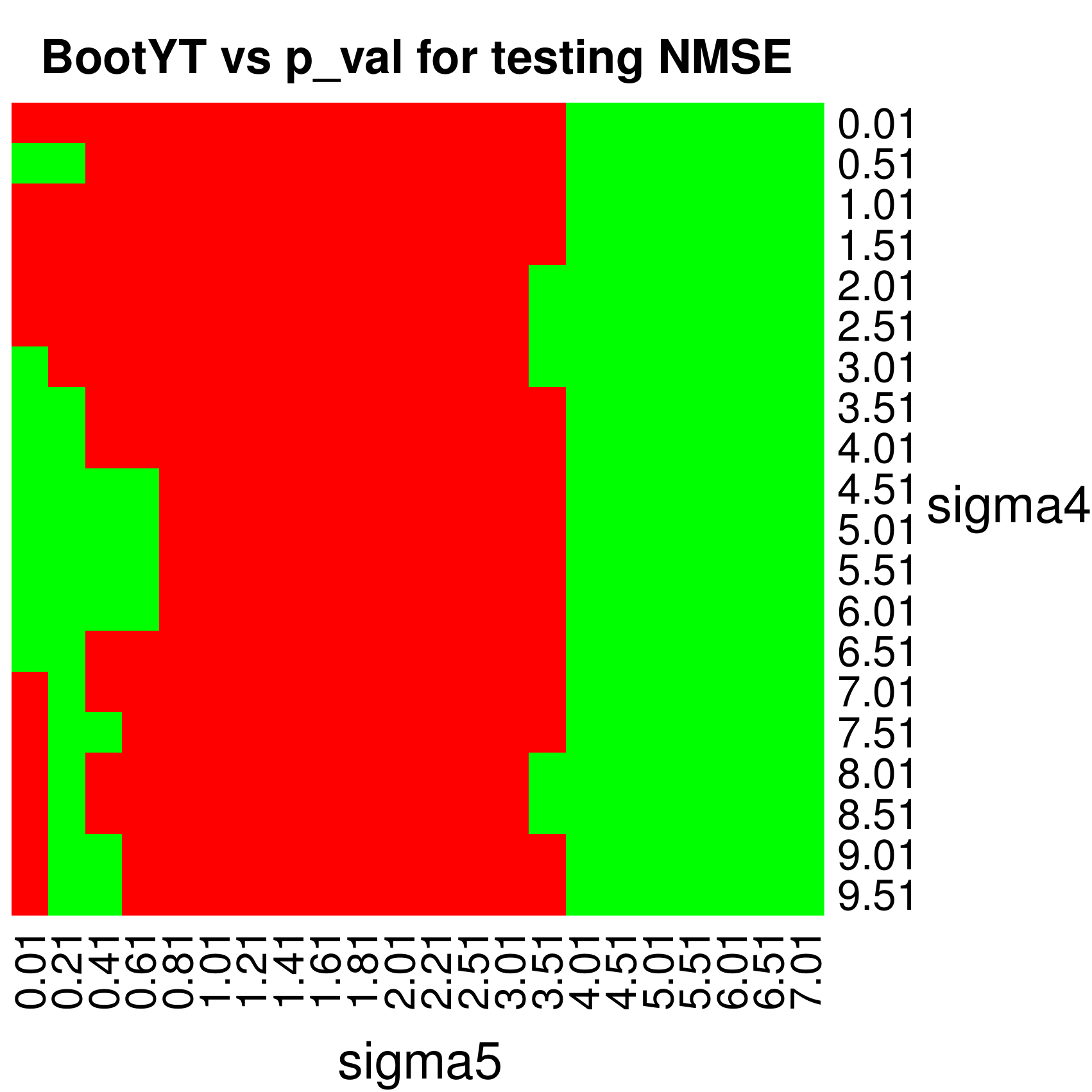}
		\caption{\textit{t}-tests results: BootYT better (red), no significant difference (green)}
	\label{fig:21}
\end{figure}

Based on these graphics, we have to notice that models build up with the bootstrap-based criterion are on average better than the trivial ones for $\sigma_5\leqslant 2.51$ while p\_val models become worse than these trivial ones for $\sigma_5> 1.81$ (Fig.\ref{fig:20}). Thus, p\_val fails to construct better models than the trivial ones when the noise level in $\mathbf{y}$ is higher than the common information standard deviation. Then, both criteria return low variances of their testing NMSE for $\sigma_5\leqslant 3.01$ so that \textit{t}-tests return consistent outcomes on this range of values. Results of these \textit{t}-tests are displayed in Fig.\ref{fig:21}.

Based on these \textit{t}-tests results, we conclude that to set up a predictive model on a dataset with significant random noise in it, our new boostrap based criterion is the one which should be advised. Note that non-significant differences for $\sigma_5\geqslant 3.51$ are due to the high increase of variances linked to the p\_val results (see Fig.\ref{fig:20}).

\subsubsection{PLS-PR: Conclusion}
\label{424}

In the case of response vector $\mathbf{y}$ linked to a Poisson distribution with a significant random noise in it, the bootstrap based criterion stands out as the only one which should be used. Indeed, others could be interpreted as increasing functions of $\sigma_5$, so that they will model the random noise in $\mathbf{y}$, leading to overfitting issues. As a direct consequence, they return models with poor predictive abilities compared to the ones our new criterion build up. 

\section{Application to a real dataset}
\label{5}

\begin{figure*}[h]
		\centering
		\subfigure{\includegraphics[trim = 0cm 0cm 0cm 0cm, clip,scale=0.23]{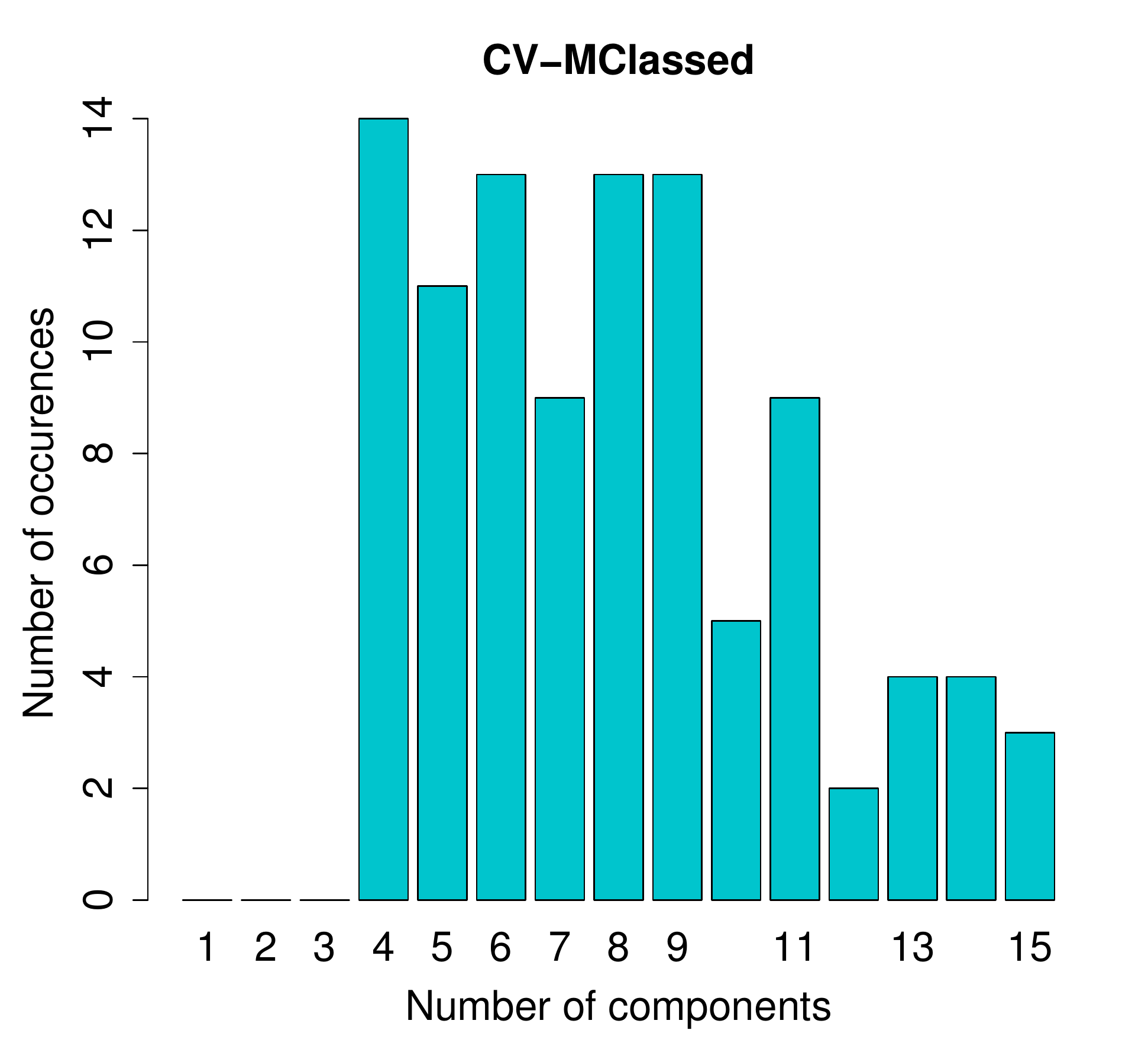}}
    \subfigure{\includegraphics[trim = -4cm 0cm 0cm 0cm, clip,scale=0.23]{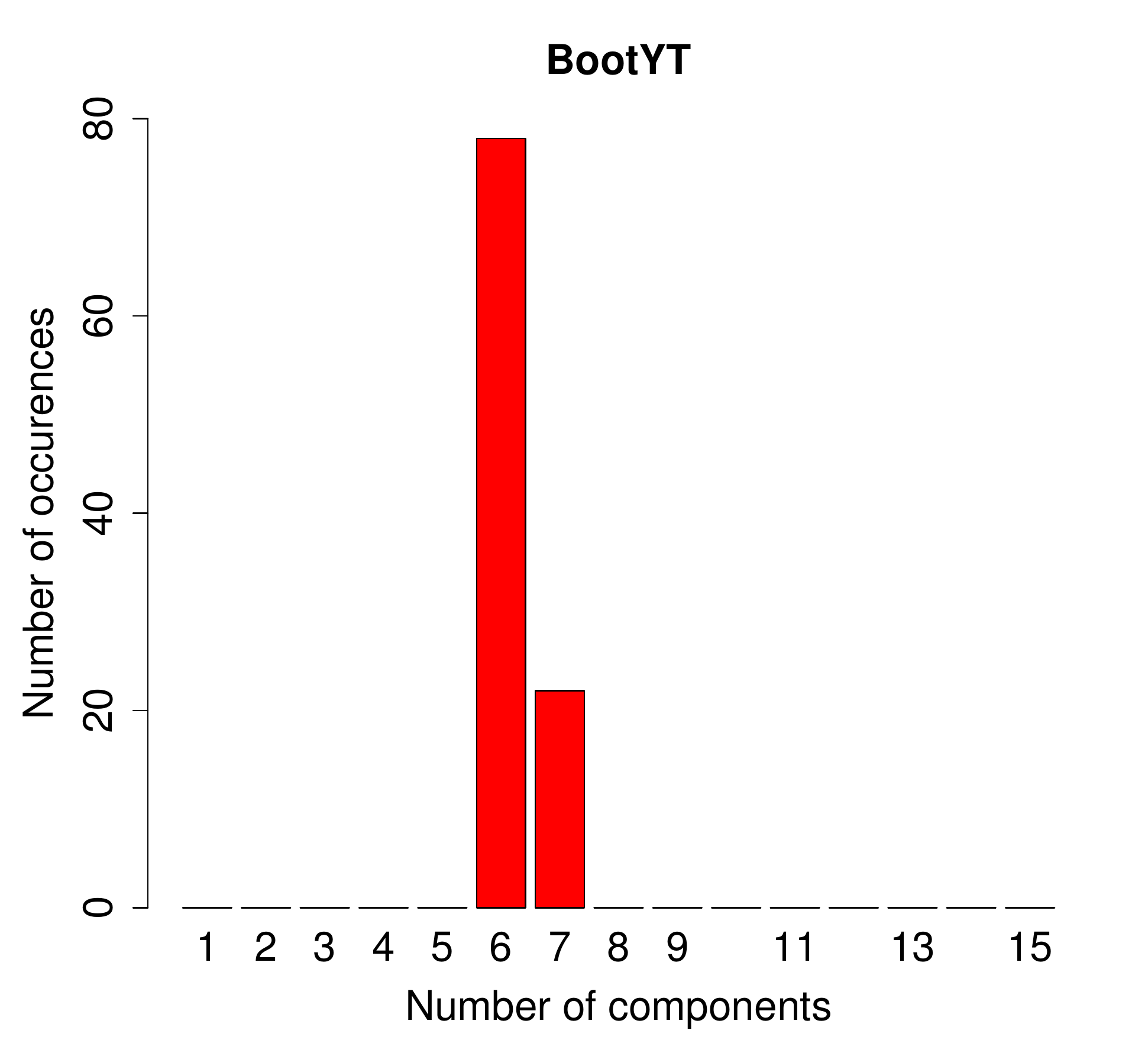}}
		\caption{Distribution of the extracted number of components with the $q=5$ CV-MClassed criterion (Left) and BootYT (Right)}
	\label{fig:22}
\end{figure*}

In this section, we focus on a allelotyping study. Our method is applied on a dataset that concerns 267 subjects with colon cancer. Measures were done on 33 microsatellites in search of an allelic imbalance that indicates an abnormal number of allele copies of a nearby gene of interest. The aim of the study was to find the microsatellites subsets that would best discriminate left and right colon tumors. Thus, the univariate response corresponds to the original localisation of a colon tumor, leading to a binary response $\mathbf{y}$, taking value 0 (respectively 1) if the localisation was on the right colon (respectively on the left). More details on it are available in \citet{weber2007allelotyping}. \\
This dataset contains missing values, so that we did a preprocessing in order to complete it by using the $R$ package \textit{mice}. As $\mathbf{y}$ is a 0-1 response, we used the three following stopping criteria in components construction: our new bootstrap-based criterion, the CV-MClassed and the p\_val ones. The corresponding extracted numbers of components are respectively equal to 6, 5 and 4. The bootstrap-based criterion selects more components than the CV-MClassed one, which does not match to what we observed in the simulations (see Fig.\ref{fig:10}).\\
In fact, results displayed in Fig.\ref{fig:10} are obtained by computing the means of extracted number of components obtained on the 100 simulated datasets for each fixed couple $\left(\sigma_4,\sigma_5\right)$. As we mentioned in part \ref{41}, results based on the CV-MClassed criterion have a high variability. To take this variability into account, we computed the CV-MClassed criterion 100 times on our dataset, leading to the distribution of the extracted number of components reported in Fig.\ref{fig:22}. Then, by computing the mean of the 100 values of extracted numbers of components, we obtained 7.99, a result similar to the simulation ones. Thus, our simulation results are validated on this real case study. We also perform the same process using our new bootstrap based criterion and report the results in Fig.\ref{fig:22}.

Based on distributions in Fig.\ref{fig:22}, the major default of the CV-MClassed criterion, namely the dependency of the extracted number of components to the way the group will be randomly formed, is clearly showed out. Thus, performing a single CV to find the number of components using this criterion must be avoided. As expected, the BootYT criterion returns stable results and conclude, in almost 80\% of cases, in 6 components. 

\begin{figure*}[ht]
       \centering
          \subfigure{\includegraphics[trim = 0cm 0cm 0cm 0cm, clip,scale=0.275]{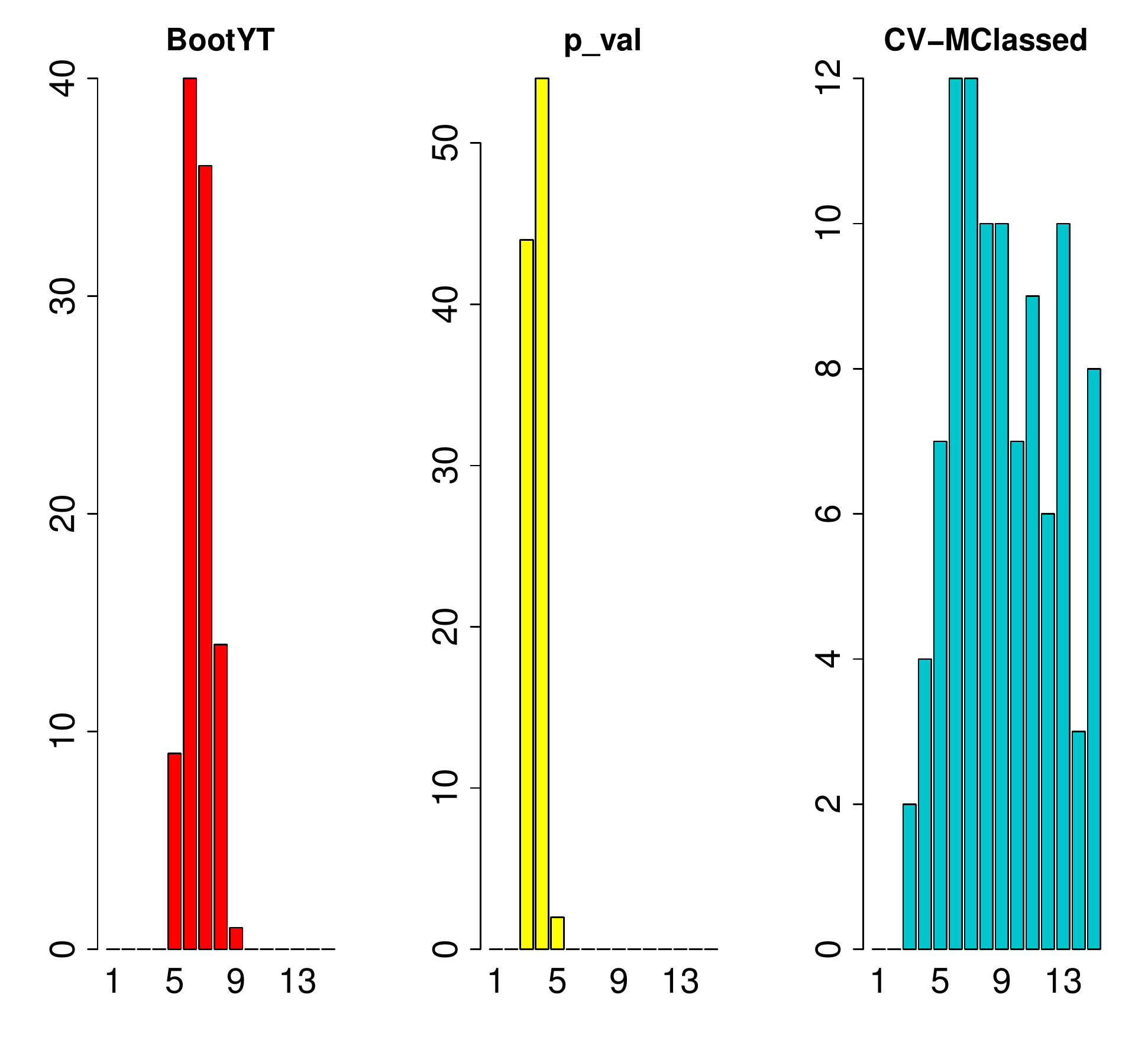}}
          \subfigure{\includegraphics[trim = -4cm 0cm 0cm 0cm, clip,scale=0.275]{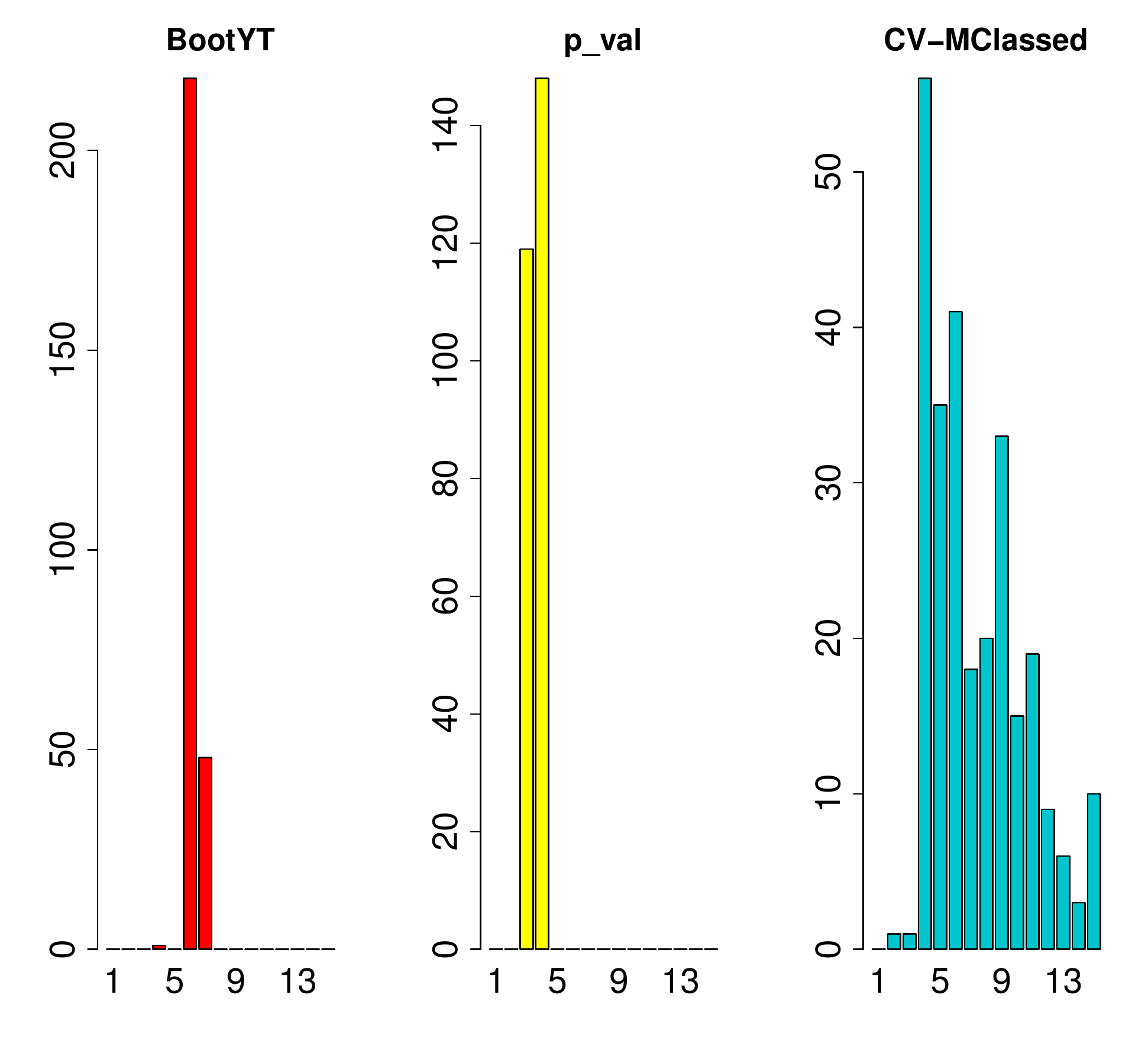}}
    \caption{Distribution of extracted number of components through bootstrap (Left) and jackknife (Right) resampling processes}
   \label{fig:23}
\end{figure*}

We also test the robustness of these three criteria through a bootstrap resampling process, with 100 boostrap iterations, and a jackknife one. These two resampling methods leads to distributions of the extracted number of components linked to each of the three compared criteria. Results are graphically reported in Fig.\ref{fig:23}.

These results confirm the high resampling robustness of our new bootstrap-based criterion compared to the CV-MClassed one. The p\_val criterion owns a comparable robustness but, based on our simulation results, with a higher bias. Based on these results and our simulation ones, we can reasonably conclude that for this dataset, the optimal number of components is 6. 

\section{Discussion}
\label{6}

We developed a new bootstrap based stopping criterion in PLS components construction, characterised by a high level of stability and robustness with respect to noise levels compared to others criteria usually used.

This bootstrap based criterion has a shortcoming in that its computational runtime is higher than the others since it requires $\left(k\times\left(\left(p_l+1\right)\times R\right)\right)$ least squares regressions per dataset. A first improvement has already be done by developing a parallel processing for this criterion. Furthermore, the development of corrected \textit{dof} in PLSGLR framework would also permit to develop a corrected BIC formulation in this framework. This corrected BIC could provide an interesting alternative to the boostrap-based criterion since it could save an important computational runtime conditionally to the fact that it would have, at least, similar properties to those we conclude in part \ref{33}. 

However, this new bootstrap-based criterion represents a reliable, consistent and robust stopping criterion in order to select the optimal number of \textit{PLS} components. It can also be performed both for PLSR and PLSGLR frameworks and allows users to test the significance of a new component with a preset risk level $\alpha$.

In the $n>p$ PLSR framework, our simulations confirm the BICdof as being an appropriate and well designed criterion. However, our new bootstrap-based criterion is an appropriate alternative in the $n<p$ case, since the BICdof criterion suffers from  high variances and overestimating issues, especially for models with low random noise levels in $\mathbf{y}$. Furthermore, both BICdof and Q2K5 criteria are more sensitive than the bootstrap-based criterion to the increasing noise level in $\mathbf{y}$ in this case.

Concerning the PLSGLR framework, our simulations results based on two specific distributions (Binomial and Poisson) lead to recommend this new bootstrap-based criterion. Indeed, in the PLS-LR case, we show that depending on the statistic we used (testing NMSE or predictive number of miss-classed values) to test its predictive ability, the bootstrap-based criterion is never significantly worse than both the CV-MClassed and p\_val criteria. Concerning results we obtained for a response vector under a Poisson distribution, the bootstrap-based criterion is the only one which returns consistent results by retaining a decreasing number of components while the random noise level in $\mathbf{y}$ increases. By adding to this fact the MSE analysis and the obtained \textit{t}-tests results, we can reasonably advise to apply this new criterion in such a framework.

\appendix
\section{Proof of Proposition \ref{prop2}}

\begin{proof} Let $n=pq+r$ be the euclidian division of $n$ by $q$. Then, a partition of the dataset for CV is composed of $r$ $\left(p+1\right)$-element subsets and $\left(q-r\right)$ $p$-element subsets, noted $A_k,\:1\leqslant k\leqslant q$. We set $q_0=0$ and let $E=\left\{q_1,\ldots,q_r\right\},\:1\leqslant q_j\leqslant q,\: \forall j\in \llbracket 1,r\rrbracket$ be a subset of $\left\{1,\ldots,q\right\}$ containing the ordered set of indices of the $\left(p+1\right)$-element subsets so that:\\
\begin{center}
$Card\left(A_k\right)=\left\{
		    \begin{array}{ll}
          p+1,\; \forall k\in E\\
					p,\ \ \  \ \ \  \forall k\in E^c
			  \end{array}	
			\right.
$
\end{center}

Let us first determine the number of distinct partitions of $\left\{1,\ldots,n\right\}$ which can be written as:

\begin{equation}
\left\{A_1,\ldots,A_{{q_1}-1},A_{q_1},A_{{q_1}+1},\ldots,A_{{q_r}-1},A_{q_r},A_{{q_r}+1},\ldots,A_q\right\}
\label{part}
\end{equation}
To lighten the formulas, let us set $\omega=\left\{q_j\in E\left|q_j-q_{j-1}\ne 1\right.\right\}$ and $m_j=q_j-q_{j-1}-1$. Then, by knowing that, for any set containing $n$ elements, the number of distinct $p$-element subsets of it that can be formed is given by $\binom {n}{p}$, we obtain that the number of distinct partitions of the form \eqref{part} is equal to:
%\small
\begin{align*}
f&(n,q)=\begin{multlined}[t]
\stackrel[j=1]{r}{\prod{}}\left[\left[\mathbb{1}_{\left\{q_j\in\omega\right\}}\stackrel[i=1]{m_j}{\prod{}}\binom{n-\left(q_{j-1}-\left(j-1\right)+\left(i-1\right)\right)p-\left(j-1\right)\left(p+1\right)}{p}\right.\right.\\
+\mathbb{1}_{\left\{q_j\notin\omega\right\}}\Bigg]\times\binom{n-\left(\left(q_j-1\right)-\left(j-1\right)\right)p-\left(j-1\right)\left(p+1\right)}{p+1}\Bigg]\\
\times\left[\mathbb{1}_{\left\{q_r\ne q\right\}}\stackrel[i=1]{q-q_r}{\prod{}}\binom{n-\left(q_r-r+\left(i-1\right)\right)p-r\left(p+1\right)}{p}+\mathbb{1}_{\left\{q_r=q\right\}}\right]\\
\end{multlined}\\
&=\begin{multlined}[t]
\stackrel[j=1]{r}{\prod{}}\left[\left[\mathbb{1}_{\left\{q_j\in\omega\right\}}\stackrel[i=1]{m_j}{\prod{}}\binom{n-\left(q_{j-1}+i-1\right)p-\left(j-1\right)}{p}+\mathbb{1}_{\left\{q_j\notin\omega\right\}}\right]\right.\\
\times\binom{n-\left(q_j-1\right)p-\left(j-1\right)}{p+1}\Bigg]\\
\times\left[\mathbb{1}_{\left\{q_r\ne q\right\}}\stackrel[i=1]{q-q_r}{\prod{}}\binom{n-\left(q_r+\left(i-1\right)\right)p-r}{p}+\mathbb{1}_{\left\{q_r=q\right\}}\right]\\
\end{multlined}\\
&=\begin{multlined}[t]
\stackrel[j=1]{r}{\prod{}}\left[\left[\mathbb{1}_{\left\{q_j\in\omega\right\}}\stackrel[i=1]{m_j}{\prod{}}\frac{\left(n-\left(q_{j-1}+i-1\right)p-\left(j-1\right)\right)!}{p!\left(n-\left(q_{j-1}+\left(i+1\right)-1\right)p-\left(j-1\right)\right)!}+\mathbb{1}_{\left\{q_j\notin\omega\right\}}\right]\right.\\
\times\frac{\left(n-\left(q_j-1\right)p-\left(j-1\right)\right)!}{\left(p+1\right)!\left(n-\left(\left(q_j+1\right)-1\right)p-\left((j+1)-1\right)\right)!}\Bigg]\\
\times\left[\mathbb{1}_{\left\{q_r\ne q\right\}}\stackrel[i=1]{q-q_r}{\prod{}}\frac{\left(n-\left(q_r+\left(i-1\right)\right)p-r\right)!}{p!\left(n-\left(q_r+\left(i+1\right)-1\right)p-r\right)!}+\mathbb{1}_{\left\{q_r=q\right\}}\right]
\end{multlined}
\end{align*}

Each second factorial term in the denominator being equal to the numerator of the following product term, we obtain that:

\begin{align*}
f(n,q)&=\begin{multlined}[t]
\stackrel[j=1]{r}{\prod{}}\left[\left[\mathbb{1}_{\left\{q_j\in\omega\right\}}\frac{\left(n-q_{j-1}p-\left(j-1\right)\right)!}{\left(p!\right)^{m_j}\left(n-\left(q_{j}-1\right)p-\left(j-1\right)\right)!}+\mathbb{1}_{\left\{q_j\notin\omega\right\}}\right]\right.\\
\times\frac{\left(n-\left(q_j-1\right)p-\left(j-1\right)\right)!}{\left(p+1\right)!\left(n-\left(\left(q_j+1\right)-1\right)p-\left((j+1)-1\right)\right)!}\Bigg]\\
\times\left[\mathbb{1}_{\left\{q_r\ne q\right\}}\frac{\left(n-q_rp-r\right)!}{\left(p!\right)^{q-q_r}\left(n-qp-r\right)!}+\mathbb{1}_{\left\{q_r=q\right\}}\right]
\end{multlined}\\
&=\begin{multlined}[t]
\stackrel[j=1]{r}{\prod{}}\left[\mathbb{1}_{\left\{q_j\in\omega\right\}}\frac{\left(n-q_{j-1}p-\left(j-1\right)\right)!}{\left(p!\right)^{m_j}\left(p+1\right)!\left(n-q_jp-j\right)!}\right.\\
+\mathbb{1}_{\left\{q_j\notin\omega\right\}}\frac{\left(n-\left(q_j-1\right)p-\left(j-1\right)\right)!}{\left(p+1\right)!\left(n-q_jp-j\right)!}\Bigg]\\
\times\left[\mathbb{1}_{\left\{q_r\ne q\right\}}\frac{\left(n-q_rp-r\right)!}{\left(p!\right)^{q-q_r}}+\mathbb{1}_{\left\{q_r=q\right\}}\right]
\end{multlined}
\end{align*}

However, by definition, if $q_j\notin\omega$ then $q_j-1=q_{j-1}$ and $m_j=0$. Furthermore, if $q_r=q$ then $\frac{\left(n-q_rp-r\right)!}{\left(p!\right)^{q-q_r}}=1$, so:

\begin{align*}
f(n,q)&=\begin{multlined}[t]
\stackrel[j=1]{r}{\prod{}}\left[\mathbb{1}_{\left\{q_j\in\omega\right\}}\frac{\left(n-q_{j-1}p-\left(j-1\right)\right)!}{\left(p!\right)^{m_j}\left(p+1\right)!\left(n-q_jp-j\right)!}\right.\\
+\mathbb{1}_{\left\{q_j\notin\omega\right\}}\frac{\left(n-q_{j-1}p-\left(j-1\right)\right)!}{\left(p!\right)^{m_j}\left(p+1\right)!\left(n-q_jp-j\right)!}\Bigg]\\
\times\left[\mathbb{1}_{\left\{q_r\ne q\right\}}\frac{\left(n-q_rp-r\right)!}{\left(p!\right)^{q-q_r}}+\mathbb{1}_{\left\{q_r=q\right\}}\frac{\left(n-q_rp-r\right)!}{\left(p!\right)^{q-q_r}}\right]\end{multlined}\\
&=\stackrel[j=1]{r}{\prod{}}\left[\frac{\left(n-q_{j-1}p-\left(j-1\right)\right)!}{\left(p!\right)^{m_j}\left(p+1\right)!\left(n-q_jp-j\right)!}\right]\times\frac{\left(n-q_rp-r\right)!}{\left(p!\right)^{q-q_r}}\\
&=\left(\frac{1}{p!}\right)^{\stackrel[j=1]{r}{\sum{}}m_j+q-q_r}\times\left(\frac{1}{\left(p+1\right)!}\right)^{r}\times n!\\
&=\left(\frac{1}{p!}\right)^{qr}\times\left(\frac{1}{\left(p+1\right)!}\right)^{r}\times n!
\end{align*}

Finally, since the order of parts creation within each of the two classes of subsets ((p+1) and p-element subsets) is not useful in distinguishing one partition from another in a CV framework, we had to divide our result by the number of distinct permutations it exists in each class, so:
\begin{equation}
f(n,q)=\frac{n!}{r!\left(q-r\right)!}\times\left(\frac{1}{\left(p+1\right)!}\right)^r\times\left(\frac{1}{p!}\right)^{q-r}
\end{equation}
This result is therefore not dependant on $E$.
\end{proof}

\section*{Acknowledgements}

The authors gratefully acknowledge the Labex IRMIA for J. Magnanensi's PhD grant.



  \bibliographystyle{elsarticle-harv} 
  \bibliography{biblio}


%\begin{thebibliography}{00}



%\end{thebibliography}
\end{document}